\documentclass{appolb}
\usepackage{epsfig,axodraw}
\usepackage{amssymb}

\usepackage{graphicx}
\usepackage{bm}
\def\be{\begin{eqnarray}}
\def\ee{\end{eqnarray}}
\def\xperp{{\bf x_{\perp}}}
\def\xpperp{{\bf x'_{\perp}}}
\def\ppperp{{\bf p'_{\perp}}}
\def\pperp{{\bf p_{\perp}}}
\def\xav{\bar{\bf x}_{\perp}}
\def\xperp{{\bf x_{\perp}}}
\def\xiperp{{\bf x_{i\perp}}}
\def\xipoperp{{\bf x_{(i+1)\perp}}}

\def\mN{{\mathcal {N}}}
\def\mT{{\mathcal {T}}}
\def\mC{{\mathcal {C}}}
\def\mL{{\mathcal {L}}}
\def\mF{{\mathcal {F}}}
\def\Eq#1{Eq.~(\ref{#1})}
\def\del{\partial}
\def\llangle{\left\langle}
\def\rrangle{\right\rangle}
\def\x{{\bf x}}

\def\w{\mathfrak{w}}

\newcommand{\slh}[1]{\rlap{\, /}\;\, #1}

\def\mN{{\mathcal {N}}}

\def\del{\partial}

\def\grad{{\bm\nabla}}

\def\Eq#1{Eq.~(\ref{#1})}

\def\w{\mathfrak{w}}

\begin{document}
\title{INTRODUCTORY LECTURES ON JET QUENCHING \\ IN HEAVY ION COLLISIONS%
\thanks{Presented at the XLVII Cracow School of Theoretical Physics, Zakopane, Poland, June 14--22, 2007. }%
}
\author{Jorge Casalderrey-Solana\footnote{JCasalderrey-Solana@lbl.gov}
\address{Lawrence Berkeley National Laboratory, Berkeley, CA 94720}
\and
Carlos A. Salgado\footnote{carlos.salgado@cern.ch}
\address{Dipartimento di Fisica, Universit\`a di Roma "La Sapienza"\\ and INFN, Roma, Italy\\ and \\
Departamento de F\'\i sica de Part\'\i culas and IGFAE, \\ Universidade de Santiago de Compostela, Spain}
}
\maketitle
\begin{abstract}
Jet quenching has become an essential signal for the characterization of the medium formed in experiments of heavy-ion collisions. After a brief introduction to the field, we present the full derivation of the medium-induced gluon radiation spectrum, starting from the diagrammatical origin of the Wilson lines and the medium averages and including all intermediate steps. The application of this spectrum to actual phenomenological calculations is then presented, making comparisons with experimental data and indicating some improvements of the formalism to the future LHC program. The last part of the lectures reviews calculations based on the AdS/CFT correspondence on the medium parameters controlling the jet quenching phenomenon.
\end{abstract}

\section{Contents of the lectures}

The lectures are organized as follows, in Sections \ref{sec:extremeQCD} and \ref{sec:hard} the properties of QCD matter and the use of heavy-ions to study these extreme conditions are briefly reviewed; Section \ref{sec:propagation} presents the general formalism to describe highly energetic particles propagating in matter which is then applied to the case of medium--induced gluon radiation in Section \ref{sec:migr}; Section \ref{sec:application} summarizes how this formalism is implemented in actual phenomenological calculations and gives a brief review of the comparison with experimental data; Section \ref{sec:strings} presents some new developments on the calculation of the transport coefficient using the Anti-de-Sitter/Conformal-Field-Theory correspondence; finally, some conclusions are presented. Readers interested on the formalism of medium-induced gluon radiation could directly jump to Section \ref{sec:propagation}. The derivation of the medium-induced gluon radiation presented here is new and entirely based on resummation of the multiple scattering diagrams which we find more intuitive.

\section{Heavy-ion collisions and extreme QCD matter}

\label{sec:extremeQCD}

QCD is the theory which describe the strong interaction. Its asymptotic states are hadrons which are colorless objects composed by quarks and gluons, the degrees of freedom of the QCD lagrangian. Confinement is a property of the strong interaction which forbids the existence of asymptotic colored states in normal conditions of temperature and densities. Quarks are fermions with fractional electric charge and very different masses. Light quark masses are ${\cal O}$(10 MeV) for $up$ and $down$ and ${\cal O}$(100 MeV) for $strange$. Heavy quark masses are $\sim$1.5 GeV for $charm$, $\sim$5 GeV for $beauty$ and $\sim$175 GeV for $top$. The masses of the hadrons formed by light quarks is much larger than the sum of the individual masses of their constituents, indicating a large dynamical origin for the first.  The spectroscopy of both light and heavy mesons and baryons is an interesting area to study the QCD dynamics. A main observation is that the chiral symmetry -- an exact symmetry of the lagrangian when the quark masses are neglected -- is not observed. This symmetry breaking is not realized by the presence of a new particle as the Higgs boson in the EW sector, but by the structure of the QCD vacuum, in which the symmetry is spontaneously broken. In this case, the associated Goldstone bosons are identified to be the pions and kaons, which are almost massless when their mass is compared with that of other hadrons.

So, two basic properties of the QCD vacuum are the confinement of quarks and gluons and the chiral symmetry breaking. One relevant question is, then,
 {\it is there a regime where these symmetries are restored?}

Another essential property of QCD is asymptotic freedom, the fact that the interaction disappears at small distances/large scales. As a consequence, QCD medium at asymptotic temperatures is predicted to be a gas of free quarks and gluons with a restoration of the symmetries mentioned above. For a null net baryon number in the medium, the transition temperature is found to be in the range $T_c\sim$ 175 -- 190 MeV -- see e.g. \cite{Hatsuda:2007rt} for a recent discussion. Most of the theoretical information we have about the transition region comes from lattice QCD calculations where different quantities can be studied\footnote{See e.g. \cite{Karsch:2001cy} for a description of the QCD thermodynamics as studied by lattice.} such as the equation of state, the chiral condensate or the free energy. 

\begin{figure}
\begin{minipage}{0.5\textwidth}
\begin{center}
\includegraphics[width=0.85\textwidth]{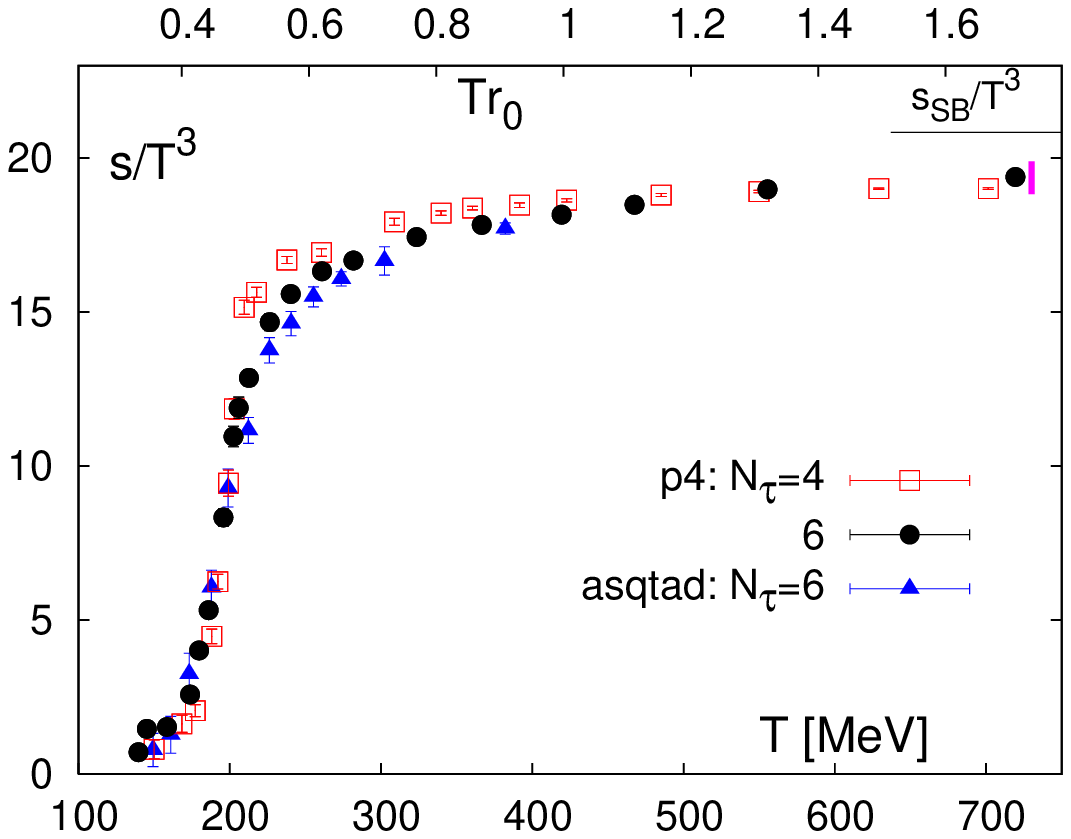}
\end{center}
\end{minipage}
\hfill
\begin{minipage}{0.5\textwidth}
\begin{center}
\includegraphics[width=0.85\textwidth]{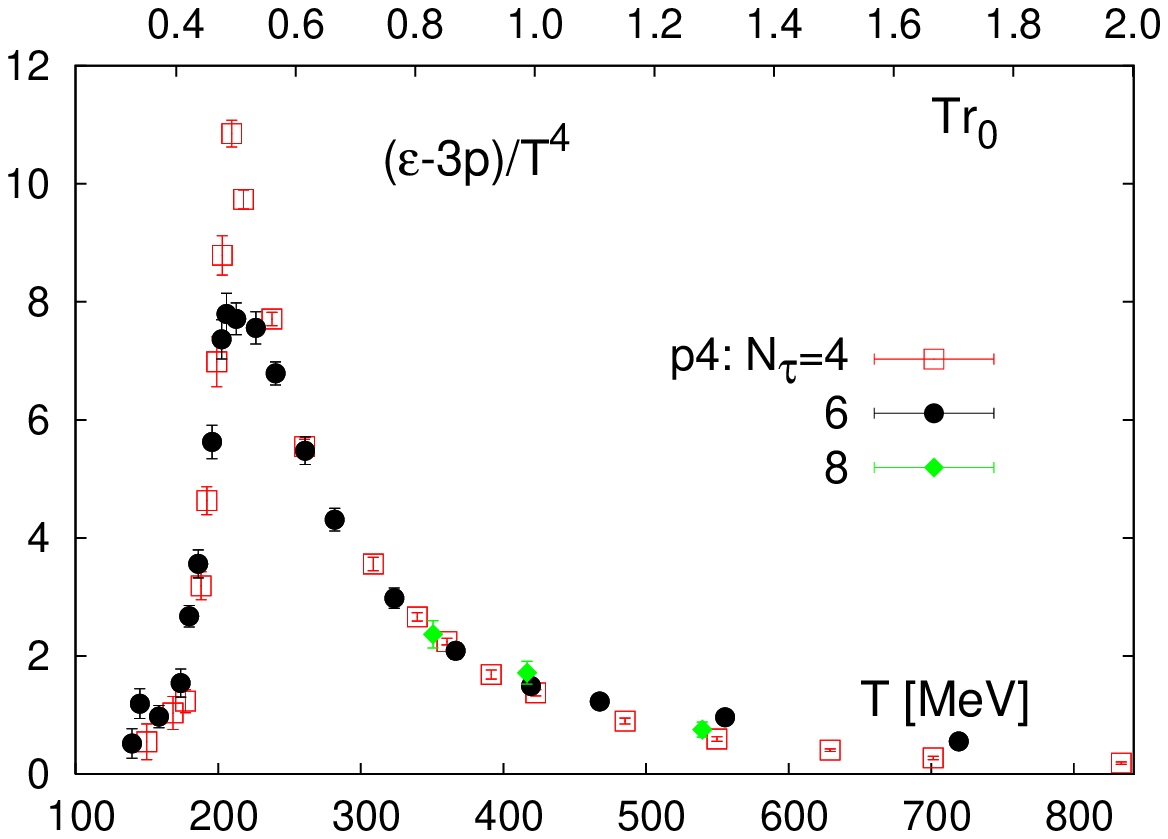}
\end{center}
\end{minipage}
\caption{Left: Entropy, $s\equiv \varepsilon+p$ in units of $T^4$ versus temperature computed in lattice. Right: Trace anomaly in units of $T^4$ as a function of the temperature. Figures from \protect\cite{Cheng:2007jq}}
\label{fig:EoS}
\end{figure}

A first example of this collective behavior is the equation of state. The pressure or the energy density measured in units of $T^4$ of an ideal gas are, according to the Stefan-Boltman law, proportional to the number of degrees of freedom in the system: so, for a free gas of pions this quantity is $N^\pi_{\rm dof}=3$, while for a free gas of quarks and gluons this quantity is much larger, $N^q_{\rm dof}=2\times 2\times 3$, $N^g_{\rm dof}=2\times 8$, counting spin, color and (two) flavor states. This different behavior is observed in lattice calculations where a jump at the transition temperature, $T_c$, appears -- see Fig. \ref{fig:EoS}. Interestingly, the lattice results also signal to a significant departure of the ideal gas behavior, $\varepsilon=3p$ at temperatures close to $T_c$ -- see Fig. \ref{fig:EoS}.

In order to understand the nature of the transition order parameters are needed. In QCD with massless quarks, the chiral condensate is the order parameter, while in the infinite mass limit the order parameter is the Polyakov Loop. The order of the transition depend on the actual value of the quark masses and most calculations agree in the absence of a real phase transition at zero baryochemical potential -- the transition would just be a rapid cross-over. 

The behavior of the potential between heavy quarks are also of interest for phenomenological applications. Although some discussion about the precise meaning and the definition of a potential exists in the case of a hot medium \cite{Mocsy:2004bv}, a screening leading to a non-confining potential is expected to appear at some point above the phase transition and, correspondingly, the heavy quark bound states would cease to exist if this temperature is reached. A prediction of the presence of a deconfined medium in heavy ion collisions is then the disappearance of heavy-quark bound states (quarkonium) and in particular the $J/\Psi$.

\subsection{Heavy-ion collisions to characterize hot QCD states}

\label{sec:hic}

The description of hadronic collisions in terms of the formation of a thermal state, which then evolves according to a hydrodynamical behavior, has an old story. In the 50's, Landau \cite{Landau:1953gs}  proposed a model in which a transient thermal state -- {\it little fireball} -- is formed by energy deposition in a small slice of space, of the size of the Lorentz-contracted nuclei, at the center of mass of the collision. The subsequent hydrodynamical evolution produce an expansion and cooling of the system up to a freeze-out temperature $T\sim m_\pi$ when the formed hadrons free stream to the detectors. A rapidity-independent version of this model was proposed by Bjorken \cite{Bjorken:1982qr} -- see e.g. \cite{Rischke:1998fq} for a review on hydrodynamical models in heavy-ion collisions. Although more sophisticated implementations are used nowadays, very demanding from the computational point of view, for the hydrodynamical description of nuclear collisions, the Bjorken model estimate gives a convenient and simple parametrization of some medium properties, as a function of the proper time $\tau$, as the energy density or the temperature of the created medium
\begin{equation}
\epsilon(\tau)\simeq\frac{\epsilon_0}{\tau^{4/3}}; \ \ \ \ T\simeq\frac{T_0}{\tau^{1/3}}
\end{equation}

Present facilities to study this active field of research -- see \cite{d'Enterria:2006su} for a recent review on the experimental situation -- are the CERN SPS, with fixed target collisions of  different nuclei up to $\sqrt{s}\simeq$ 20 GeV/A; the RHIC at Brookhaven, a dedicated collider which reaches much larger energies, $\sqrt{s}=$200 GeV/A; and in the near future the LHC at CERN, with an important program of nuclear collisions which will provide a spectacular jump in the collision energy upto $\sqrt{s}=$ 5500 GeV/A. Here, the quoted figures refer to the total available energy in center of mass of the collision divided by the number of nucleons in the nuclei -- {\it energy per nucleon} -- so that it can be directly compared to a corresponding proton-proton collision energy.

The lifetime of the hot medium produced in heavy ion collisions is so small, of the order of its own transverse size, that only indirect probes are available to characterize its properties. Here we will only focus on hard probes and specially on the study of the high transverse momentum part of the spectrum of produced particles. {\it Jet quenching} is the generic name under which the corresponding effects on high-$p_\perp$ particles are known. The original proposal \cite{Bjorken:1982tu}, a suppression of the the large transverse momentum yields due to the energy lost by interaction of the fast partons with the medium has been observed at RHIC \cite{RHIC} and partially also at the SPS \cite{SPS}.

\section{Hard probes to heavy-ion collisions}
\label{sec:hard}

Hard processes are those involving large momentum exchanges. Asymptotic freedom allows, in these conditions, to perform calculations by a perturbative expansion in terms of $\alpha_s(Q^2)$, where $Q$ is the large scale of the process, as, e.g. a large mass or a large transverse momentum of the produced particles. This perturbative expansion is, however, computed in terms of quarks and gluons, the degrees of freedom of the QCD lagrangian, and some extra information is usually needed to connect it with the initial and final hadrons appearing as asymptotic states. A crucial simplification of the problem is possible thanks to the factorization theorems of QCD which allow to separate long- and short-distance contributions to the cross section in the form\footnote{For simplicity we have made all the scales the same, although, strictly speaking, the factorization, renormalization and fragmentation scales can be different.}
\begin{equation}
\sigma^{AB\to h}=
f_A(x_1,Q^2)\otimes f_B(x_2,Q^2)\otimes \sigma(x_1,x_2,Q^2)\otimes D_{i\to h}
(z,Q^2)\, ,
\label{eqhard}
\end{equation}
where the short-distance perturbative cross section, $\sigma(x_1,x_2,Q^2)$, is computable in powers of $\alpha_s(Q^2)$ and the long-distance terms are non-perturbative quantities involving scales ${\cal O}(\Lambda_{\rm QCD})$. More specifically, the proton/nuclear parton distribution functions (PDF), $f_A(x,Q^2)$, encode the partonic structure of the colliding objects and can be interpreted as the probability to find a parton (quark or gluon) inside the hadron with fraction of momentum $x$; the fragmentation functions (FF), $D(z,Q^2)$, describe the hadronization of the parton $i$ into a final hadron $h$ with a fraction of momentum $z$. 

Although the PDFs and the FFs are non-perturbative quantities, their evolution in $Q^2$ can be computed by the Dokshitzer-Gribov-Lipatov-Altarelli-Parisi (DGLAP) equations \cite{dglap}. This equations resum higher orders in $\alpha_s$ which, due to enhanced contributions in some regions of phase-space, cannot be neglected. For example, we will see Section \ref{sec:migr} that the spectrum of gluons radiated with energy fraction $z$ out of a quark produced in a hard process is 
\begin{equation}
\frac{dI}{dz dt}=\frac{\alpha_s}{2\pi^2}\frac{1}{t}P(z)\simeq \frac{\alpha_s}{\pi^2}\frac{1}{t}\frac{C_R}{z}
\label{eq:split1}
\end{equation}
with $t$ the invariant mass of the produced pair. This radiation is divergent in the infrared and collinear limits giving large ${\cal O}(\alpha_s\log t)$ terms to the cross section which need to be taken into account. These terms are resummed by the DGLAP evolution equations\footnote{To simplify the notation, we include only one flavor, see, e.g., \cite{ellis} for a complete description.}
\begin{equation}
\frac{\partial D(x,t)}{\partial \log t} =\int_x^1\frac{dz}{z}\frac{\alpha_s}{2\pi}\ P(z)\, D\left(\frac{x}{z},t\right).
\label{eq:DGLAP}
\end{equation}
where $D(x,t)$ can be the PDFs or the FFs and $P(z)$ are the splitting functions. At the lowest order (LO) these equations admit a probabilistic interpretation in which a parton shower is formed by subsequent branching of the initial quark or gluon. We will come back to this interpretation in Section \ref{sec:application}.

These equations provide an excellent description of experimental data as deep inelastic lepton-proton scattering (DIS) or jets.

\subsection{Hard processes as probes of the medium}

The hard process described by the perturbative cross section $\sigma(x_1,x_2,Q^2)$ in (\ref{eqhard}) takes place in a very short time, ${\cal O}(1/Q)$, and is basically unchanged in nuclear collisions. The interest of hard processes as probes of the medium is that the long distance parts are modified when they communicate with the extension of the medium. Measuring these modifications allow for a characterization of the medium properties -- see e.g. \cite{Salgado:2007rs} for a recent review.

A conceptually simple example is the $J/\Psi$, whose production cross section can be written as
\begin{equation}
\sigma^{hh\to J/\Psi}=
 f_i(x_1,Q^2)\otimes f_j(x_2,Q^2)\otimes
\sigma^{ij\to [c\bar c]}(x_1,x_2,Q^2)
 \langle {\cal O}([c\bar c]\to J/\Psi)\rangle\, ,
\end{equation}
where now $ \langle {\cal O}([c\bar c]\to J/\Psi)\rangle$ describes the hadronization of a $c\bar c$ pair in a given state (for example a color octet) into a final $J/\Psi$. In the case that the pair is produced inside a hot medium this long-distance part is modified. Actually, the potential between the pair is screened in a hot medium and the hadron is dissolved, making $ \langle {\cal O}([c\bar c]\to J/\Psi)\rangle\to 0$. The experimental observation of this effect is a disappearance of the $J/\Psi$ in nuclear collisions \cite{Matsui:1986dk}. This suppression has been discovered in experiments at the CERN SPS \cite{jpsiSPS} and then also at RHIC \cite{jpsiRHIC}. The interpretation of the experimental data is, however, not simple; one of the reasons being the lack of good theoretical control over the modification of a purely non-perturbative quantity as the hadronization of a $c\bar c$ pair into a charmonium state.

From the computational point of view, a theoretically simpler case is the modification of the {\it evolution} of fragmentation functions of high-$p_T$ particles due to the presence of a dense or finite--temperature medium. Here, highly energetic partons, produced in a hard process, propagate through the produced matter, loosing energy by medium-induced gluon radiation. This phenomenon is generically known as {\it jet quenching} and its implementation into the perturbative formula (\ref{eqhard}) will be discussed in the next sections.

\subsection{Nuclear parton distribution functions}

 A good knowledge of the PDFs is essential in any calculation of hard processes. The usual way of obtaining these distributions is by a global fit of data on different hard processes 
(mainly deep inelastic scattering, DIS) to obtain a set of parameters for the initial, non-perturbative, input $f(x,Q^2_0)$ to be evolved by DGLAP equations \cite{dglap}.

In the nuclear case, the initial condition, $f_A(x,Q^2_0)$, is modified compared to the proton. Moreover, at small enough $x$, non-linear corrections to the evolution equations \cite{Gribov:1984tu, Mueller:1985wy, Kovchegov:1999yj, jimwlk} are expected to become relevant. Global DGLAP analyses, paralleling those for free protons are available \cite{Eskola:1998iy,Eskola:1998df,Hirai:2001np,Hirai:2004wq,deFlorian:2003qf,Eskola:2007my,Hirai:2007sx}. These studies fit the available data on DIS and Drell-Yan with nuclei providing the needed benchmark for additional mechanisms.

\begin{figure}[tbh]
\centering
\centering\includegraphics[width=0.7\textwidth]{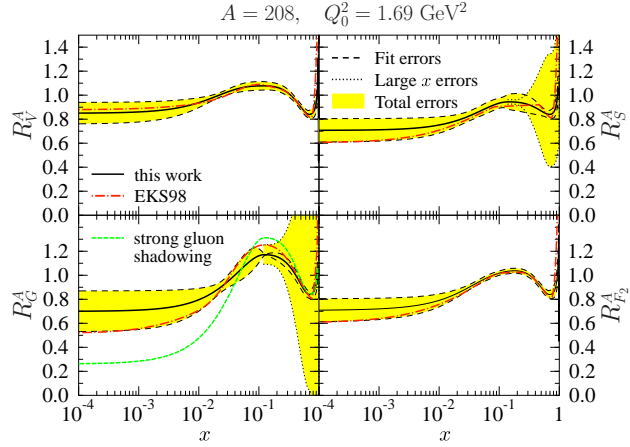}
\caption{Ratios of nuclear to free proton PDFs for different flavors at the initial scale $Q_0^2$=1.69 GeV$^2$ from \protect\cite{Eskola:2007my} with error estimates. The green line in the gluon panel is an attempt to check the strongest gluon shadowing supported by present data. }
\label{fig:nPDF}
\end{figure}

 The DGLAP analyses of nuclear PDFs from \cite{Eskola:2007my}  is shown in Fig. \ref{fig:nPDF}, including the corresponding error estimates. An important issue, partially visible in Fig. \ref{fig:nPDF}, is that present nuclear DIS and DY data can only constrain the distributions for $x\gtrsim 0.01$ in the perturbative  region. By chance, this region covers most of the RHIC kinematics, so that, the description of e.g. $J/\Psi$-suppression or inclusive particle production in  dAu collisions as given by the nuclear PDFs can be taken as a check of universality of these distributions. These checks present a quite reasonable agreement with data \cite{Vogt:2004hf}, but some extra suppression for the inclusive yields at forward rapidities is probably present. The strong gluon shadowing plotted in Fig. \ref{fig:nPDF} improves the situation at forward rapidities without worsening the fit of DIS or DY data -- $\chi^2/{\rm dof}<1$. Whether a DGLAP analysis can accommodate all sets of data is an open question, but the finding in Ref. \cite{Eskola:2007my} are encouraging. A suppression at forward rapidities was also predicted in terms of saturation of partonic densities \cite{satur}. 

\section{Particle propagation in matter}

\label{sec:propagation}

To describe the jet quenching phenomenon, we start from a high-$p_\perp$ quark or gluon produced in an elementary hard collision which subsequently interacts with the surrounding matter. So, the first question we need to address is how a highly energetic particle propagates through a dense medium. In this section we will present the most widely used formalism based on a semiclassical approach in which the changes in the medium configuration due to the passage of the fast particle -- recoil -- are neglected. In this approximation the medium can be considered as a background field. 

A convenient formulation of the problem is in terms of Wilson lines
\begin{equation}
W({\bf x})={\cal P}\exp\left [ig\int dx_+ A_-(x_+,{\bf x})\right]
\label{eq:wilson}
\end{equation}
describing the propagation of a particle through a medium  field $A_-(x_+{\bf x})$. Its origin will be explained in the next subsection. Here we introduce the light cone variables 
\begin{equation}
x_\pm=\frac{1}{\sqrt{2}}\left(x_0\pm x_3\right) \ \ \ \ \ 
p_\pm=\frac{1}{\sqrt{2}}\left(p_0\pm p_3\right)
\label{eq;lightconevar}
\end{equation}
So that the scalar product is
\begin{equation}
p\cdot x=p_+x_-+p_-x_+-{\bf p_\perp\cdot x_\perp}
\end{equation}
and the rapidity
\begin{equation}
y=\frac{1}{2}{\rm ln}\left[\frac{p_0+p_3}{p_0-p_3}\right]=\frac{1}{2}{\rm ln}\left[\frac{p_+}{p_-}\right]
\end{equation}

\subsection{Wilson lines, eikonal approximation}
\label{sec:eikonal}

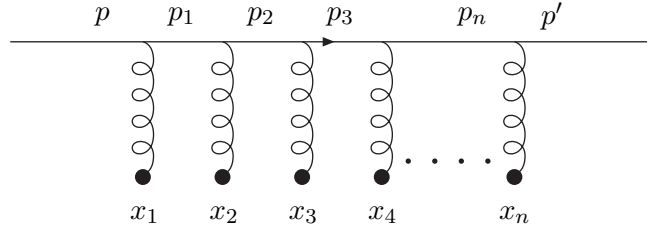
\begin{figure}
\begin{center}
\begin{picture}(300,80)(0,0)
\ArrowLine(30,70)(270,70)
\Gluon(80,70)(80,20){4}{4}
\Gluon(110,70)(110,20){4}{4}
\Gluon(140,70)(140,20){4}{4}
\Gluon(170,70)(170,20){4}{4}
\Gluon(220,70)(220,20){4}{4}
\Vertex(80,19){3}
\Vertex(110,19){3}
\Vertex(140,19){3}
\Vertex(170,19){3}
\Vertex(220,19){3}
\Text(81,5)[]{$x_1$}
\Text(111,5)[]{$x_2$}
\Text(141,5)[]{$x_3$}
\Text(171,5)[]{$x_4$}
\Text(221,5)[]{$x_n$}
\Vertex(180,25){1}
\Vertex(190,25){1}
\Vertex(200,25){1}
\Vertex(210,25){1}
\Text(65,80)[]{$p$}
\Text(235,80)[]{$p'$}
\Text(95,80)[]{$p_1$}
\Text(125,80)[]{$p_2$}
\Text(155,80)[]{$p_3$}
\Text(205,80)[]{$p_n$}
\end{picture}
\caption{A multiple scattering eikonal trajectory}
\label{fig:multiple}
\end{center}
\end{figure}

A simple derivation of the Wilson line is obtained in terms of multiple scatterings, providing a clear physical picture of the eikonal propagation. Consider the diagram of Fig. \ref{fig:multiple}, where static centers of scattering are placed at $x_1,\ x_2,\ ...\ x_n$. Let us fix that the quark is moving in the positive $x_3$ direction, i.e. the large component of the momentum is $p_+$.

The contribution to the S-matrix of one scattering is
\begin{equation}
S_1(p',p)=\int d^4x\ {\rm e}^{i (p'-p)\cdot x} \   \bar u(p')\ ig A_\mu^a(x) T^a\gamma^\mu\ u(p)
\label{eq:1scat}
\end{equation}
Taking the eikonal limit, $p\simeq p'$, $\frac{1}{2}\sum_\lambda \bar u^\lambda(p)\gamma^\mu u^\lambda(p)=2p^\mu$ and $p^\mu A_\mu^a\simeq 2p_+A_-^a$. In order to proceed, we will assume that the fields have a small dependence on the small coordinate $x_-$: due to the Lorenz contraction, the medium can be seen as a small sheet in this coordinate. Putting all together one obtains \footnote{Note that we neglect the $p_-$ component. In the eikonal approximation
$p'_-<<\left|\pperp\right|$, however, the phase factor $i p'_- x_+$ is 
potentially enhanced by the medium lenght. Thus, we are implicitilly assuming
that the medium is smaller than the coherence length 
$p_-L\approx \mu^2 L/2 p_+<<1$. In the next section we will see how this assumption can be relaxed.}
\begin{equation}
S_1(p',p)\simeq 2\pi\delta(p'_+-p_+) 2p_+\int d{\bf x_\perp} {\rm e}^{-i {\bf x_\perp(p'_\perp-p_\perp)} }\left[ ig\int dx_+A_-(x_+,{\bf x_\perp})\right],
\label{eq:1scatfinal}
\end{equation}
where we have singled out with brackets the contribution of the field which will exponentiate to give the Wilson line and the color matrix has been omitted for clarity.

The contribution with two scatterings is given by 
\begin{eqnarray}
S_2(p',p)&=&\int  \frac{d^4p_1}{(2\pi)^4}d^4x_1d^4x_2\ {\rm e}^{i (p_1-p)\cdot x_1} {\rm e}^{i (p'-p_1)\cdot x_2}   \bar u(p')\ ig A_{\mu_1}^{a_1}(x_1)T^{a_1}\gamma^{\mu_1} \times\nonumber \\
&  \times&i\frac{\slh p_1}{p_1^2+i\epsilon}i gA_{\mu_2}^{a_2}(x_2) T^{a_2}\gamma^{\mu_2}\ u(p)  
\label{eq:2scatt}
\end{eqnarray}
Applying the Dirac equation one can write in the eikonal limit $\bar u(p)\gamma_{\mu_1}{\slh p_1}\gamma_{\mu_2}u(p')\simeq (2p_+)^2 g_{\mu_1 -}g_{\mu_2 -}$ giving
\begin{equation}
S_2(p',p)= -ig^2  (2p_+)^2\int \frac{d^4p_1}{(2\pi)^4}d^4x_1d^4x_2\
\frac{{\rm e}^{i (p_1-p)\cdot x_1+i (p'-p_1)\cdot x_2}}{p_1^2+i\epsilon} A_-(x_1) A_-(x_2) ,
\label{eq:2scatt2}
\end{equation}
where again the color matrices have been omitted as will be omitted in the following. In the high-energy limit, the integrals in $d^4 p_1$ can be performed:
\begin{eqnarray}
\int dp_{1-}\frac{{\rm e}^{i(x_{1+}-x_{2+})p_{1-}}}{2p_{1+}p_{1-}+i\epsilon}&=&-\Theta(x_{2+}-x_{1+})\frac{2\pi i}{2p_+}\label{eq:kminuspole}\\
\int dp_{1+}{\rm e}^{i(x_{1-}-x_{2-})p_{1+}}&=&2\pi\delta(x_{1-}-x_{2-})\\
\int d^2p_{1\perp}{\rm e}^{-i{\bf (x_{1\perp}-x_{2\perp})p_{1\perp}}}&=&(2\pi)^2\delta({\bf x_{1\perp}-x_{2\perp}})
\label{eq:deltaxperp}
\end{eqnarray}
Doing the $x_-$-integration as in the case of one scattering one obtains
\begin{equation}
S_2(p',p)\simeq 2\pi\delta(p'_+-p_+) 2p_+\int d{\bf x_\perp} {\rm e}^{-i {\bf x_\perp(p'_\perp-p_\perp)} }\frac{1}{2}{\cal P}\left[ ig\int dx_+A_-(x_+,{\bf x_\perp})\right]^2,
\label{eq:2scattfinal}
\end{equation}
where again the second term in the expansion of (\ref{eq:wilson}) appears. In order to obtain (\ref{eq:2scattfinal}) we have used that
\begin{equation}
\int dx_1 dx_2 ... dx_n \Theta(x_2-x_1)....\Theta(x_n-x_{n-1})A(x_1)...A(x_n)=\frac{1}{n!}{\cal P}\left[\int dx A(x)\right]^n
\end{equation}
where ${\cal P}$ means path-ordering of the fields $A(x)$. 

The generalization to n-scattering centers follows the same lines and the corresponding contribution is given by 
\begin{equation}
S_n(p',p)\simeq 2\pi\delta(p'_+-p_+) 2p_+\int d{\bf x_\perp} {\rm e}^{-i {\bf x_\perp(p'_\perp-p_\perp)} }\frac{1}{n!}{\cal P}\left[ ig\int dx_+A_-(x_+,{\bf x_\perp})\right]^n,
\label{eq:nscattfinal}
\end{equation}
So that the total S-matrix is 
\begin{equation}
S(p',p)=\sum_{n=0}^\infty S_n(p',p)\simeq 2\pi\delta(p'_+-p_+) 2p_+\int d{\bf x_\perp} {\rm e}^{-i {\bf x_\perp(p'_\perp-p_\perp)} }
W({\bf x_\perp}),
\label{eq:smattot}
\end{equation}
with $W({\bf x_\perp})$ given by eq. (\ref{eq:wilson}). Similar expressions are obtained in the case of a fast gluon of momentum $k$ with the only replacement of the external fields in (\ref{eq:wilson}), now in the adjoint representation -- see e.g. \cite{Hebecker:1999ej}. In order to see this, the external fields coupled to three gluon vertices (four gluon vertex are subleading in energy) are written in the adjoint representation 
\begin{equation}
A_{\cal A}=A^a(T^a_{\cal A})^{bc}=-iA^a f^{abc}
\end{equation}
The sum of gluon polarization vectors can be written as
\begin{equation}
\sum_{i=1}^2 \epsilon_{(i)}^\mu \epsilon_{(i)}^\nu=g^{\mu\nu}-\frac{m^\mu k^\nu}{m\cdot k}-\frac{k^\mu m^\nu}{m\cdot k}
\label{eq:gmunudec}
\end{equation}
with $m$ a light-like vector with non-zero minus component, $m=(0,1,0,0)$ and the polarization vectors defined by $\epsilon\cdot k=\epsilon\cdot m=0$; $\epsilon^2=-1$ 
\begin{equation}
\epsilon_{(i)}=\left(0,\frac{{\bf k_\perp\cdot \epsilon}_{(i)\perp}}{k_+},{\bf \epsilon}_{(i)\perp}\right)
\end{equation}
and ${\bf \epsilon}_{(1)\perp}=(1,0)$, ${\bf \epsilon}_{(2)\perp}=(0,1)$. This prescription allows to simplify the structure of the gluon line since, due to gauge invariance, only the first term in 
eq. (\ref{eq:gmunudec}) contributes in the high-energy limit when the propagator is in between two three-gluon vertices. So that, one can write for each gluon propagator
\begin{equation}
g^{\mu\nu}\simeq \sum_{i=1}^2 \epsilon_{(i)}^\mu \epsilon_{(i)}^\nu
\label{eq:gmunusimp}
\end{equation}
and now the contribution from each vertex $V_{\mu\nu\sigma}$ can be simplified as
\begin{equation}
\epsilon_{(i')}^\mu(k') V_{\mu\nu\sigma}(-k',k,k'-k)A_{\cal A}^\sigma \epsilon_{(i)}^\nu(k)\simeq-i2gk_+A_{{\cal A}-}\delta_{ii'}
\end{equation}
which has a structure similar to the quark case allowing for a resummation to obtain a Wilson line now with fields in the adjoint representation. We will denote this gluon line with a superscript, $W^A({\bf x_\perp})$.

Eq. (\ref{eq:smattot}) describes the scattering of a quark (or gluon, with the changes explained above) in a medium with a given field configuration, determined by the scattering centers. The effect of the medium is to induce color rotation at each scattering center without changing the helicity (polarization) of the quark (gluon), which loses a negligible amount of energy and propagates in straight lines. In order to have the final answer for a physical cross section, these medium configurations need to be averaged within an ensemble describing the medium. This issue will be discussed later.

\subsection{Relaxing the eikonal approximation}

\label{sec:relaxing}

In some cases, the restrictions applied in the above formulation need to be relaxed  to allow small changes in the transverse position of the propagating particle. This is the case, for instance, of the medium--induced gluon radiation, where the gluon position follows Brownian motion in the transverse plane. The eikonal Wilson line is now replaced by the propagator
\begin{equation}
G(b,a)=\int  {\cal D}{\bf r}(x_+)\exp\left\{i\frac{p_+}{2}\int dx_+\left[\frac{d{\bf r}}{dx^+}\right]^2\right\}W({\bf r})
\label{eq:propag}
\end{equation}
This change can be derived in the multiple soft scattering presented in the previous section. In order to do that, we have to keep the subleading $p_\perp^2$ terms in the poles of the propagators. So, now the integration in $p_{i-}$ reads
\begin{equation}
\int dp_-\frac{{\rm e}^{ip_-(x_{i+}-x_{(i+1)+})}}{2p_+p_--p_\perp^2+i\epsilon}=-i\frac{2\pi}{2p_+}\Theta(x_{(i+1)+}-x_{i+}){\rm e}^{i\frac{p_\perp^2}{2p_+}((x_{i+}-x_{(i+1)+})}
\end{equation}
and instead of (\ref{eq:deltaxperp}) the integration in $p_{i\perp}$ is gaussian, giving
\begin{eqnarray}
\int \frac{d^2p_{i\perp}}{(2\pi)^2}\ {\rm e}^{i\frac{p_\perp^2}{2p_+}((x_{i+}-x_{(i+1)+})}
{\rm e}^{-i{\bf p_{i\perp}}({\bf x_{i\perp}-x_{(i+1)\perp}})}=\nonumber \\
=\frac{p_+}{2\pi i(x_{i+}-x_{(i+1)+})}\exp\left\{-i\frac{p_+}{2}\frac{({\bf x_{i\perp}-x_{(i+1)\perp}})^2}{x_{i+}-x_{(i+1)+}}\right\}
\label{eq:patheps}
\end{eqnarray}
Eq.(\ref{eq:patheps}) is the Feymann propagator of a free particle that 
propagates in the transverse plane from $\xiperp$ at time $x_{i+}$
to  $\xipoperp$ at time $x_{(i+1)+}$ (see e.g. \cite{feynmhibbs}) $G_0(\xipoperp-\xiperp;x_{(i+1)+}-x_{i+})$. Thus, Eq. (\ref{eq:patheps}) can be 
expressed as
\begin{equation}
G_0(\xipoperp-\xiperp;x_{(i+1)+}-x_{i+})
=\int \mathcal{D}\xperp(x_p)
\exp\left\{
     i\frac{p_+}{2}\int dx_+ \left[\frac{d\xperp}{dx_+}\right]^2
    \right\}                                       
\end{equation}
where the paths $\xperp(x_+)$ connect the two endpoints of the propagator.

The scattering matrix now reads
\begin{eqnarray}
S(p',p)=2\pi\delta(p'_+-p_+)2p_+
       \sum^{\infty}_{n=0} 
       \int \mathcal {P} \prod^n_{i=0} dx_{i+}d\xiperp ig A(x_{i+},\xiperp)\times\nonumber\\
          \times                       G_0(\Delta \xipoperp;\Delta x_{(i+1)+})
                                 ig A(x_{(i+1)+},{\bf x}_{(i+1)\perp})
\end{eqnarray}
where $\Delta \xipoperp=\xipoperp-\xiperp$, 
$\Delta x_{(i+1)+}=x_{(i+1)+}-x_{i+}$. This expression  can be reorganized as
\begin{eqnarray}
S(p',p)=2\pi\delta(p'_+-p_+)2p_+\int d{\bf x_\perp}{\rm e}^{-i{\bf x_\perp(p'_\perp-p_\perp)}}\times\nonumber\\ \times\int{\cal D}{\bf r}(x_+)\exp\left\{i\frac{p_+}{2}\int dx_+\left[\frac{d{\bf r}}{dx_+}\right]^2\right\}W({\bf r})
\label{eq:noneikonal}
\end{eqnarray}
with $W({\bf x_\perp})$ given by (\ref{eq:wilson}). 

Eq. (\ref{eq:noneikonal}) describes the propagation of a highly energetic particle through a medium when changes in the transverse position are allowed. The propagation takes phases both from the motion in the transverse plane and the color rotation by interaction with the medium field.

\subsection{The medium averages}
\label{sec:medav}

The S-matrices derived in the previous sections are valid for a given configuration of the fields and should be averaged over the proper ensemble of the medium field configurations. Several prescriptions have been used in the literature. Here we will present two of them used for the calculation of the jet quenching. Notice first that any physical quantity must contain, at least, the medium average of two Wilson lines since the average is done at the level of the cross section where only colorless states  are allowed. In general, we will be interested in quantities like
\begin{equation}
\frac{1}{N} {\rm Tr}\langle W^{\cal y}({\bf x_\perp})W({\bf y_\perp})\rangle.
\label{eq:wilsonav}
\end{equation}
Where the trace and the $1/N$ term correspond to average the initial color indices -- correspondingly, a factor $1/(N^2-1)$ would appear in case of Wilson lines in the adjoint representation.

\begin{figure}
\begin{center}
\begin{picture}(300,80)(0,0)
\ArrowLine(100,70)(30,70)
\ArrowLine(30,60)(100,60)
\Gluon(55,70)(55,20){4}{5}
\Gluon(70,70)(70,20){4}{5}
\ArrowLine(200,70)(130,70)
\ArrowLine(130,60)(200,60)
\Gluon(155,60)(155,20){4}{4}
\Gluon(170,60)(170,20){4}{4}
\ArrowLine(300,70)(230,70)
\ArrowLine(230,60)(300,60)
\Gluon(255,70)(255,20){4}{5}
\Gluon(270,60)(270,20){4}{4}
\Oval(65,20)(3,15)(0)
\Oval(165,20)(3,15)(0)
\Oval(265,20)(3,15)(0)
\end{picture}
\caption{Different contributions to the dipole cross section}
\label{fig:dipol}
\end{center}
\end{figure}
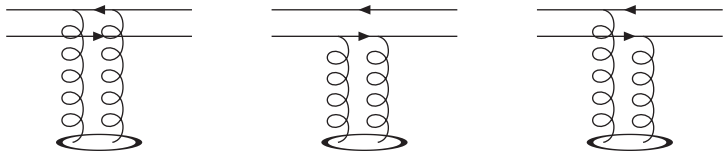

Two main approximations to the averages (\ref{eq:wilsonav}) are used in jet quenching phenomenology, the multiple soft scattering approximation and the opacity expansion. Both can be understood in the multiple scattering picture we are employing. The main assumption is that the centers of scattering are independent, i.e. no color flow appears between scattering centers separated more than a distance $\lambda\sim 1/\mu$, $\mu$ being a typical scale in the medium as the Debye screening length. So, we want to calculate 
\begin{eqnarray}
\frac{1}{N} {\rm Tr}\langle W^{\cal y}({\bf x_\perp})W({\bf y_\perp})\rangle
=\frac{1}{N} {\rm Tr}\Big\langle \exp\{-ig\int dx_+A_-^{\cal y}(x_+,{\bf x_\perp})\}\times\nonumber\\
\times\exp\{ig\int dx_+A_-(x_+,{\bf y_\perp})\}\Big\rangle
\label{eq:medaverage}
\end{eqnarray}
Expanding the exponents and taking the contribution from one scattering center, the leading contribution is quadratic in the fields -- the linear contributions cancels due to the color trace
\begin{eqnarray}
{\cal P}\Big\langle 1+\frac{1}{2}(ig)^2\left[\int dx_+A_-^{\cal y}(x_+,{\bf x_\perp})\right]^2+\frac{1}{2}(ig)^2\left[\int dx_+A_-(x_+,{\bf y_\perp})\right]^2\nonumber \\-(ig)^2\left[\int dx_+A_-^{\cal y}(x_+,{\bf x_\perp})\right]\left[\int dx_+A_-(x_+,{\bf y_\perp})\right]\Big\rangle
\label{eq:opacity}
\end{eqnarray}
These terms have a clear diagramatic interpretation, they correspond to the two-scattering diagrams represented in Fig. \ref{fig:dipol}. The first two do not resolve the transverse dimension of the dipole and are sometimes called {\it contact terms}. Writing the field as the Fourier transform of a field amplitude produced by a scattering center at position $(x_{i+},{\bf x_{i\perp}})$
\begin{equation}
gT^bA^b_-(x_+,{\bf x_\perp})=\int\frac{d^2{\bf q}}{(2\pi)^2}{\rm e}^{i({\bf x_\perp-x_{n\perp}}){\bf q}}T^ba^b_-({\bf q})\delta(x_+,x_{i+}),
\end{equation}
the medium averages $\langle...\rangle$ are done by integrating in the longitudinal $x_{i+}$ and transverse coordinates $\bf x_{n\perp}$ of the scattering centers and adding color configurations. For example, the third term in (\ref{eq:opacity}) gives
\begin{eqnarray}
\int dx_+ dx_{i+}d{\bf x_{i\perp}}\int \frac{d^2{\bf q_1}}{(2\pi)^2} \frac{d^2{\bf q_2}}{(2\pi)^2} {\rm e}^{-i({\bf x_\perp-x_{i\perp}}){\bf q_1}}{\rm e}^{i({\bf y_\perp-x_{i\perp}}){\bf q_2}}\times\nonumber\\ \times a^*_-({\bf q_1})a_-({\bf q_2})\delta(x_+-x_{i+})
=\int dx_+\int \frac{d^2{\bf q}}{(2\pi)^2} {\rm e}^{i({\bf y_\perp-x_\perp}){\bf q}}\left |a({\bf q})\right |^2
\end{eqnarray}
Putting all together, eq. (\ref{eq:opacity}) can be written in terms of the cross section of a dipole, where the quark and the antiquark are located at ${\bf y_\perp}$ and ${\bf x_\perp}$ respectively, with one of the scattering centers of the medium.  
\begin{equation}
\mbox{\protect (\ref{eq:opacity})}=1-C_F\int \frac{d^2{\bf q}}{(2\pi)^2} \left |a({\bf q})\right |^2
(1-{\rm e}^{i({\bf y_\perp-x_\perp}){\bf q}})=1-\frac{1}{2}\sigma({\bf y_\perp-x_\perp})
\label{eq:dipolcs}
\end{equation}
The factor $C_F$ comes from the color average and the traces over the matrices 
\begin{equation}
\frac{1}{N}\sum_{a,b}{\rm Tr}T^aT^b=C_F=\frac{N^2-1}{2N}=\frac{4}{3}
\label{eq:colortrace}
\end{equation}
The corresponding factor for a gluon would be $C_A=N=3$.

There is a subtlety in the derivation we have just presented. Strictly speaking, the contact terms are not included in the multiple scattering derivation of the Wilson lines made before, where each scattering center contributes with a factor $A_-$ and path ordering appears. What we have implicitly assumed is that scattering centers separated by distances smaller than $\lambda\sim1/\mu$ resolve color correlations while those separated by more than $\lambda$ do not. This is usually implemented by considering the contribution of one scattering center by defining a density $n(x_+)=\sum_i\delta(x_+-x_{i+})$ and taking the continuum by an integration in $x_+$
\begin{equation}
\frac{1}{N} {\rm Tr}\langle W^{\cal y}({\bf x_\perp})W({\bf y_\perp})\rangle_{\rm 1\  scatt}\simeq 1-\frac{C_F}{2}\int dx_+n({x_+})\sigma({\bf y_\perp-x_\perp})
\label{eq:1storder}
\end{equation}
Eq. (\ref{eq:1storder}) is the first order in an opacity expansion of the medium, the sum of all orders exponentiate and the average (\ref{eq:medaverage}) can be written as
\begin{equation}
\frac{1}{N} {\rm Tr}\langle W^{\cal y}({\bf x_\perp})W({\bf y_\perp})\rangle\simeq \exp\left\{-\frac{C_F}{2}\int dx_+n({x_+})\sigma({\bf y_\perp-x_\perp})\right\}
\label{eq:multipleq}
\end{equation}
Eqs. (\ref{eq:1storder}) and (\ref{eq:multipleq}) are the two main medium averages used in the literature of jet quenching. In order to proceed, the functional form of the dipole cross section needs to be specified. In the opacity expansion a Yukawa-type elastic scattering center with Debye screening mass $\mu$ is usually taken in (\ref{eq:dipolcs})
\begin{equation}
\left |a({\bf q})\right|^2=\frac{\mu^2}{\pi({\bf q^2+\mu^2})}.
\end{equation}
When the number of scattering centers is very large, all of them need to be resummed and the first orders of the opacity are not enough. In this conditions, it is convenient to take the dipole cross section at leading logarithmic accuracy \cite{Nikolaev:1990ja} and write the small distance component of the cross section
\begin{equation}
\sigma({\bf r})\simeq C {\bf r}^2
\end{equation}
The proportionality factor $C$ with the squared dipole size is usually taken to be constant and defines the transport coefficient $\hat q(\xi)\equiv 2\sqrt{2}n(\xi)C$, encoding all the information about the dynamical properties of the medium. This is the main parameter to be determined by fits to experimental data and to be compared with theoretical calculations. The Wilson line averages define this parameter by\footnote{The factor $\sqrt{2}$ is included here as the transport coefficient is usually defined in ordinary coordenates, where the longitudinal distance for a $\xi=z\simeq x_+/\sqrt{2}$}
\begin{equation}
\frac{1}{N^2-1} {\rm Tr}\langle W^{A{\cal y}}({\bf x_\perp})W^A({\bf y_\perp})\rangle\simeq \exp\left\{-\frac{1}{4\sqrt{2}}\int dx_+\hat q({x_+})({\bf x_\perp-y_\perp})^2\right\}.
\label{eq:multiple}
\end{equation}
Here we have considered a gluon instead of a quark as the fast particle traversing the medium. This is the most relevant quantity as we will see in the next sections.

It is worth noticing that similar medium averages are used in the study of deep inelastic scattering (DIS) with nuclei, where a phenomenon of saturation of partonic densities is expected. In this case, the convention is to call the relevant parameter saturation scale $Q_{\rm sat}$ and the Wilson line average is given by
\begin{equation}
\frac{1}{N} {\rm Tr}\langle W^{{\cal y}}({\bf x_\perp})W({\bf y_\perp})\rangle\simeq \exp\left\{-\frac{1}{4}Q^2_{\rm sat}({\bf x_\perp-y_\perp})^2\right\}.
\label{eq:multipleqsat}
\end{equation}
Now the saturation scale contains not only the dynamical information about the target but also the geometry. A simple model for this saturation scale gives \cite{Kovchegov:1998bi}
\begin{equation}
Q_{\rm sat}^2=\frac{8\pi^2\alpha_sN}{N^2-1} \frac{A^{1/3}}{R_0^2} xG(x,1/({\bf x_\perp-y_\perp})^2)
\label{eq:satKM}
\end{equation}
Where the factor $A^{1/3}$ comes from the proportionality with the nuclear medium -- the radius for a nucleus is approximately $R_A\simeq R_0 A^{1/3}$, $R_0\simeq 1.2$ fm. In (\ref{eq:satKM}), $xG(x,Q^2)$ is the gluon distribution of a nucleon inside a nucleus giving an energy dependence to the saturation scale which, as we have said, is usually neglected for the case of $\hat q$.

\subsection{A simple application: the dipole model}

The above rules allow us to write the cross section for the scattering of a (colorless) dipole of size $r^2={\bf x_\perp^2}$ with a medium. Taking the (forward scattering) amplitude as $iT=1-S$ and applying the optical theorem, 
$\sigma=2{\rm Im} T$
\begin{equation}
\sigma_{\rm tot}^{\rm dip}(r)= 2-\frac{2}{N}{\rm Tr}\langle W^{\cal y}({\bf x_\perp})W({\bf 0})\rangle\simeq
2\left[1-\exp\left\{-\frac{1}{4}Q_{\rm sat}^2\ r^2\right\}\right]
\end{equation}

This dipole cross section finds a direct application in the calculation of the deep inelastic scattering at small values of the Bjorken variable $x$. The dipole model tells us that in these conditions, the collision can be seen as a two-step process in which the virtual photon from the lepton splits into a $q\bar q$ dipole and then this fluctuation scatters with the hadronic target. The problem now reduces to compute the photon-dipole vertex $\Psi^{\gamma*}_{T,L}$ for a transversely or longitudinally polarized photon (see e.g. \cite{Kovchegov:1999kx}) which factorizes to give the cross section for the virtual photon-hadron as
\begin{equation}
\sigma^{\gamma*h}_{T,L}(x,Q^2)=\int d^2{\bf r}\int_0^1 dz\, |\Psi^{\gamma*}_{T,L}|^2\sigma_{\rm tot}^{\rm dip}({\bf r},x)
\label{eq:diph}
\end{equation}
Eq. (\ref{eq:diph}) is extensively used in the phenomenology of DIS to include saturation effects -- simpler to include in configuration space as we do here. As the photon wave function is known, the problem reduces, in general, to compute the dipole cross section for different medium averages and including different resummation techniques for the non-linear terms.

\subsection{\label{RMB}Relation with the momentum broadening}

From the parton S matrix one can readily compute the momentum broadening. After passing though the medium, the particle distribution is given by
\be
\frac{d{\cal N}}{d^2{\bf p}'_\bot}\propto \int dp'_+ \delta (p'^2-m^2)
\frac{1}{N}{\rm Tr} \left<
\left|S(p',p)\right|^2
\right>
\ee
And the transverse momentum broadening is given by
\be
\left<{\bf p }^2_\bot\right>=\frac{1}{\cal N}\int d^2{\bf p}'_\bot {\bf p'}^2_\bot \frac{d{\cal N}}{d^2{\bf p}'_\bot}
\ee
From \Eq{eq:noneikonal} we find 
\be
{\cal N}\left<{\bf p }^2_\bot\right> \propto
\int d\ppperp  \int d\xperp d\xpperp \ppperp^2 e^{i\ppperp\left(\xperp-\xpperp\right)}
\int \mathcal{D} {\bf r} \mathcal{D} {\bf r'}
\nonumber
\\
\times\exp\left\{ i \frac{p_+}{2} 
     \int dx_+\left(\left[\frac{d{\bf r}}{d x_+}\right]^2-\left[\frac{d{\bf r'}}{d x_+}\right]^2\right)
    \right\}
 \frac{1}{N}{\rm Tr}\left<W^\dagger({\bf r'}) W({\bf r})\right>
\ee
A simple manipulation leads to
\be
{\cal N}\left<{\bf p }^2_\bot\right> &\propto&
\lim_{\Delta \xperp\rightarrow 0 } -\grad^2_{\Delta \xperp} 
\int \mathcal{D} {\bf r} \mathcal{D} {\bf r'}
\exp\left\{ i \frac{p_+}{2} 
     \int dx_+\left(\left[\frac{d{\bf r}}{d x_+}\right]^2-\left[\frac{d{\bf r'}}{d x_+}\right]^2\right)
    \right\}
\nonumber
\\
 & & \times \frac{1}{N}{\rm Tr}\left<W^\dagger({\bf r'}) W({\bf r})\right>
\ee
In the high energy limit, $p_+>>\mu$, we can approximate the path integral 
by the classical path with smallest action, $d{\bf r}/dx_+=0$. Performing  a 
similar manipulation for the normalization we obtain
\be
\left<{\bf p }^2_\bot\right>
&=&-\frac{1}{{\rm Tr} \left<W^\dagger(\xav) W(\xav)\right>}
\times\nonumber\\
&&
\grad^2_{\Delta \xperp} 
{\rm Tr}\left<W^\dagger\left(\xav-\frac{\Delta \xperp}{2}\right) 
              W        \left(\xav+\frac{\Delta \xperp}{2}\right)\right>
\ee
with $\xav=(\xperp+\xpperp)/2$. If the medium is large in the transverse
direction, the dependence in $\xav$ drops.

The expression derived above is general. In the particular case of 
the multiple soft scattering, we find 
\be
\left<{\bf p }^2_\bot\right> =\frac{1}{\sqrt{2}}\int dx_+ \hat{q}(x_+)
\ee
Thus, we can interpret the $\hat{q}$ parameter as the momentum broadening
per unit length \cite{Baier:1996sk}.

Note also that within this approximation, in a homogeneous medium 
the number particle distribution at a certain transverse position 
after passing though a medium of lenght L is
\be
\mN
\left(\xperp\right)=\int \frac{d\pperp}{2\pi} e^{-i\xperp \pperp} 
\mN\left(\pperp\right)\propto {\rm Tr} \left<W^{\dagger}(0)W(\xperp)\right>
\ee
Thus, the particle distribution follows a diffusion equation in transverse 
space:
\be
\label{hatqmb}
\partial_+ {\cal N}= \frac{1}{4\sqrt{2}}\, \hat{q}\, \nabla^2_{{\bf p}_\perp} {\cal N}.
\ee
From this fact, the interpretation of $\hat{q}$ as a jet transport parameter
is clear.

\section{The medium--induced gluon radiation}
\label{sec:migr}

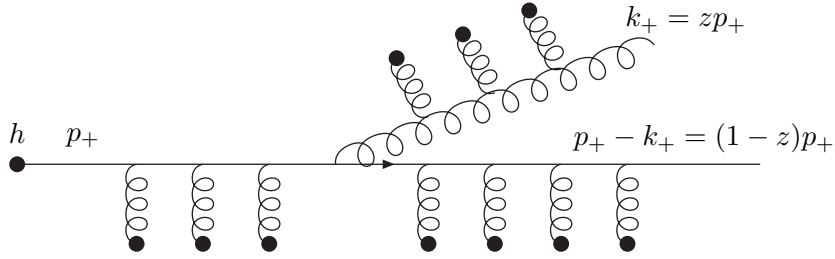
\begin{figure}
\begin{center}
\begin{picture}(300,90)(0,0)
\ArrowLine(10,30)(290,30)
\Gluon(55,0)(55,30){4}{3}
\Gluon(80,0)(80,30){4}{3}
\Gluon(105,0)(105,30){4}{3}
\Gluon(130,30)(250,75){5}{10}
\Gluon(165,0)(165,30){4}{3}
\Gluon(190,0)(190,30){4}{3}
\Gluon(215,0)(215,30){4}{3}
\Gluon(240,0)(240,30){4}{3}
\Gluon(165,48)(153,70){4}{3}
\Gluon(190,57)(178,79){4}{3}
\Gluon(215,66)(203,88){4}{3}
\Vertex(55,0){3}
\Vertex(80,0){3}
\Vertex(105,0){3}
\Vertex(165,0){3}
\Vertex(190,0){3}
\Vertex(215,0){3}
\Vertex(240,0){3}
\Vertex(153,70){3}
\Vertex(178,79){3}
\Vertex(203,88){3}
\Vertex(10,30){3}
\Text(10,42)[]{$h$}
\Text(35,40)[]{$p_+$}
\Text(270,40)[]{$p_+-k_+=(1-z)p_+$}
\Text(263,85)[]{$k_+=zp_+$}
\end{picture}
\caption{The medium-induced gluon radiation diagram.}
\label{fig:medind}
\end{center}
\end{figure}

With the formalism developed in the last sections we can now compute the medium-induced gluon radiation needed for jet quenching studies. In this formalism, the fast particle is produced at a given point inside the medium where the hard process, $h$, takes place, see Fig. \ref{fig:medind}. Then this particle and the emitted gluon suffer multiple scattering described by the path integral propagators  (\ref{eq:propag}). We will work in the approximation $p_+\gg k_+\gg k_\perp$ but keep terms in $k_\perp^2/k_+$ as explained in the previous section.

\subsection{The gluon emission vertex}

\begin{figure}
\begin{center}
\begin{picture}(300,80)(0,0)
\GCirc(20,40){10}{0.5}
\Line(30,40)(120,40)
\Gluon(50,40)(120,70){4}{5}
\Text(40,30)[]{$p_i$}
\Text(100,30)[]{$p_f$}
\Text(90,73)[]{$k$}
\GCirc(190,40){10}{0.5}
\Line(200,40)(300,40)
\Gluon(240,40)(290,70){4}{5}
\Gluon(225,40)(225,10){3}{3}
\Vertex(225,10){3}
\Text(210,30)[]{$p_i$}
\Text(280,30)[]{$p_f$}
\Text(235,25)[]{$q$}
\Text(260,73)[]{$k$}
\Text(10,58)[]{${\cal M}_h$ }
\Text(180,58)[]{${\cal M}_h$ }
\Text(70,5)[]{(a)}
\Text(250,5)[]{(b)}
\end{picture}
\caption{Radiation diagrams after a hard process represented by the blob.}
\label{fig:vertex}
\end{center}
\end{figure}
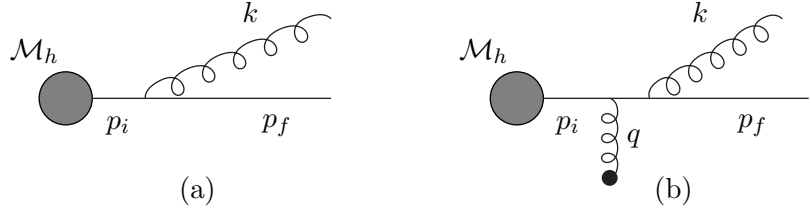

Let us first compute the amplitude for the radiation when the gluon does not interact after it is emitted. We start with  the case when there is no scattering either of the quark, i.e. we consider the case of the radiation off a quark which has been produced in a hard process, $h$, with amplitude ${\cal M}_h$ -- see  Fig. \ref{fig:vertex} -- we fix ${\bf p}_{i\perp}=0$ to have
\begin{equation}
{\cal M}_{\rm rad}^{(0)}={\cal M}_{h}(p_{f}+k) i(igT^a)\frac{2 p_f\cdot\epsilon^a}{(p_f+k)^2} u(p_f)\simeq {\cal M}_{h}(-2gT^a)\frac{\bf k_\perp\cdot\epsilon_\perp}{{\bf k_\perp}^2} u(p_f)
\label{eq:rad1}
\end{equation}
where we have used the high-energy approximation to write
\begin{equation}
\frac{p_f\cdot\epsilon}{p_f\cdot k}\simeq\frac{p_i\cdot\epsilon}{p_i\cdot k}\simeq 2\,\frac{\bf k_\perp\cdot\epsilon_\perp}{{\bf k_\perp}^2}
\label{eq:epsapprox}
\end{equation}

The spectrum of radiated gluon is computed as the radiation over the {\it elastic} cross sections, where by elastic we mean the hard process ${\cal M}_h$. Introducing the relevant phase-space factors we find the double differential spectrum of radiated gluons in the vacuum 
\begin{equation}
k_+\frac{dI^{\rm vac}}{dk_+ d^2{\bf k}_\perp}\simeq \frac{\alpha_s C_R}{\pi^2}\frac{1}{{\bf k}_\perp^2}
\label{eq:vacrad}
\end{equation}
where $C_R=C_F=(N^2-1)/2N$ for quarks and $C_R=C_A=N$ for gluons correspond to the average over the color states -- see eq. (\ref{eq:colortrace}). Eq. (\ref{eq:vacrad}) is the leading $1/z$ contribution to the vacuum radiation spectrum. For the case where one scattering is present -- Fig. \ref{fig:vertex} (b) -- the amplitude is\footnote{For clarity, we omit the term ${\cal M}_h$ which cancels when the spectrum is computed -- see eq. (\ref{eq:vacrad}) -- if the momenta $q$ are soft enough as we will standardly assume.}
\begin{equation}
{\cal M}_{\rm rad}^{(1)}
\simeq -2 g\int dx_+dx_\perp i  g T^bA^b_-(x)T^a \frac{\bf k_\perp\cdot\epsilon_\perp}{{\bf k_\perp}^2}\ {\rm e}^{-i(p_f+k)_\perp\cdot x_\perp}u(p_f)
\end{equation}
This expression contains a factorization of the radiation amplitude (\ref{eq:rad1}) and the collision term which can be identified as the first term in the expansion of the Wilson line. The corresponding contribution resumming an arbitrary number of scatterings for a gluon emitted outside the medium is then
\begin{equation}
{\cal M}^{\rm rad}_q
\simeq -2 g\int dx_\perp W({\bf x}_\perp,x_{0+},L_+) T^a\frac{{\bf k_\perp\cdot\epsilon}^a_\perp}{{\bf k_\perp}^2}\ {\rm e}^{-i(p_f+k)_\perp\cdot x_\perp} u(p_f)
\label{eq:ampq}
\end{equation}
Where $x_{0+}$ and $L_+$ are the positions where the medium begins and ends in light-cone variables -- so that the $x_+$-integral in the Wilson line has these limits -- and the subscript $q$ is included to signal that the gluon does not interact with the medium. 

\begin{figure}
\begin{center}
\begin{picture}(300,80)(0,0)
\Line(40,5)(120,5)
\Gluon(60,5)(100,40){4}{4}
\Gluon(100,40)(260,40){4}{15}
\Gluon(100,42)(100,80){4}{5}
\Gluon(130,42)(130,80){4}{5}
\Gluon(160,42)(160,80){4}{5}
\Gluon(190,42)(190,80){4}{5}
\Gluon(230,42)(230,80){4}{5}
\Vertex(100,80){3}
\Vertex(130,80){3}
\Vertex(160,80){3}
\Vertex(190,80){3}
\Vertex(230,80){3}
\Text(50,13)[]{$p_n$}
\Text(100,13)[]{$p_{n+1}$}
\Text(250,30)[]{$k$}
\end{picture}
\caption{Radiation vertex when the gluon reinteracts with the medium.}
\label{fig:vertexglue}
\end{center}
\end{figure}
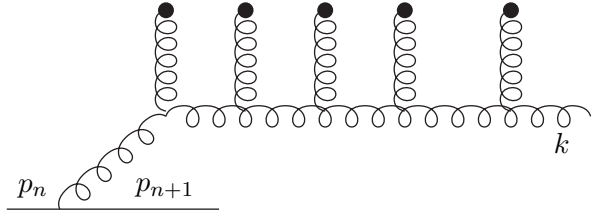

The case when the emitted gluon interacts with the medium can be computed similarly and the corresponding contribution from the vertex would give ${{\bf k}_{1\perp}/{\bf k}_{1\perp}}^2$, which is now an internal integration variable. In order to deal with it, we will write the delta function for the momentum conservation in the radiation vertex as
\begin{equation}
(2\pi)^4\delta^4(-k_1-p_{n+1}+p_n)=\int d^4 y\ {\rm e}^{i(-k-p_{n+1}+p_n)\cdot y}.
\label{eq:delta}
\end{equation}
and integrate over the three momenta $k_1$, $p_{n+1}$, $p_n$. By doing this, $y$ signals now the spatial position of the radiation vertex. Let us define the contribution from a gluon radiated at $y$ from a quark with momentum $p_n$ before the splitting and $p_{n+1}$ after it -- see Fig. \ref{fig:vertexglue} -- as\footnote{We have used here the steps explained in Section \ref{sec:eikonal}, in particular eq. (\ref{eq:gmunusimp}),  to simplify the triple gluon vertices in the high-energy limit}
\begin{eqnarray}
V_g^{\mu(n)}=\int d^4 y\ {\rm e}^{-i(p_{n+1}-p_n)\cdot y}\prod_{j=1}^{n}\int \frac{d^4 k_j}{(2\pi)^4}d^4x_j \epsilon_{(i)}^\mu(k_1)\left[-2g\frac{k_{j+}A_-(x_j)}{k_{j}^2+i\epsilon}\right]\times\nonumber\\
\times \exp\left\{-i(k_{j+1}-k_j)\cdot x_j\right\}\exp\left\{-i k_1\cdot y\right\}.\hspace{0.3cm}
\label{eq:greendelta}
\end{eqnarray}
(we write $k_{n+1}\equiv k_f$ for the final momentum of the gluon). Writing $\epsilon_-={\bf k}_{1\perp}\cdot {\bf \epsilon}_\perp/k_{1+}$ and proceeding as explained in section \ref{sec:relaxing} it is easy to see that the $-$-component of (\ref{eq:greendelta}), which will be used later, is 
\begin{eqnarray}
V_g^{-(n)}\simeq\int d^4 y\  d^2{\bf x}_{1\perp}dx_{1+} \ {\rm e}^{-i(p_{n+1}-p_n)\cdot y}\frac{i}{k_+}\epsilon_\perp\cdot\frac{\partial}{\partial {\bf y}_\perp} G_0({\bf y_\perp},y_+;{\bf x}_{1\perp},x_{1+})\nonumber \\ \times igA_-(x_1)\prod_{j=2}^{n}\int d^2{\bf x}_{j\perp}dx_{j+}
G_0({\bf x_{j-1\perp}},x_{j-1+};{\bf x}_{j\perp},x_{j+})
 igA_-(x_j){\rm e}^{i{\bf k}_{f\perp}\cdot{\bf x}_{n\perp}}\hspace{0.2cm}
 \end{eqnarray}
so that the sum over scattering centers gives\footnote{We include here the case when the gluon does not interact once it is emitted {\it inside} the medium, otherwise a factor $G_0({\bf y_\perp},y_+;{\bf x}_{n\perp},x_{n+}) $ should be subtracted.}
\begin{equation}
V_g^-\simeq\frac{i}{k_+}\int d^4 y\ {\rm e}^{-i(p_{n+1}-p_n)\cdot y}\epsilon_\perp\cdot\frac{\partial}{\partial {\bf y}_\perp}\int d^2{\bf x}_\perp G({\bf y_\perp},y_+;{\bf x}_{n\perp},x_{n+}) ){\rm e}^{-i{\bf k}_{f\perp}\cdot{\bf x}_{\perp}}
\label{eq:greendelta2}
\end{equation}
And we define for future convenience
\begin{equation}
{\cal A}_-(y)\equiv\frac{i}{k_+}\epsilon_\perp\cdot\frac{\partial}{\partial {\bf y}_\perp}\int d^2{\bf x}_\perp G({\bf y_\perp},y_+;{\bf x}_{n\perp},x_{n+}) ){\rm e}^{-i{\bf k}_{f\perp}\cdot{\bf x}_{\perp}}
\end{equation}

In the way eq. (\ref{eq:greendelta}) has been defined, the amplitude for the process $q\to qg$ corresponds to make the replacement $A(y)\to {\cal A}(y)$ for one of the scattering centers in the derivation of the quark Wilson line described in Section \ref{sec:eikonal}, so that the calculation is now similar to the one performed before. More specifically, the amplitude of Fig. \ref{fig:medind} can be written as
\begin{eqnarray}
{\cal M}_g^m=\int\prod_{i=0}^m\left[\frac{d^4 p_i}{(2\pi)^4}d^4x_i\right] {\rm e}^{-ip\cdot x_1}\prod_{j=0}^n\left[(-2gT^{a_j})\frac{A^{a_j}(x_j)\cdot p_j}{p_j^2+i\epsilon}{\rm e}^{ip_j\cdot(x_j-x_{j+1})}\right]\times\nonumber\\
\times (-2gT^b)\frac{{\cal A}^b(x_{n+1})\cdot p_{n+1}}{p_{n+1}^2+i\epsilon}{\rm e}^{ip_{n+1}\cdot(x_{n+1}-x_{n+2})}\times\nonumber\\
\times \prod_{k=n+2}^m \left[(-2gT^{a_k})\frac{A^{a_k}(x_k)\cdot p_k}{p_k^2+i\epsilon}{\rm e}^{ip_k\cdot(x_k-x_{k+1})}\right]{\rm e}^{ip_f\cdot x_m} u(p_f)\hspace{0.3cm}
\end{eqnarray}
The structure in poles and phases is the same as the one studied in Section \ref{sec:eikonal}. The poles give again an ordering in the longitudinal variable $x_+$, so that the position of the radiation contribution ${\cal A_-}(x_+,{\bf x_{\perp}})$ is ordered with the rest of the external fields $A_-(x_+,{\bf x_\perp})$ to give
\begin{eqnarray}
{\cal M}_g^{\rm rad}=\int_{x_{0+}}^{L_+} dx_+\int d{\bf y}_\perp {\rm e}^{i(p_f-p_i)_\perp\cdot y_\perp}{\cal P}\exp\left[ig\int_{x_{0+}}^{x_+} dy_+ A(y_+,{\bf y}_\perp)\right]\times\nonumber\\i2gT^b{\cal A}^b(x_+,{\bf y}_\perp){\cal P}\exp\left[ig\int_{x_+}^{L_+} dy_+ A(y_+,{\bf y}_\perp)\right]
\end{eqnarray}
We have made explicit the color matrix $T^b$ at the radiation vertex while it is included as a redefinition of the external fields in the rest of the cases as done in previous sections. We are interested in the case where the quark is completely eikonal, so, we fix ${\bf y}_\perp=0$ to get
\begin{eqnarray}
{\cal M}_g^{\rm rad}=-\frac{2 g}{k_+}\int_{x_{0+}}^{L_+} dx_+\int d{\bf x}^2_\perp {\rm e}^{-i{\bf k}_\perp\cdot{\bf x}_\perp}
W({\bf 0};x_{0+},x_+)\times\nonumber\\ \times T^b\epsilon_\perp\cdot\frac{\partial}{\partial{\bf y}_\perp}G^b({\bf y}_\perp=0,x_+;{\bf x_\perp},L_+) W({\bf 0};x_+,L_+)
\label{eq:ampg}
\end{eqnarray}

The total amplitude for the medium-induced gluon radiation is then the sum of (\ref{eq:ampq}) and (\ref{eq:ampg})
\begin{equation}
{\cal M}^{\rm rad}={\cal M}_q^{\rm rad}+{\cal M}_g^{\rm rad}\, .
\label{eq:amp}
\end{equation}
We will now compute the spectrum of radiated gluons in the presence of a medium, including all the relevant color factors to perform the medium averages.

\subsection{\label{migr}The medium-induced gluon radiation}

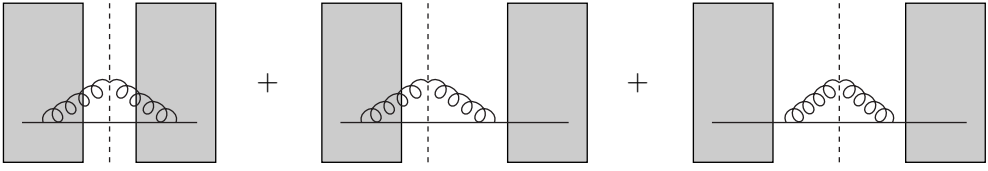
\begin{figure}
\begin{center}
\begin{picture}(350,90)(0,0)
\GBoxc(10,40)(30,60){0.8}
\GBoxc(60,40)(30,60){0.8}
\GBoxc(130,40)(30,60){0.8}
\GBoxc(200,40)(30,60){0.8}
\GBoxc(270,40)(30,60){0.8}
\GBoxc(350,40)(30,60){0.8}
\Line(2,25)(68,25)
\Gluon(10,25)(35,40){3}{4}
\Gluon(35,40)(60,25){3}{4}
\Line(122,25)(208,25)
\Gluon(130,25)(155,40){3}{4}
\Gluon(155,40)(180,25){3}{4}
\Line(262,25)(358,25)
\Gluon(290,25)(310,40){3}{4}
\Gluon(310,40)(330,25){3}{4}
\Text(95,40)[]{$+$}
\Text(235,40)[]{$+$}
\DashLine(35,10)(35,70){2}
\DashLine(155,10)(155,70){2}
\DashLine(310,10)(310,70){2}
\end{picture}
\caption{The three contributions to the squared amplitude of the medium-induced gluon radiation. The dashed line is the cut indicating the final outgoing particles.}
\label{fig:medind2}
\end{center}
\end{figure}

The locality of the medium averages -- see also below -- allows for a simple diagramatical interpretation, in which three different cases appear when the amplitude (\ref{eq:amp}) is squared depending on the position of the radiation vertex:  when the gluon is emitted inside the medium in both amplitude and conjugate amplitude; when it is emitted inside the medium in amplitude and outside the medium in conjugate amplitude; and finally when the gluon is emitted outside the medium in both amplitude and conjugate amplitude -- see Fig. \ref{fig:medind2}.  We take the case that $x_+<\bar x_+$ to obtain -- see eqs. (\ref{eq:comp})--(\ref{eq:qqtog})

\begin{eqnarray}
\langle|{\cal M}_{a\to bc}|^2\rangle=\frac{g^2}{N^2-1}2{\rm Re}\Bigg[\frac{1}{k_+^2}\int_{x_{0+}}^{L_+} dx_+\int_{x_+}^{L_+} d\bar x_+\int d{\bf x}d{\bf \bar x}\
e^{i{\bf k}_\perp({\bf x}-{\bf \bar{x}})}\times\nonumber\\
 \Big\langle W_{aa_1}({\bf 0},x_{0+},x_+)
T_{a_1b_1}^{c_1}\frac{\partial}{\partial {\bf y}}G_{c_1c}({\bf y}=0,x_+;{\bf x},L_+)W_{b_1b}({\bf 0},x_+,L_+)\times\nonumber\\
W_{b\bar b_1}^{\cal y}({\bf 0},\bar x_+,L_+)
\frac{\partial}{\partial {\bf \bar y}}G_{c\bar c_1}({\bf \bar x},L_+;{\bf \bar y}=0,\bar x_+)T^{\bar c_1}_{\bar b_1\bar a_1}W^{\cal y}_{\bar a_1a}({\bf 0},x_{0+},\bar x_+)\Big\rangle-\nonumber\\
-\frac{2}{k_+}\frac{\bf k_\perp}{k_\perp^2}\int_{x_{0+}}^{L_+} dx_+\int d{\bf x}{\rm e}^{i{\bf k}_\perp{\bf x}}\Big\langle W_{aa_1}({\bf 0},x_{0+},x_+)T_{a_1b_1}^{c_1}\times\nonumber\\ \frac{\partial}{\partial {\bf y}}G_{c_1c}({\bf y}=0,x_+;{\bf x},L_+)
\times W_{b_1b}({\bf 0},x_+,L_+)T^{c_1}_{b\bar a_1}W^{\cal y}_{\bar a_1 a}({\bf 0},x_{0+},L_+)\Big\rangle\Bigg]+\nonumber\\ +\frac{4g^2C_R}{k_\perp^2}\hspace{0.5cm}
\label{eq:medind0}
\end{eqnarray}
Where we have written an eikonal Wilson line $W({\bf x_\perp},x_{1+},x_{2+})$ for each quark propagation between positions $x_{1+}-x_{2+}$, the gluon propagators\footnote{Notice a small change of notation here on the order of the variables for the propagators $G$ to more easily follow the lines in the figures.} by $G({\bf x_\perp},x_{1+};{\bf y_\perp},x_{2+})$ and the corresponding vertex factors as determined in the previous section. We have also explicitly included the color indices. 

\begin{figure}
\begin{center}
\begin{picture}(150,120)(0,0)
\GBoxc(60,55)(140,90){0.9}
\Line(10,55)(140,55)
\Line(10,45)(140,45)
\Gluon(40,55)(140,90){3}{10}
\Gluon(60,45)(140,10){-3}{10}
\Text(12,60)[]{$a$}
\Text(12,38)[]{$a$}
\Text(138,61)[]{$b$}
\Text(138,38)[]{$b$}
\Text(35,60)[]{$a_1$}
\Text(55,38)[]{$\bar a_1$}
\Text(75,61)[]{$b_1$}
\Text(97,38)[]{$\bar b_1$}
\Text(50,70)[]{$c_1$}
\Text(78,27)[]{$\bar c_1$}
\Text(138,20)[]{$c$}
\Text(138,100)[]{$c$}
\DashLine(42,5)(42,105){2}
\DashLine(62,5)(62,105){2}
\Text(42,110)[]{$x_+$}
\Text(62,110)[]{$\bar x_+$}
\end{picture}
\caption{The color structure of the medium averages in eq. (\protect\ref{eq:medind0}).}
\label{fig:medind3}
\end{center}
\end{figure}
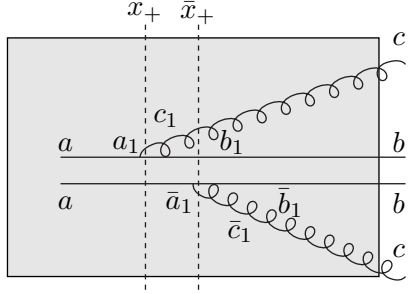

All the medium averages can be written in terms of the Wilson loop average (\ref{eq:multiple}) for gluons. To see how this works, it is useful to draw the conjugate amplitude as the Wilson line for the corresponding antiparticle in the amplitude -- see Fig. \ref{fig:medind3}.

Taking into account the composition of propagators with $x_+<z_+<y_+$
\begin{equation}
G({\bf x_\perp},x_+;{\bf y_\perp},y_+)=\int d{\bf z_\perp}G({\bf x_\perp},x_+;{\bf z_\perp},z_+)G({\bf z_\perp},z_+;{\bf y_\perp},y_+)
\label{eq:comp}
\end{equation}
One can make the substitutions
\begin{eqnarray}
 W_{aa_1}({\bf 0},0,x_+)W^{\cal y}_{\bar a_1a}({\bf 0},0,\bar x_+)\longrightarrow W^{\cal y}_{\bar a_1a_1}({\bf 0},x_+,\bar x_+)\label{eq:line1}\\
 W_{b_1b}({\bf 0},x_+,L)W_{b\bar b_1}^{\cal y}({\bf 0},\bar x_+,L)\longrightarrow W_{b_1b}({\bf 0},x_+,\bar x_+)
 \label{eq:line2}
\end{eqnarray}
Where we have used that in the averages
\begin{eqnarray}
\Big\langle W_{aa_1}({\bf 0},0,x_+)W^{\cal y}_{\bar c a}({\bf 0},0,x_+)\Big\rangle=\delta_{a_1\bar c}\\
\Big\langle W_{c b}({\bf 0},\bar x_+,L)W_{b\bar b_1}^{\cal y}({\bf 0},\bar x_+,L)\Big\rangle=\delta_{c \bar b_1}
\end{eqnarray}
The locality (in $x_+$) of the medium averages we have presented in the previous sections imply that the only non-zero contribution come from Wilson lines which overlap in $x_+$.
Eqs. (\ref{eq:line1}) and (\ref{eq:line2}) tell us that the contribution from the quark rescattering outside the longitudinal range $(x_+,\bar x_+)$ cancel when doing the average. The remaining contribution from the quark lines can, in fact, be written as a Wilson line in the adjoint representation -- i.e. as if the quark-antiquark propagation in the segment $(x_+,\bar x_+)$ would be a gluon propagation because -- see the Appendix:
\begin{equation}
T_{ab_1}^{c_1}W^{\cal y}_{a\bar a_1}({\bf 0},x_+,\bar x_+)T_{\bar b_1\bar a_1}^{\bar c_1}W_{b_1\bar b_1}({\bf 0},x_+,\bar x_+)\longrightarrow W^{A}_{c_1\bar c_1}({\bf 0},x_+,\bar x_+)
\label{eq:qqtog}
\end{equation}
With these simplifications, only possible in the case that the quark is taken as completely eikonal, eq. (\ref{eq:medind0}) can be written in terms only of Wilson lines in the adjoint representation where the medium averages (\ref{eq:multiple}) for gluons can be used.
\begin{eqnarray}
\langle|{\cal M}_{a\to bc}|^2\rangle=\frac{g^2}{N^2-1}2{\rm Re}\Bigg[\frac{1}{k_+^2}\int_{x_{0+}}^{L_+} dx_+\int_{x_+}^{L_+} d\bar x_+\int d{\bf x}d{\bf \bar x}
e^{i{\bf k}_\perp({\bf x}-{\bf \bar{x}})}
\times\nonumber\\ 
\times{\rm Tr}\Big\langle W^A({\bf 0},x_+,\bar x_+)\frac{\partial}{\partial {\bf y}}G({\bf y}=0,x_+;{\bf x},{L_+})
\frac{\partial}{\partial {\bf \bar y}}G({\bf \bar x},{L_+};{\bf \bar y}=0,\bar x_+)\Big\rangle-\nonumber\\
-\frac{2}{k_+}\frac{\bf k_\perp}{k_\perp^2}\int_{x_{0+}}^{L_+} dx_+ \int d{\bf x} e^{i{\bf k}_\perp{\bf x}}
{\rm Tr}\Big\langle W^A({\bf 0},x_+,L_+)\frac{\partial}{\partial {\bf y}}G({\bf y}=0,x_+;{\bf x},{L_+})\Big\rangle\Bigg]
\nonumber\\
+\frac{4 C_R g^2}{k_\perp^2},\ \ \ 
\label{eq:medind1}
\end{eqnarray}
Using the relation eq. (\ref{eq:comp}) we have the average of only two Wilson lines at each segment in Fig. \ref{fig:medind3} -- we write explicitly the two transverse components, then ${\bf y}$ and ${\bf \bar y}$ will be put to ${\bf 0}$ again:
\begin{eqnarray}
{\rm Tr}\left\langle W^A({\bf y},x_+,\bar x_+)G({\bf y},x_+;{\bf x},\bar x_+)\right\rangle\label{eq:av1}\\
{\rm Tr}\left\langle G({\bf y},\bar x_+;{\bf x},L_+)G(\bar {\bf x},L_+;\bar{\bf y},\bar x_+)\right\rangle\label{eq:av2}\\
{\rm Tr}\left\langle W^A({\bf y},x_+,L_+)G({\bf y},x_+;{\bf x},L_+)\right\rangle\label{eq:av3}
\end{eqnarray}
The averages (\ref{eq:av1}) and (\ref{eq:av2}) -- taking $\bar x_+=L_+$ -- give
\begin{eqnarray}
\frac{1}{N^2-1}{\rm Tr}\left\langle W^A(\tilde{\bf y},x_+,\bar x_+)G({\bf y},x_+;{\bf x},\bar x_+)\right\rangle=\nonumber\\
=\int {\cal D}{\bf r}\exp\left[\frac{ip_+}{2}\int_{x_+}^{\bar x_+} d\xi \,\dot{\bf r}^2(\xi)\right]\exp\left[-\frac{1}{2}n(\xi)\sigma({\bf r})\right]
\label{eq:avsec1}
\end{eqnarray}
with the boundary conditions ${\bf r}(x_+)={\bf y}-\tilde{\bf y}$ and ${\bf r}(\bar x_+)={\bf x}-\tilde{\bf y}$. The other average gives \cite{Zakharov:1998sv,Wiedemann:1999fq,MehtarTani:2006xq}
\begin{eqnarray}
\frac{1}{N^2-1}{\rm Tr}\left\langle G({\bf y},\bar x_+;{\bf x},L_+)G(\bar {\bf x},L_+;\bar{\bf y},\bar x_+)\right\rangle=\nonumber\\
=\int{\cal D}{\bf r}_1\int {\cal D} {\bf r}_2 \exp\left\{i\int_{x_+}^{\bar x_+}d\xi\left[\frac{p_+}{2}\left(\dot{\bf r}^2_1(\xi)-\dot{\bf r}^2_2(\xi)\right)+\frac{i}{2}n(\xi)\sigma({\bf r}_1-{\bf r}_2)\right]\right\}=\nonumber\\
=\int{\cal D}{\bf u}\int {\cal D} {\bf v} \exp\left\{i\int_{x_+}^{\bar x_+}d\xi\left[\frac{p_+}{2}\dot{\bf u}(\xi)\cdot\dot{\bf v}(\xi)+\frac{i}{2}n(\xi)\sigma({\bf v(\xi)})\right]\right\}\hspace{0.5cm}
\end{eqnarray}
where we have made the change of variables 
\begin{eqnarray}
{\bf u}(\xi)={\bf r}_1+{\bf r}_2\nonumber\\
{\bf v}(\xi)={\bf r}_1-{\bf r}_2
\end{eqnarray}
The integration in ${\bf u}$ gives a $\delta$-function constraining ${\bf v}$ to a straight line
\begin{equation}
{\bf v}_s(\xi)={\bf v}(\bar x_+)\frac{\xi-x_+}{\bar x_+-x_+}+{\bf v}(x_+)\frac{\bar x_+-\xi}{\bar x_+-x_+}
\end{equation}
the result is \cite{Zakharov:1998sv}
\begin{eqnarray}
\frac{1}{N^2-1}{\rm Tr}\left\langle G({\bf y},\bar x_+;{\bf x},L_+)G(\bar {\bf x},L_+;\bar{\bf y},\bar x_+)\right\rangle=\nonumber\\
\left(\frac{p_+}{2\pi\Delta\xi}\right)^2\exp\left\{\frac{ip_+}{2\Delta\xi}\left[({\bf y}-{\bf x})^2-(\bar{\bf y}-\bar{\bf x})^2\right]-\frac{1}{2}\int d\xi n(\xi)\sigma({\bf w}(\xi))\right\}
\end{eqnarray}
with
\begin{equation}
{\bf w}(\xi)=({\bf y}-\bar{\bf y})\frac{L_+-\xi}{\Delta\xi}+({\bf x}-\bar{\bf x})\frac{\xi-\bar x_+}{\Delta\xi}\, ;\ \ \ \Delta\xi=L_+-\bar x_+
\label{eq:avsec2}
\end{equation}
Eq. (\ref{eq:avsec1}) and (\ref{eq:avsec2}) allow to compute the average of the first term in (\ref{eq:medind1})
\begin{eqnarray}
\int d{\bf x}d\bar {\bf x}\frac{{\rm e}^{i{\bf k}_\perp(\bar{\bf x}-{\bf x})}}{N^2-1}{\rm Tr}\left\langle W^A(\tilde{\bf y},x_+,\bar x_+)G({\bf y},x_+;{\bf x},L_+)G(\bar{\bf y},\bar x_+;\bar {\bf x},L_+)\right\rangle=\nonumber\\
=\frac{1}{N^2-1}\int d{\bf x}d\bar {\bf x}d{\bf z}\,{\rm e}^{i{\bf k}_\perp(\bar{\bf x}-{\bf x})}\ {\rm Tr}\left\langle W^A(\tilde{\bf y},x_+,\bar x_+)G({\bf y},x_+;{\bf z},\bar x_+)\right\rangle\times\nonumber\\
\times{\rm Tr}\left\langle G({\bf z},\bar x_+;{\bf x},L_+)G(\bar {\bf x},L_+;\bar{\bf y},\bar x_+)\right\rangle=\nonumber\\
=\int d{\bf z}\,{\rm e}^{i{\bf k}_\perp\cdot {\bf z}} \ {\cal K}({\bf y}-\tilde{\bf y},x_+;{\bf z}+\bar {\bf y}-\tilde{\bf y},\bar x_+)\exp\left[-\frac{1}{2}\int d\xi n(\xi)\sigma({\bf z})\right]\hspace{0.3cm}
\end{eqnarray}
Where we have defined
\begin{equation}
{\cal K}\left({\bf r}(x_+),x_+;{\bf r}(\bar x_+),\bar x_+\right)=\int {\cal D}{\bf r}\exp\left[\int_{x_+}^{\bar x_+}d\xi\left(i\frac{p_+}{2} \dot{\bf{r}}^2-\frac{1}{2}n(\xi)\sigma({\bf r})\right)\right]
\label{eq:kprop0}
\end{equation}
Putting all together, we obtain the radiation spectrum in the presence of a medium. 
\begin{eqnarray}
k_+\frac{dI}{dk_+ d^2{\bf k_\perp}}=\frac{\alpha_S C_R}{(2\pi)^2 k_+}2{\rm Re}\int_{x_{0+}}^{L_+} dx_+\int d^2{\bf x}\ e^{-i{\bf k_\perp\cdot x}}\times\nonumber\\\times \Bigg[\frac{1}{k_+}\int_{x_+}^{L_+} d\bar x_+\ e^{-\frac{1}{2}\int_{x_+}^{L_+} d\xi n(\xi) \sigma({\bf x})}\frac{\partial}{\partial{\bf y}}\cdot\frac{\partial}{\partial{\bf x}}{\cal K}({\bf y}=0,x_+;{\bf x},\bar x_+)-\nonumber\\
-2\frac{\bf k_\perp}{{\bf k}_\perp^2}\cdot \frac{\partial}{\partial {\bf y}}{\cal K}({\bf y}=0,x_+;{\bf x},{L_+})\Bigg]+\frac{\alpha_S C_R}{\pi^2}\frac{1}{{\bf k}_\perp^2}
\label{eq:MIGR}
\end{eqnarray}

\subsection{The multiple soft scattering approximation}

In section \ref{sec:medav} we have presented different average procedures to solve (\ref{eq:MIGR}) with (\ref{eq:kprop0}). In the multiple soft scattering approximation, valid for opaque media, the dipole cross section is approximated by its quadratic term and the medium averages of two Wilson lines are given by (\ref{eq:multiple}). In this approximation, the path integrals (\ref{eq:kprop0}) correspond to a harmonic oscillator of imaginary frequency 
\begin{equation}
{\cal K}\left({\bf r}(x_+),x_+;{\bf r}(\bar x_+),\bar x_+\right)=\int {\cal D}{\bf r}\exp\left[i\frac{p_+}{2} \int_{x_+}^{\bar x_+}d\xi\left(\dot{\bf{r}}^2+i\frac{\hat q(\xi)}{2\sqrt{2} p_+} {\bf r}^2\right)\right]
\label{eq:kprop}
\end{equation}
In order to proceed, we need to say something about the temporal dependence of the transport coefficient. Two classes of media have been studied: a static medium in which $\hat q(\xi)=\hat q$ is a constant; an expanding medium in which the density of scattering centers is expected to produce a dilution as $\hat q(\xi)=\hat q_0(\xi_0/\xi)^\alpha$, with $\alpha$ characterizing the speed of the dilution, and $\alpha=1$ for the Bjorken scaling scenario -- see Section \ref{sec:hic}. Explicit solutions for this path integral and the corresponding spectrum (\ref{eq:MIGR}) are given in the Appendix. In the next sections we will present numerical calculations of these spectra, study their properties and explain how they are included in present phenomenology of jet quenching in heavy ion collisions.

\subsection{Numerical results and heuristic discussion: static medium}

\begin{figure}
\begin{minipage}{0.5\textwidth}
\begin{center}
\includegraphics[width=0.85\textwidth]{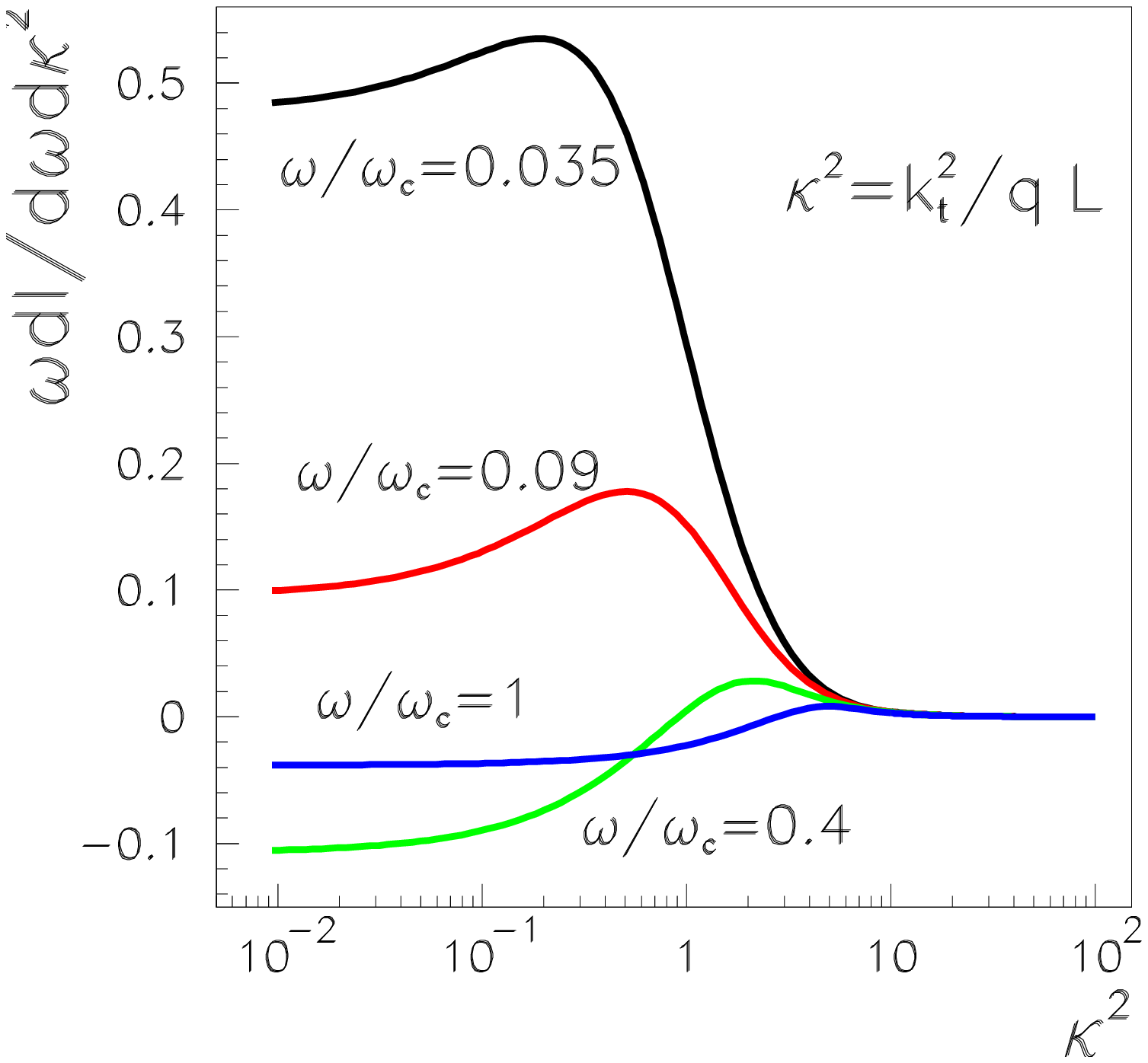}
\end{center}
\end{minipage}
\hfill
\begin{minipage}{0.5\textwidth}
\begin{center}
\includegraphics[width=0.85\textwidth]{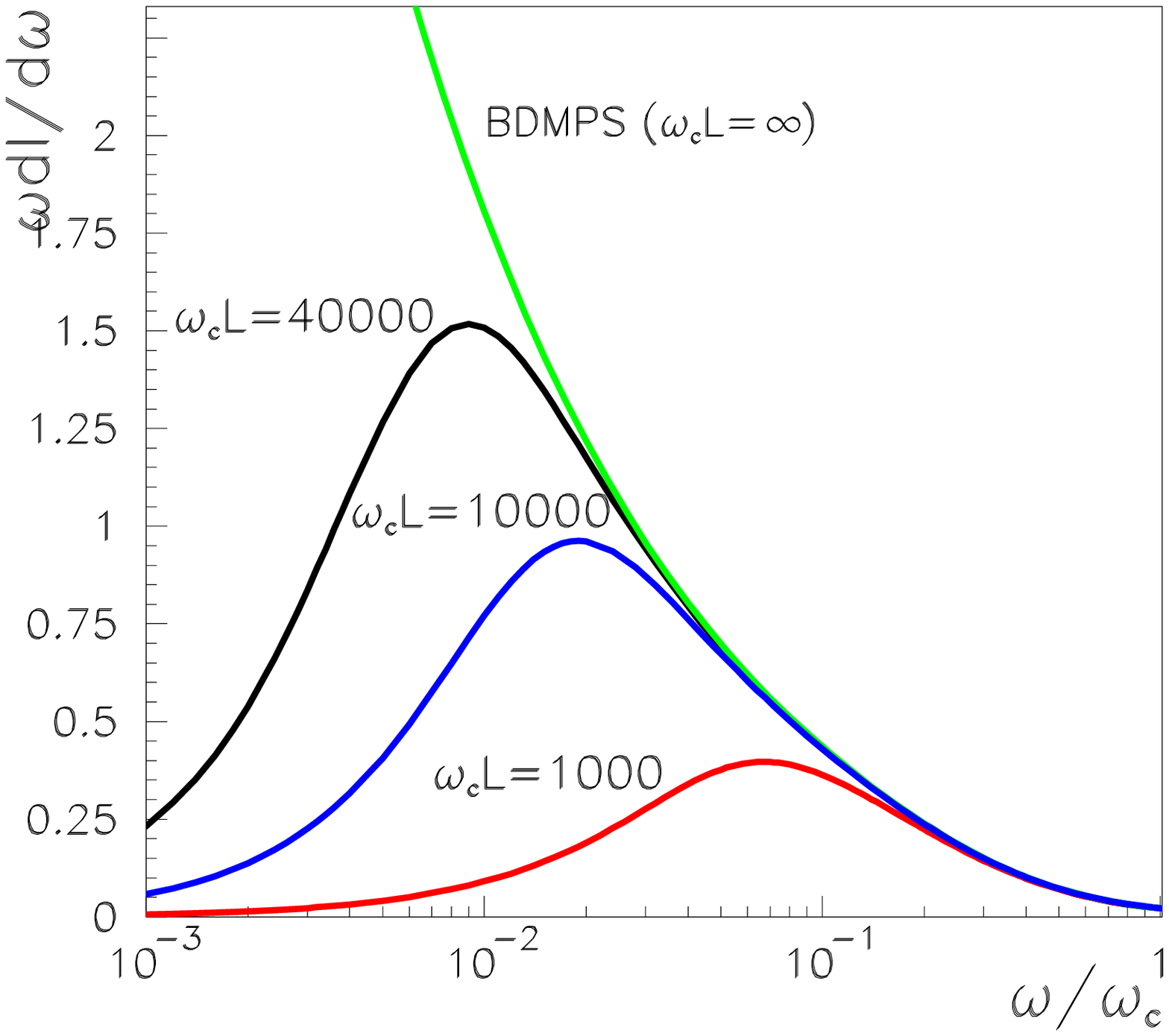}
\end{center}
\end{minipage}
\caption{Left: numerical results for the medium induced gluon radiation spectrum (\protect\ref{eq:MIGR}) of a quark in a static medium as a function of the dimensionless variables (\ref{eq:dimvar}). Right: Same but integrated in $k_t<\omega$}
\label{fig:static}
\end{figure}

In Fig. \ref{fig:static} we present the results for the double-differential medium-induced gluon radiation spectrum for a quark traversing a static medium. The results are given as a function of the variables
\begin{equation}
\omega_c\equiv\frac{1}{2}\hat q\, L^2{\hspace{20pt}} \kappa^2\equiv\frac{k_\perp^2}{\hat qL}
\label{eq:dimvar}
\end{equation}
One important feature of the spectrum is the presence of small-$k_\perp$ and large-$\omega$ cuts which can be understood by the formation time of the gluon
\begin{equation}
t_{\rm form}\simeq \frac{2\omega}{k_\perp^2}.
\label{eq:form}
\end{equation}
The presence of this coherence length can be traced back to the non-eikonal terms in the propagators (\ref{eq:noneikonal}). Recalling that these terms come from keeping $k_\perp^2/2p_+$ corrections in the phases and translating light-cone to ordinary variables
\begin{equation}
\prod{\rm e}^{i\frac{k_\perp^2}{2p_+}(x_{i+}-x_{(i+1)+})}\simeq {\rm e}^{i\frac{k_\perp^2}{2\omega}L},
\end{equation}
these contributions define the coherence time (\ref{eq:form}). When $t_{\rm form}\ll L$ multiple incoherent collisions are present and parametrically the spectrum is proportional to $L/t_{\rm form}$. In the opposite limit, when $t_{\rm form}\gg L$ the gluon formation time is much larger than the medium size and the whole medium acts as  a single scattering center. As a result, a reduction of the gluon radiation is produced in the last case. This is the generalization to QCD of the Landau-Pomeranchuk-Migdal effect \cite{Zakharov:1996fv, Baier:1998yf, Gyulassy:1993hr}. The numerical effect appears clearly in Fig. \ref{fig:static} as a suppression of the spectrum for small values of $\kappa^2$. An important consequence is that the spectrum is not collinear divergent (i.e. it can be safely integrated to $k_\perp=0$) nor infrared divergent (i.e. it can be integrated to $\omega=0$) as can be seen in Fig. \ref{fig:static}. In contrast the vacuum part of the spectrum (\ref{eq:MIGR}) present both collinear and soft divergences
\begin{equation}
\frac{dI^{\rm vac}}{d\omega d^2k_\perp}=\frac{\alpha_s C_R}{\pi^2}\frac{1}{k_\perp^2}\frac{1}{\omega}
\end{equation}
The position of the infrared cut in the $k_\perp$-integrated spectrum -- Fig. \ref{fig:static} Right -- can also be understood in terms of the formation time: integrating the spectrum in the kinematically allowed region $0<k_t<\omega$, and noticing\footnote{This can be estimated by taking $\langle k^2_t\rangle\sim \hat q t_{\rm form}$, using eq. (\protect\ref{eq:form}) $\langle k^2_t\rangle\sim\sqrt{\hat q\omega}$} that $\langle k^2_t\rangle\sim\sqrt{\hat q \omega}$ then a suppression of the spectrum for $\omega\lesssim \hat q^{1/3}$ should appear.

\subsection{Expanding medium}

\begin{figure}
\begin{center}
\includegraphics[width=0.6\textwidth]{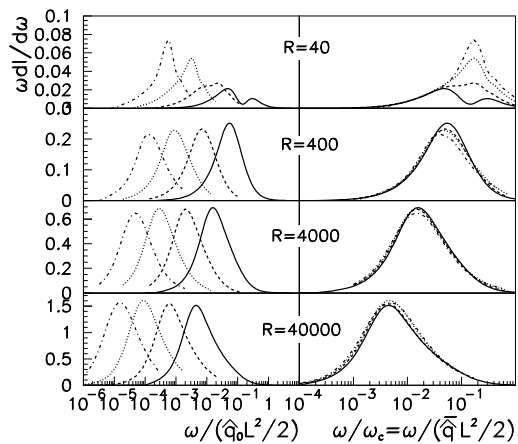}
\end{center}
\caption{The gluon energy distribution for different values of the expansion parameter $\alpha=$0, 0.5, 1 and 1.5 and the corresponding scaling using Eq. (\ref{eq:scal}). Fig. from \protect\cite{Salgado:2003gb}.}
\label{fig:scal}
\end{figure}

In the physical situation of a heavy-ion collision, the medium formed is rapidly evolving and expanding both longitudinally and (as data seem to indicate) also transversely. A simple way of including at least part of the effect of this evolution in the formalism is to consider a medium which is diluting with time, so that the density of scattering centers decreases with some power-like behavior $n(\xi)\sim 1/\xi^\alpha$. The corresponding transport coefficient will follow a similar behavior $\hat q(\xi)=\hat q_0 (\xi_0/\xi)^\alpha$ as discussed before. In the Appendix the corresponding spectra for different values of $\alpha$ are computed. For practical applications, however, it is very convenient to use a scaling law which relates any expanding medium to an equivalent static scenario with time-averaged transport coefficient 
\begin{equation}
\bar{\hat q}=\frac{2}{L^2}\int_{\xi_0}^{L+\xi_0} d\xi(\xi-\xi_0)\hat q(\xi).
\label{eq:scal}
\end{equation}
The quality of this scaling for different values of $\alpha$ is presented in Fig. \ref{fig:scal}. Although some deviations appear for small values of $R\equiv \omega_cL$ they are, in practice, of limited relevance as they occur only when the radiation is anyway small -- small lengths and/or $\hat q$.

\subsection{Massive quarks}

It is now also simple to introduce the case of massive quarks into the formalism in the limit $M\ll E$. In order to do so, let us observe that when solving the poles in $p_{i-}$ in each quark propagator, eq. (\ref{eq:kminuspole}) now reads
\begin{equation}
\int dk_-\frac{{\rm e}^{i(x_{1+}-x_{2+})k_-}}{2p_+k_-+i\epsilon}=-\Theta(x_{2+}-x_{1+})\frac{2\pi i}{2p_+}\exp\left\{i\frac{M^2}{p_+}(x_{1+}-x_{2+})\right\}.
\end{equation}
This introduces only a multiplicative factor in the eikonal trajectories. This factor depends on the energy of the quark. The relevant $+$-component of the momentum is thus $p_+$ before the gluon splitting and $(1-z)p_+$ after the gluon splitting. So, the total contribution from the amplitude times the complex conjugate amplitude to be included in the integrand of  (\ref{eq:MIGR}) is
\begin{eqnarray}
\exp\left\{i\frac{M^2}{p_+}(x_{0+}-x_+)\right\}\times
\exp\left\{i\frac{M^2}{(1-z)p_+}(x_+-L_+)\right\}\times\nonumber\\
\exp\left\{-i\frac{M^2}{p_+}(x_{0+}-\bar x_+)\right\}\times
\exp\left\{-i\frac{M^2}{(1-z)p_+}(\bar x_+-L_+)\right\}
\end{eqnarray}
which, expanding $1/(1-z)\simeq 1-z$ and writing in terms of the gluon energy is just  \cite{Armesto:2003jh}
\begin{equation}
\exp\left\{i\frac{x^2M^2}{k_+}(x_+-\bar x_+)\right\}
\label{eq:masscor}
\end{equation}
So, in the limit of small-$z$ and $M\ll E$, the leading mass correction to the medium-induced gluon radiation corresponds to include the multiplicative factor (\ref{eq:masscor}) inside the integrand of  (\ref{eq:MIGR}) -- see also \cite{Armesto:2003jh,Dokshitzer:2001zm, Zhang:2003wk, Djordjevic:2003zk}.

\begin{figure}
\begin{center}
\includegraphics[width=0.7\textwidth]{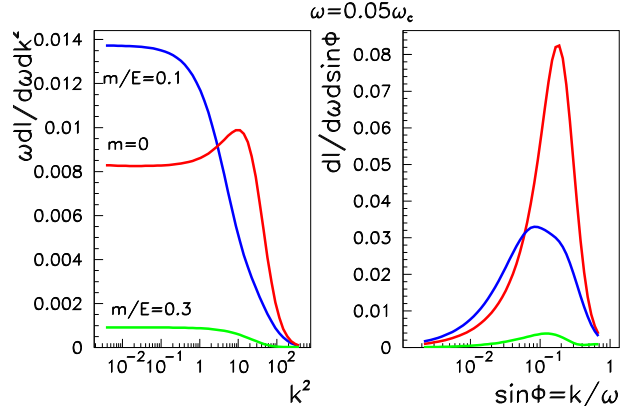}
\end{center}
\caption{$k_t^2$ (left) and angular (right) spectra of medium--induced radiated gluons off a massless (red), charm (blue) and bottom (green) quarks \cite{Armesto:2003jh}.}
\label{fig:specmas}
\end{figure}

In Fig. \ref{fig:specmas} we compare the spectra of gluons radiated off massless and massive quarks.
The effect of including quark masses into the propagators is to reduce the amount of radiation. This is known in the vacuum under the name of {\it dead cone effect}, as the mass terms in the propagators define a minimum angle $\theta_{dc}\sim M/E$ below which the radiation is suppressed. In the medium the situation is a bit more complicated because as we have discussed, formation time effects already suppress the radiation for small angles. In this case, two competing effects exist, on the one hand, there is a genuine suppression produced by mass terms in the propagators; on the other hand, the formation time is now smaller $t_{\rm form}\sim 2\omega/(M^2+k_\perp^2)$ and the LPM suppression appears at smaller values of $k_\perp$, enhancing the radiation at small angles as compared with the massless case. This effect is, however, restricted to a limited region of phase space, negligible for most observables, and the net effect of the mass terms is to reduce the amount of radiation. The energy loss of heavy quarks is, in this manner, smaller than that for light quarks and the formalism predicts an one-to-one correspondence between the two. This smaller radiation translates, in general, into smaller suppression of high-$p_\perp$ heavy than light quarks.


\section{Application of the formalism for jet quenching studies}

\label{sec:application}

As we discussed in Section \ref{sec:hard}, the cross section for producing a hadron $h$ at high transverse momentum $p_T$ and rapidity $y$ can be written in the factorized form
\begin{equation}
\frac{d\sigma^{AB\to h}}{dp_T^2 dy}=\int \frac{dx_2}{x_2}\frac{dz}{z}x_1f_{i/A}(x_1,Q^2)x_2f_{j/B}(x_2,Q^2)\frac{d\hat\sigma^{ij\to kl}}{d\hat t} D^{\rm vac/med}_{k\to h}(z,\mu_F^2)
\label{eq:hard}
\end{equation}
where a summation over flavors is implicit -- see e.g. \cite{Eskola:2002kv} for an explicit expression including kinematical limits, etc. For proton-proton collisions, parton distribution functions, $f_{i/p}(x,Q^2)$, and vacuum fragmentation functions, $D^{\rm vac}_{k\to h}(z,\mu_F^2)$, are known from global fits to experimental data  using the DGLAP evolution equations \cite{dglap}. In the nuclear case, the PDFs are known from similar global fits \cite{Eskola:1998iy,Eskola:1998df,Hirai:2001np,Hirai:2004wq,deFlorian:2003qf,Eskola:2007my,Hirai:2007sx}. On the other hand, the FF, $D^{\rm med}_{k\to h}(z,\mu_F^2)$, contain the information about the medium we want to study. In the following we explain how the formalism of medium-induced gluon radiation can be implemented in (\ref{eq:hard}) to compute the medium modification of the fragmentation functions and how to use them to characterize the medium properties.

\subsection{Medium-modified fragmentation functions}

The theoretical description of the fragmentation of a high-$p_T$ particle in the presence of a medium is not completely known from first principle calculations and some degree of modeling is needed. One possibility is to include all the modifications of the fragmentation functions into a modified splitting function in the DGLAP evolution equations. For a simplified discussion let us take into account only gluon FF -- including other flavors translate into a summation over flavors -- whose DGLAP evolution is 
\begin{equation}
t\frac{\partial}{\partial t} D_{g\to h}(x,t)=\int_x^1\frac{dz}{z}\frac{\alpha_s}{2\pi}\ P_{gg}(z)\, D_{g\to h}\left(\frac{x}{z},t\right).
\label{eq:DGLAP1}
\end{equation}
where $t\equiv\mu_F^2$ and $P_{gg}(z)$ is the splitting function describing the probability that a daughter gluon has been radiated from a parent gluon with fraction of momentum $z$. The probabilistic interpretation of the DGLAP evolution is more clear from its equivalent (at LO) integral formuation (see e.g. \cite{ellis})
\begin{equation}
D(x,t)=\Delta(t)D(x,t_0)+\Delta(t)\int_{t_0}^t \frac{dt_1}{t_1}
\frac{1}{\Delta(t_1)} \int \frac{dz}{z} \, P(z)
D\left(\frac{x}{z},t_1\right).
\label{eq:dglapsud}
\end{equation}
The first term on the right-hand side in this expression corresponds  to the contribution with no splittings between $t_0$ and $t$ while the second one gives the evolution when some finite amount of radiation is present. The evolution is controlled by the Sudakov form factors
\begin{equation}
\Delta(t)=\exp{\left[-\int_{t_0}^{t} {dt^\prime \over t^\prime}
\int dz {\alpha_s(t^\prime,z)
\over 2 \pi} P(z,t^\prime)\right]},
\label{eq:sudakovs}
\end{equation}
with the interpretation of the probability of no resolvable branching between the two scales $t$ and $t_0$.  The definition of the Sudakov form factors and its probabilistic
interpretation depend on the cancellation of the different divergencies
appearing in the corresponding Feynman diagrams, see e.g.
\cite{Dokshitzer:1978hw}. Although such cancellation has never been proved on general grounds for partons re-scattering in a medium,  it has been found in \cite{Wang:2001if} that, under certain assumptions, all the medium effects can be included in a redefinition of the splitting function
\begin{equation}
P^{\rm tot}(z)= P^{\rm vac}(z)+\Delta P(z,t),
\label{eq:medsplit}
\end{equation}
This possibility was exploited in \cite{Armesto:2007dt} where the additional term in the splitting probability is just taken from the medium-induced gluon radiation by comparing the leading contribution in the vacuum case -- see eq. (\ref{eq:split1})
\begin{equation}
\Delta P(z,t)\simeq \frac{2 \pi  t}{\alpha_s}\, 
\frac{dI^{\rm med}}{dzdt} ,
\label{medsplit}
\end{equation}
Implementing (\ref{eq:medsplit}) and (\ref{medsplit}) into (\ref{eq:dglapsud}) the medium-modified fragmentation functions can be computed -- see also \cite{Borghini:2005em} for a related approach. Only the initial conditions of the evolution need to be specified. In \cite{Armesto:2007dt} the KKP set of FF \cite{Kniehl:2000fe} was used for the vacuum as well as for the medium at the initial scale $Q^2_0$, i.e. $D^{\rm med}(x,Q^2_0)=D^{\rm vac}(x,Q^2_0)$. In this model all the medium-effects are built during the evolution. The motivation for this ansatz is the following: in hadronic collisions, particles produced at high enough transverse momentum hadronize outside the medium. So, this assumes that the non-perturbative hadronization is not modified by the medium, whose effect is only to modify the perturbative associated radiation\footnote{For processes with different kinematic conditions  \cite{Airapetian:2007vu} this assumption could not hold, but the negligible effects seen in dAu data at RHIC \cite{RHIC} indicate that this is a reasonable assumption for particle production at high-p$_t$ in nuclear collisions.}. All present radiative energy loss formalisms rely on this assumption.

The fact that the medium-induced gluon radiation is infrared and collinear finite allows for a simplification of this formulation, valid when $E\sim Q\gg 1$. Under these conditions, eqs. (\ref{eq:dglapsud}) and (\ref{eq:sudakovs}) can be written as \cite{Armesto:2007dt}
\begin{equation}
D^{\rm med}(x,t)=p_0D^{\rm vac}(x,t)+\int \frac{d\epsilon}{1-\epsilon}p(\epsilon)D^{\rm vac}\left(\frac{x}{1-\epsilon},t\right)
\label{eq:mffqw}
\end{equation}
with the probability of energy loss given by a Poisson distribution
\begin{eqnarray}
p_0=\exp\left[-\int_0^\infty d\omega\int_0^\omega dk_\perp\frac{dI^{\rm med}}{d\omega dk_\perp}\right] \hspace{0.3cm} \label{eq:p0}\\
p(\Delta E)=p_0\,\sum_{k=1}^\infty \frac{1}{k!}\int\left[\prod_{i=1}^k d\omega_i\int_0^{\omega_i}dk_\perp\frac{dI^{\rm med}(\omega_i)}{d\omega dk_\perp}\right]\delta\left(\sum_{j=1}^k \omega_j-\Delta E\right)\hspace{0.2cm}\label{eq:qwfin}
\end{eqnarray}
and $\epsilon=\Delta E/E$. The total distribution is normally written as 
\begin{equation}
P(\Delta E)=p_0\delta(\Delta E)+p(\Delta E)
\label{eq:qwtot}
\end{equation}

\begin{figure}
\begin{minipage}{0.5\textwidth}
\begin{center}
\includegraphics[width=0.85\textwidth]{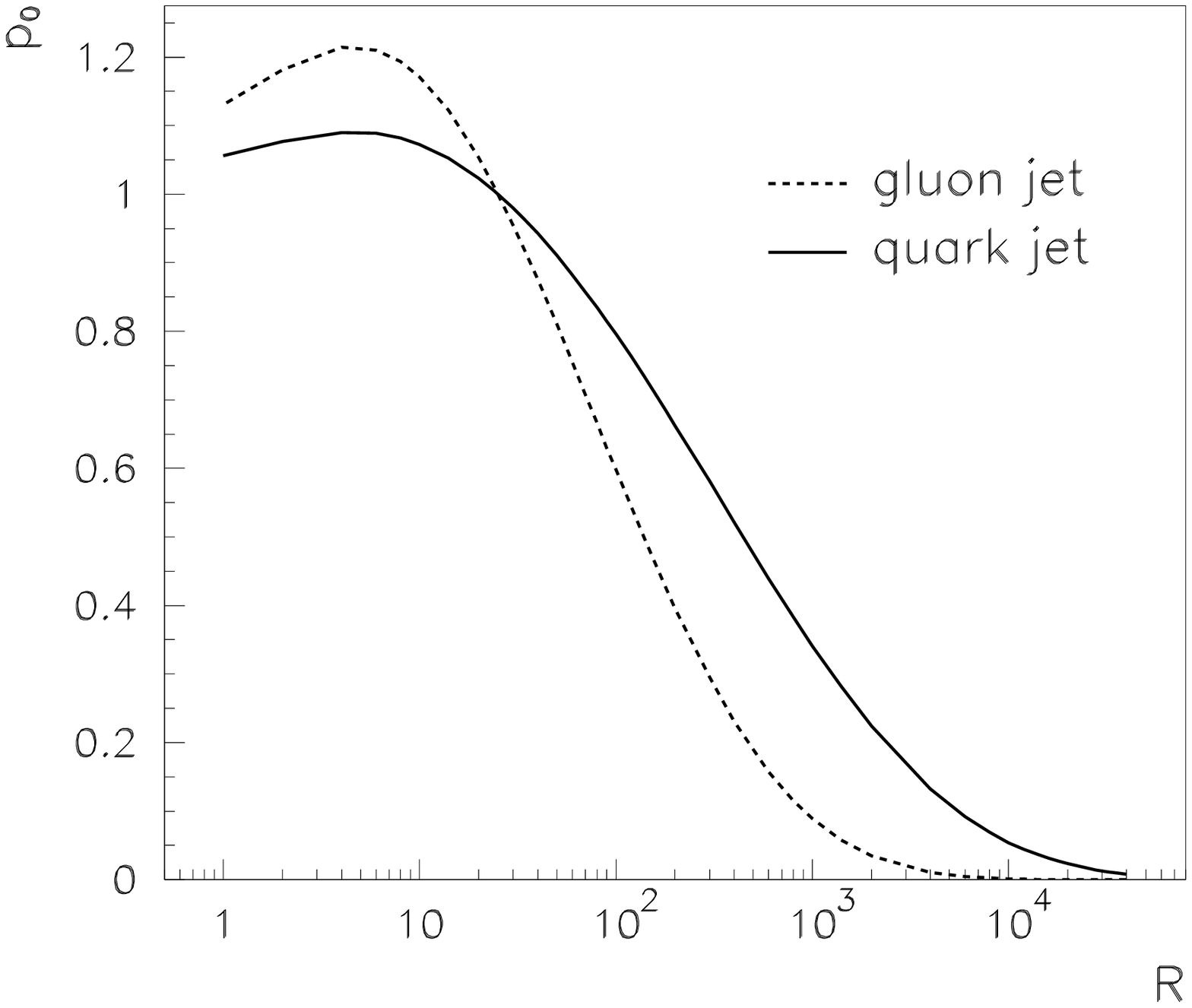}
\end{center}
\end{minipage}
\hfill
\begin{minipage}{0.5\textwidth}
\begin{center}
\includegraphics[width=0.85\textwidth]{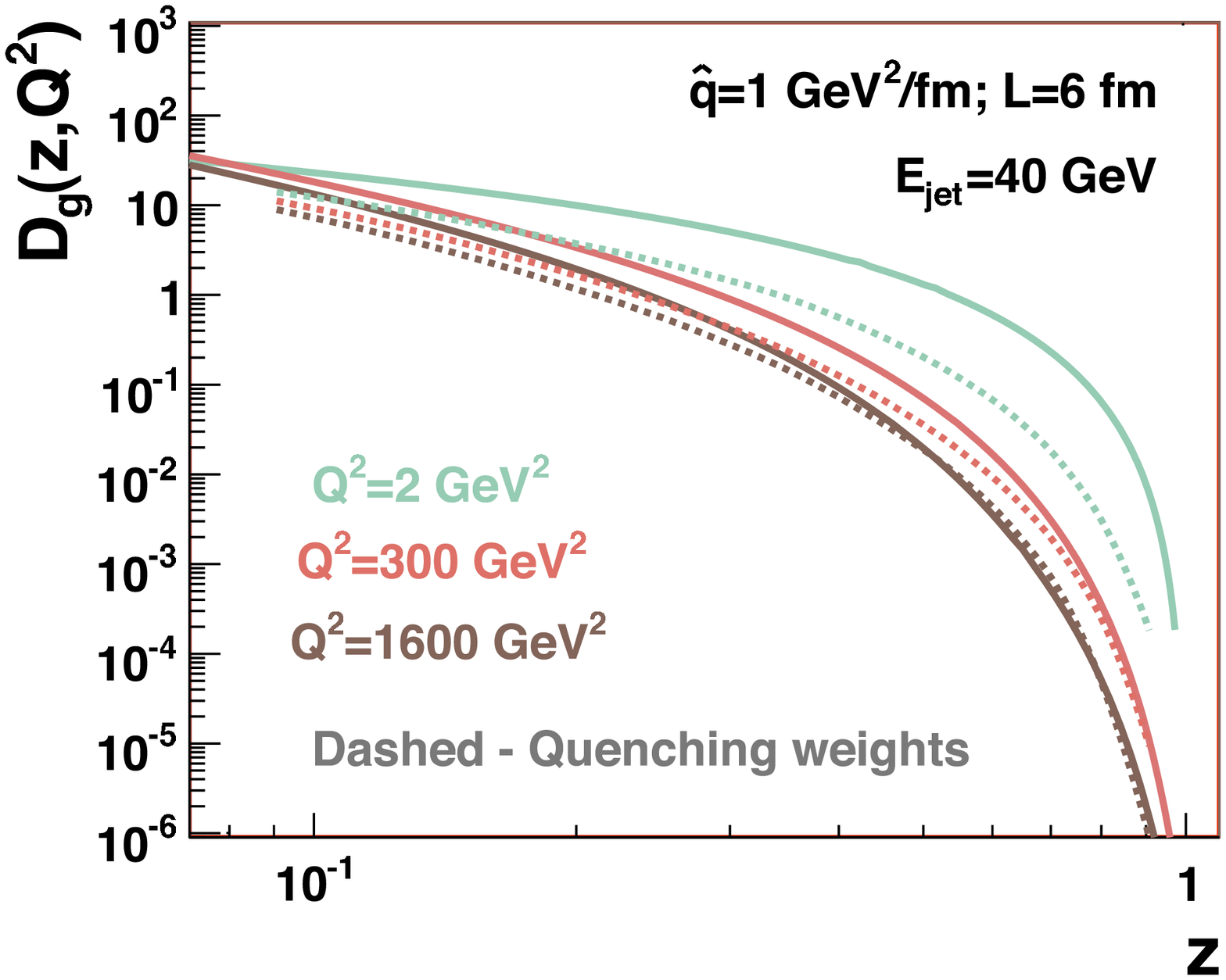}
\end{center}
\end{minipage}
\caption{Left: The probability that a parton loses no energy when traveling through the medium (\protect\ref{eq:p0}) as a function of the variable $R\equiv\omega_c L=\frac{1}{2}\hat q L^3$ -- Fig. from \cite{Salgado:2003gb}. Right: Medium-modified fragmentation functions computed in a DGLAP approach \protect\cite{Armesto:2007dt} and compared with the quenching weights results explained in the text.}
\label{fig:ff}
\end{figure}

Eqs. (\ref{eq:p0}), (\ref{eq:qwfin}) and (\ref{eq:qwtot}) are normally called {\it quenching weights}. They have been first proposed in \cite{Baier:2001yt}  and together with eq. (\ref{eq:mffqw}) constitute the basis of most of the phenomenology of jet quenching in heavy-ion collisions. Notice that the probability that a parton loses no energy when traveling through the medium, given by $p_0$, can be large if the medium length and/or the transport coefficient are small. In Fig. \ref{fig:ff} we plot this probability and the corresponding medium-modified fragmentation functions. Also plotted are the results from \cite{Armesto:2007dt}  when the medium effects are included at every individual splitting -- notice that for the case of hadronic collisions, the most relevant virtuality is precisely that of the order of the initial parton energy, $\mu_F\sim E$.

A similar formalism is used for heavy quarks, for which, as we have seen, the energy loss is predicted to be smaller due to suppressed radiation controlled by mass terms (\ref{eq:masscor}). 

\subsection{The suppression of the inclusive high-$p_T$ yields}

Fig. \ref{fig:ff} shows clearly that one of the effects of the additional medium-induced radiation on particles produced at high transverse momentum is the softening of the spectrum due to the energy loss of the leading particle. The typical values of the fraction of momentum in hadronic collisions from  (\ref{eq:hard})  are $z\sim 0.5-0.7$. The softening of the FF translates, in this way, in a suppression in the production of particles at high-$p_T$ in the presence of a medium.

\begin{figure}
\begin{minipage}{0.5\textwidth}
\begin{center}
\includegraphics[width=0.75\textwidth,angle=-90]{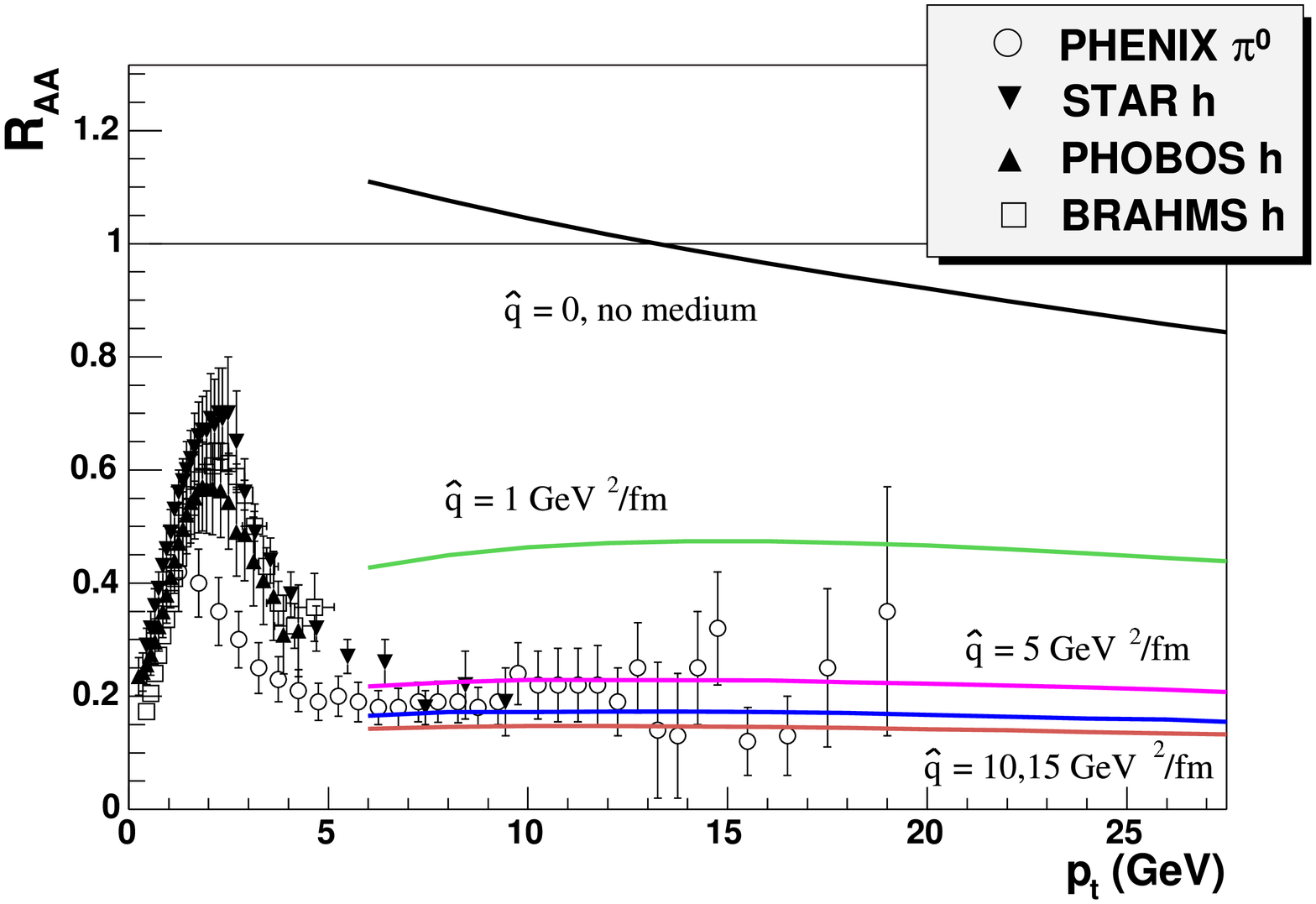}\\
\vskip 0.4cm
\centerline{  }
\end{center}
\end{minipage}
\hfill
\begin{minipage}{0.5\textwidth}
\begin{center}
\includegraphics[width=0.7\textwidth,angle=-90]{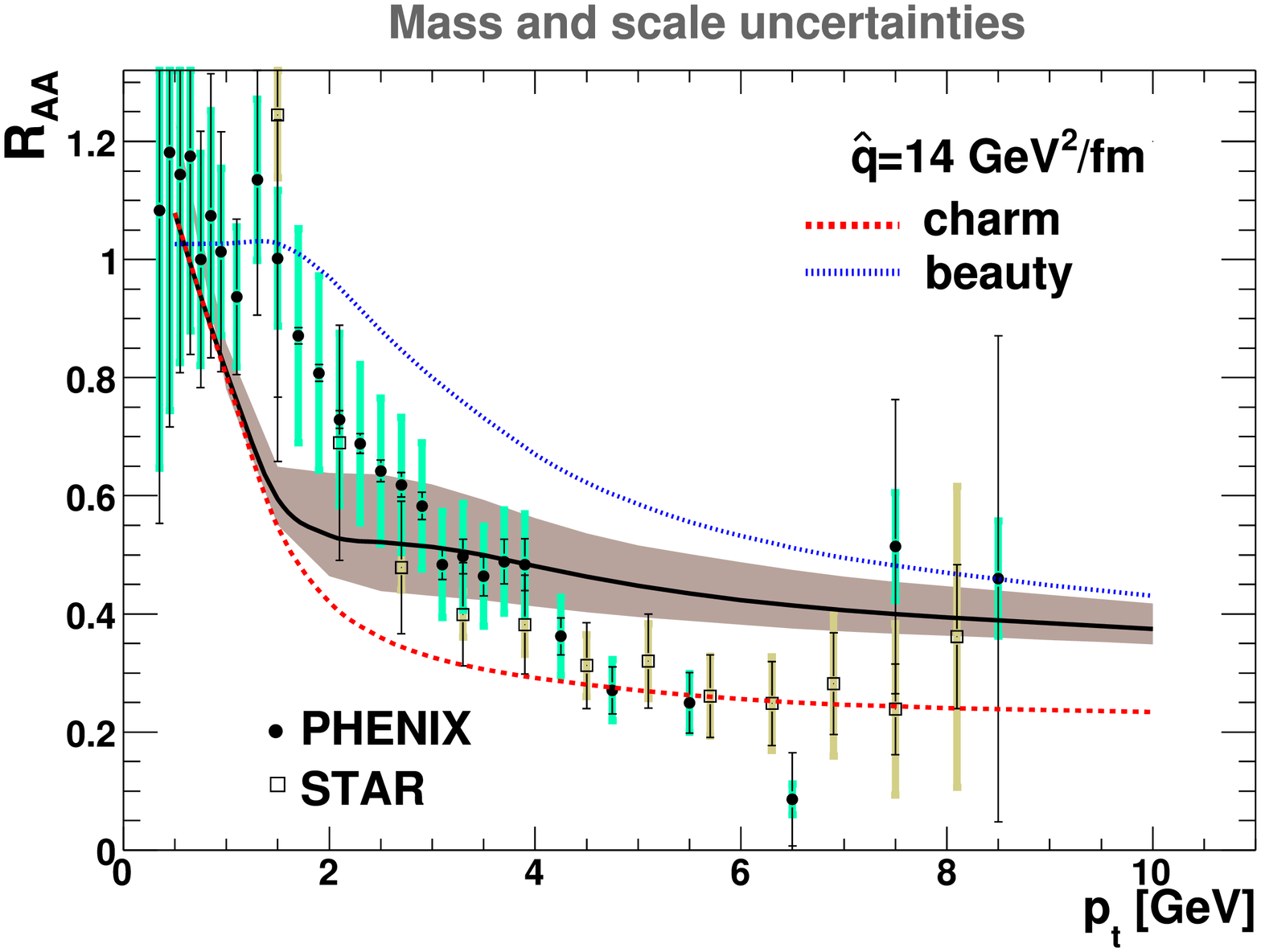}
\end{center}
\end{minipage}
\caption{Left: Nuclear modification factor, $R_{AA}$, for light hadrons in central
AuAu collisions \protect\cite{Eskola:2004cr}. Data from \protect\cite{Adcox:2001jp}. Right: $R_{AA}$ for non-photonic electrons with the corresponding uncertainty from the perturbative benchmark on the relative $b/c$ contribution \protect\cite{Armesto:2005mz}. Data from \cite{Abelev:2006db,Adare:2006hc}}
\label{fig:supp}
\end{figure}

In the massless case, the only parameters in the medium-induced gluon radiation are the transport coefficient and the length of the traversed medium. The later is given by the geometry of the system while the formed is a fitting parameter -- all the medium properties are encoded in $\hat q$, so that measuring it we will learn about the properties of the medium.  Although the geometry could, at first sight, seem a trivial feature in the calculation, different geometries lead to different values of the extracted properties of the medium. In Fig. \ref{fig:supp} the suppression computed with a static medium and medium density given by a Wood-Saxon parametrization is presented. The fit to the light meson suppression leads to quite large values of the transport coefficient \cite{Eskola:2004cr,Dainese:2004te}
\begin{equation}
\hat q\simeq 5\,\dots\,15\, {\rm GeV^2/fm}
\end{equation}
but determined with a large uncertainty. This feature can be understood as due to the dominance of surface emission: those particles produced close to the surface have also large probability to exit the medium essentially unaffected -- see $p_0$ in Fig. \ref{fig:ff}. In order to reproduce the large suppression observed, the value of $\hat q$ needs to be large, but at some point increasing this value translates only into a small reduction of the skin from which the particles abandon the medium unaffected. It's worth mentioning here that the use of more sophisticated medium profiles, as given e.g. by hydrodynamical simulations, could lead to slightly smaller results \cite{Renk:2006pk}.

\begin{figure}[h]
\begin{minipage}{0.5\textwidth}
\begin{center}
\includegraphics[width=0.8\textwidth,angle=-90]{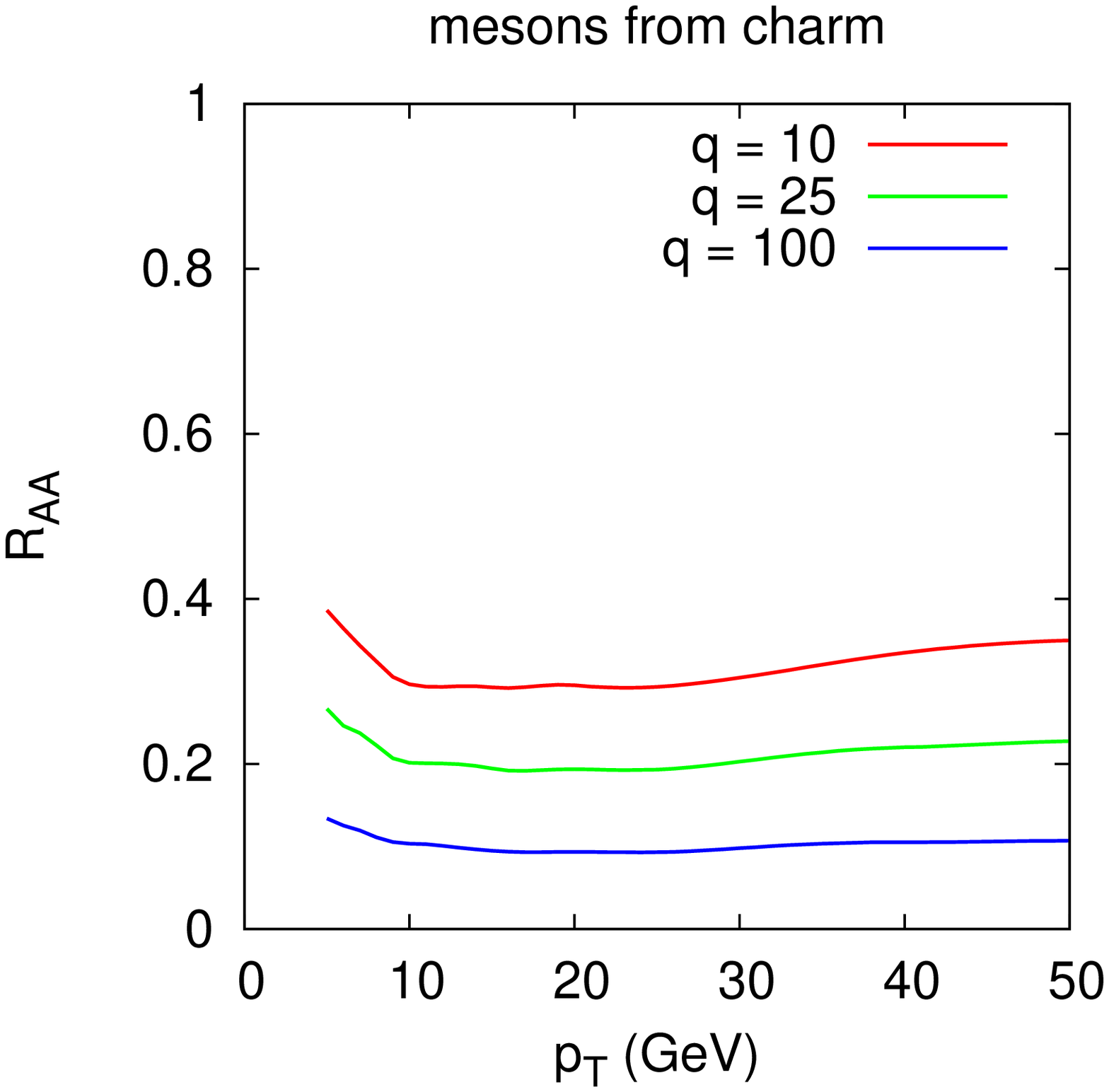}
\end{center}
\end{minipage}
\hskip -0.5cm
\begin{minipage}{0.5\textwidth}
\begin{center}
\includegraphics[width=0.8\textwidth,angle=-90]{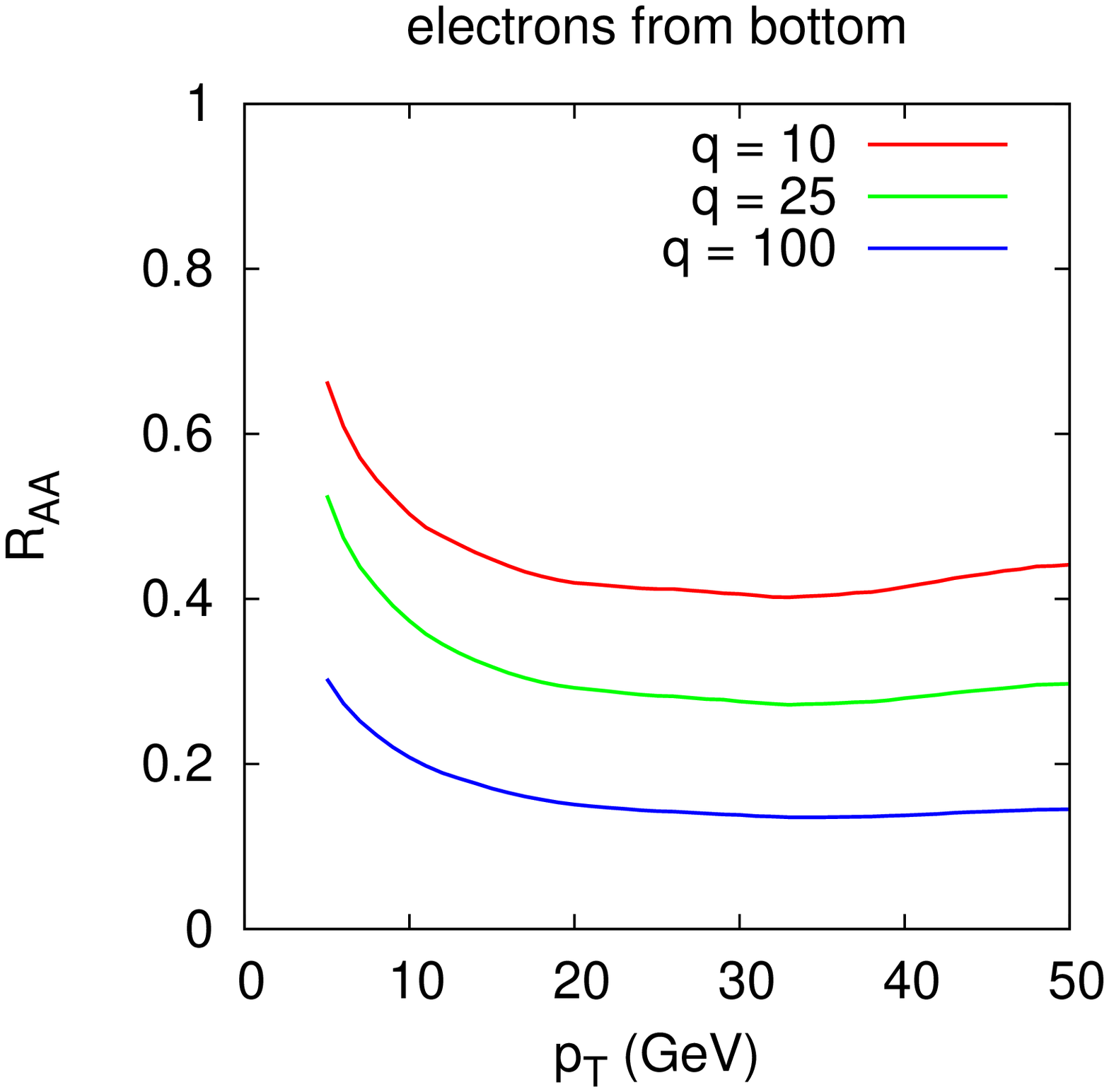}
\end{center}
\end{minipage}
\caption{$R_{AA}$ for D's (left) and for electrons coming from bottom decays (right) at $y=0$ for 10\% PbPb collisions at $\sqrt{s}=5.5$ TeV/A, for different $\hat{q}$ (in GeV$^2$/fm). Fig. from \protect\cite{Abreu:2007kv}.}
\label{fig:raamas}
\end{figure}

In the massive case, once the value of $\hat q$ is known, the formalism provides a prediction with no extra free parameter. At present, the suppression for charm or bottom mesons has not been directly measured. These quantity is indirectly measured  through the non-photonic electrons, expected to be mainly originated from semileptonic decays of the corresponding heavy mesons. In Fig. \ref{fig:supp} the suppression from charm and beauty is presented. Also presented is the combination to the final electron yield from charm and beauty decays as given by the Fixed-Order-Next-to-Leading-Log (FONLL) description of the heavy quark production  \cite{Cacciari:2005rk} . The band is just the effect of varying the masses $m_c$ and $m_b$ as well as the corresponding renormalization scales as standardly uncertainties are computed in perturbative QCD calculations. The description of the data is reasonable but an experimental separation between the charm and beauty contributions will lead to a definite answer on whether other effects could be present. RHIC experiments will be upgraded to address this issue. The LHC experiments have capabilities to distinguish charm and bottom -- see some predictions in Fig. \ref{fig:raamas}.

\subsection{Jets in heavy-ion collisions}

The amount of information about the medium extracted by measuring inclusive particle suppression is limited as we have just seen. More differential observables could lead to a more accurate determination of the transport coefficient and to pindown the interplay with geometry. Relevant questions are {\it where} and {\it how} the energy of the initial parton is lost by interactions with the medium. For that, reconstructing the whole history of the parton while traveling through the medium would clearly be an ideal situation. Fortunately, some experimental handle on this is provided by jet measurements.

Jets are extended objects composed of collimated bunches of particles contained into a reduced region of phase space extending in azimuthal angle, $\Delta\Phi$, and rapidity, $\Delta y$. A jet can be characterized by the opening angle $R=\sqrt{\Delta\Phi^2+\Delta\eta^2}$ defining a cone around the jet axis where the energy is deposited, but more sophisticated definitions are possible \cite{Cacciari:2005hq}. The description of jets in the vacuum, as e.g. in experiments of $e^+e^-$ annihilation, is one of the most precise tests of perturbative QCD. With different degrees of refinement, a jet is computed as the output of a parton branching process starting by a highly virtual quark or gluon produced in the elementary partonic collision. Each branching is controlled by the splitting function, known at LO, NLO and NNLO. 

In the medium case, with modified splitting probabilities, the jet properties are modified. We have already presented modifications to the longitudinal structure as given by the fragmentation functions. Jet properties are, however, richer and the transverse structure is expected to present a broadening produced by the multiple scattering in the medium. How the parton branching process is modified by the presence of a medium is still not known and, again, some modelization is needed. Clearly, the modification of the splitting function (\ref{eq:medsplit}) provides a suitable model for implementation in, e.g. Monte Carlo generators. Simple estimates of the effect were performed in Ref. \cite{Salgado:2003rv}  where the jet energy redistribution as given by the medium-induced gluon radiation (\ref{eq:MIGR}) was studied. 

One of the main experimental issues in jet measurements is the calibration of the jet energy -- all the energy from particles belonging to the same jet need to be reconstructed. Two sources of error appear with opposite dependence on the definition of the jet size $R$: i) out-of-cone fluctuations, which decrease for larger jet sizes, due to radiation happening to angles larger than $R$ ; ii) background fluctuations, which increase for larger $R$, consist on energy from other sources which enters the jet cone. Clearly, the high multiplicity environment of a heavy ion collision complicates the background subtraction which is under study by the different collaboration at the LHC \cite{Carminati:2004fp, Alessandro:2006yt, D'Enterria:2007xr,atlas}. Noticing that most of the particles in the background are produced at small transverse momentum, the simplest way of controlling them is by introducing small-$p_T$ cuts in the measurement.

At RHIC, real jet measurements in AuAu collisions are not possible due to the limited kinematic reach in transverse energy of the jet and the associated large background. For this reason, two- and three-particle correlation analysis are performed as an alternative way of studying the jet structure. Here, one "trigger" particle is chosen within some range of $p_T^{\rm trig}$ and correlated with "associated" particles at a relative angle $\Delta\Phi$ in a range of $p_T^{\rm assoc}$ usually smaller than the one for the trigger particle. A typical hard process, with two back-to-back partons in the transverse plane, will appear in the final state as a two gaussian-like structures one centered at $\Delta\Phi=0$, corresponding to the trigger particle, and one at $\Delta\Phi=\pi$, corresponding to the parton in the opposite direction.

An important step forward is the first measurement of two particle azimuthal correlations at large transverse momentum, with negligible combinatorial background \cite{Adams:2006yt}. By choosing $8<p_T^{\rm trig}<15$ GeV the correlations of associated particles with $p_T^{\rm assoc}> 6$ GeV present the typical two-jet event topology both in pp, dAu and central AuAu collisions. These data support the picture of a very opaque medium with large energy losses, but with a broadening of the associated soft radiation hidden underneath the cut-off. Lowering this transverse momentum cut-off needs of a good control on the background subtraction, but the different collaborations agree in the presence of non-trivial angular structures \cite{Adler:2005ee}: the two-particle-correlation signal around the direction opposite to the trigger particle presents a dip in central collisions, in striking contrast with the typical Gaussian-like shape in proton-proton or peripheral AuAu collisions. The origin of this structure is still unclear. In the presence of an ordering variable (as virtuality or angular ordering in the vacuum parton shower) the implementation of (\ref{eq:medsplit}) with the usual Sudakov form factors (\ref{eq:sudakovs}) produce similar angular structures for energies $\omega\lesssim 2\hat q^{1/3}\sim 3$ GeV in central AuAu \cite{Polosa:2006hb}. At present, the most widely accepted interpretation is in terms of shock wave formation due to the hydrodynamical behavior of the energy deposition in the medium \cite{conical}. Cherenkov gluons have also been proposed as a possible source of these non-trivial angular structures \cite{cherenkov}. Theoretical improvements on the used models, together with new experimental data, would help elucidating the origin of this interesting effect, which takes place, however, in a rather complicate kinematical region between the soft and the hard parts of the spectra.

\subsection{Is there an energy dependence on $\hat q$?}

In all the application we have discussed $\hat{q}$ has been treated as a constant. 
This is a good approximation in the description of high transverse momentum particles
at RHIC. However, the LHC will explore jets in a much broader range of energies which could allow to study the energy dependence of the momentum broadening. 

As we have seen in subsection \ref{sec:medav}, $\hat{q}$ can be related to saturation
physics. In fact, the momentum broadening of a probe propagating through the plasma 
can be expressed as  
\be
\hat{q}_R\approx\frac{C_R}{N}\frac{Q^2_s}{L} \,
\ee
where R is the color representation of the probe 
$Q^2_s$ is the saturation scale of the medium and L is the length. 
The quenching spectrum depends on the gluon broadening. The saturation scale 
depends on the value of $x$ probed in the interaction of the particle with the medium

In a thermal medium,  the saturation scale for typical thermal particles is given
by $\mu_D\sim g T$  \cite{CasalderreySolana:2007sw}. As the energy of the 
probe increases, the saturation scale increases because of evolution. In \cite{CasalderreySolana:2007sw}
it is argued that the typical x probed by a parton propagating in the medium is 
\be
x\sim\frac{Q^2_s}{E T} \,
\ee
where E is the energy of the probe. Thus, the saturation scale grows and so does $\hat{q}$.
Since saturation is a non-linear process, it is not surprising that a non linear dependence
of the $Q^2_s$ with the medium length appears also.

\begin{figure}
\begin{center}
\includegraphics[width=0.6\textwidth]{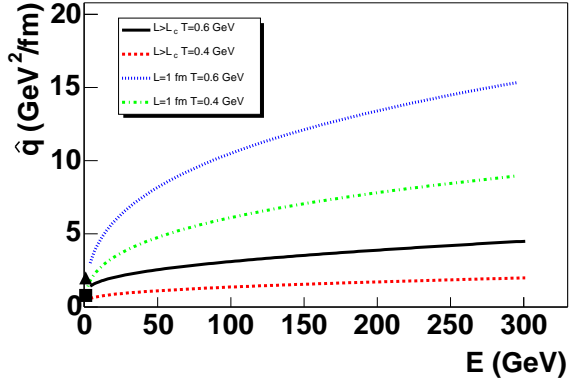}
\end{center}
\caption{ Jet quenching parameter $\hat{q}$
as a function of the jet energy for different path lengths.
The square (triangle) marks the the value of $\hat{q}$ for thermal particle 
at $T=0.4 $ GeV ($T=0.6 $ GeV).
$L_c=1/xT$ is the coherence length, {\it i. e.} the length bellow which the 
particle scatters coherently with the medium. Above this length, no 
dependence on L is observed. Figure from \cite{CasalderreySolana:2007sw}.}
\label{qhatLplot}
\end{figure}

This effect has been studied in \cite{CasalderreySolana:2007sw}, where a simplified 
treatment of the evolution, based on the leading logarithmic approximation was 
performed. Their findings are summarized in Fig \ref{qhatLplot}. A significant dependence 
on the jet energy of $\hat{q}$
is observed which could be measured in the LHC. As also proposed in \cite{CasalderreySolana:2007sw},
the broadening could be measured directly by studying the jet acoplanarity as a function
of the jet energy,  which constitutes a new observable for heavy ion collisions.

\section{New developments}

\label{sec:strings}

In Section \ref{sec:migr} we have seen that the spectrum of radiated
gluon can be expressed in terms of the expectation value of two 
adjoint lightlike Wilson lines. The phenomenological applications of the formalism
described in Section \ref{sec:application} are based in a particular model 
of this object, based on perturbation theory. As we have seen, the
value of $\hat{q}$ extracted from the data is $3-4$ times larger than 
what one expects from perturbation theory. This fact suggests that
strong coupling
non-perturbative effects are important.

The computation of these effects in QCD is hard. Lattice QCD gives non-perturbative
access to equilibrium properties. However, it is very complicated to extract dynamical
quantities from the lattice, since the computations are performed in euclidean time. What 
is more, the object we are interested in involves lightlike trajectories, which is an intrinsic
Minkowski property.  

This difficulty has lead  recently to significant interest in 
strong coupling computation in $\mN=4$ Super-Symmetric Yang Mills
in the infinity number of colors limit. This 
is, like QCD, a $SU(N_c)$ gauge theory. However, unlike QCD, it is 
a supersymetic theory with no conformal anomaly. It also posses a 
different matter content, since apart from the common 
gluons, it contains fermions and scalar in the adjoint color representation. 
Despite of all these differences, computations in this theory are of interest
due to a recent theoretical development, the AdS/CFT correspondence. This 
correspondence allows to address the dynamical (real time) properties of 
the gauge theory at strong coupling in a regime in which traditional 
non perturbative methods, such as lattice, are not applicable. 

Even though the relation between QCD and $\mN=4$ SYM is still questionable
there are experimental (strong elliptic flow and particle suppression) and 
theoretical indications (lattice computations of the coupling constant) that 
suggest that the relevant physics of QCD close to the phase transition 
are those of strong coupling. Thus, the computations within the framework 
described below are of great value, since the AdS/CFT correspondence is the 
only method known to address these problems in a dynamical gauge theory. One of the
great success of this approach is the computation of $\eta/s=1/4\pi$. Current
fits of hydrodynamical models to the elliptic flow yield a value of the 
shear viscosity close  to that predicted from the
AdS/CFT correspondence. This is a remarkable achievement for a theory in 
principle so different from QCD. Below we will review another computation 
in which a similar success has been obtained, the jet quenching parameter,
and the momentum diffusion coefficient. 

\subsection{Brief introduction to the AdS/CFT correspondence}
In this subsection we will introduce the basic concepts of the
AdS/CFT correspondence \cite{Maldacena:1997re} necessary for the computation of 
lightlike Wislon lines within this framework. It is not the 
scope of this lectures to give a systematic introduction to this
subject, which can be found in excellent reviews such as 
\cite{Aharony:1999ti} \cite{D'Hoker:2002aw}.

In its weakest form, the AdS/CFT correspondence states that  
type II B Supergravity (SUGRA) in a particular gravitational background,
the Anti de Sitter space times a five dimensional sphere $AdS_5\times S_5$, is dual to the 
large number of colors ($N_c \rightarrow \infty$),
strong coupling limit of $\mN=4$ SYM, a conformal field theory. SUGRA is a supersymmetric theory of gravity in 10 dimension and it can be obtained as the low energy 
effective action of Type II B string theory. From this point of view, 
SUGRA is the effective theory that describes the string excitation with 
energy $E^{2}<<1/\alpha'$, and $\alpha'$ the string tension.

The  $AdS_5$ space is a five dimensional space with constant (negative) curvature, $R^{-1}$. 
A convenient choice for the  $AdS_5\times S_5$ background metric is given by 
\be
\label{vac}
ds^2=-\frac{r^2}{R^2}\left(-dt^2+dx^2_1+dx^2_2+dx^2_3\right) +
      \frac{R^2}{r^2} dr^2 + R^2 d\Omega_5 \, ,
\ee
where $d\Omega_5$ is the metric of the five dimensional sphere. As can be seen, 
in the boundary, $r\rightarrow \infty$, the metric tends to the usual Minkowsky 
metric (times a prefactor). 

The duality maps the operators of the field theory to the boundary values of 
the SUGRA fields. In the gravity theory, all the dynamics are described
by the classical equations of motion of those fields
with the boundary conditions dictated by the correspondence. The expectation value 
of the operators in the field theory are obtained by evaluating the action of the
gravity field on the classical solution. The coupling constant of the gauge theory,
$g_{YM}$,
is related to the string tension and the curvature as
\be
R^2=\sqrt{g^2_{YM}N_c} \alpha'=\sqrt{\lambda} \alpha' \, ,
\ee 
where we have defined the 't Hooft coupling $\lambda=g^2_{YM} N_c$. The SUGRA approximation is valid if
$R^2>> \alpha'$, thus, the 't Hooft coupling $\lambda\rightarrow\infty$.

In order to describe the field theory at finite temperature, it is necessary to 
introduce a ``black brane''. From the SUGRA point of view, this object is analog to 
an extended black hole which leads to a 3 dimensional horizon at a given value of the
coordinate $r_0$. In the presence of this object, the metric \Eq{vac} is modified
\be
\label{thermal}
ds^2=\frac{r^2}{R^2}\left(-fdt^2+dx^2_1+dx^2_2+dx^2_3\right) +
      \frac{R^2}{r^2}\frac{1}{f} dr^2 + R^2 d\Omega_5 \, ,
\ee
where $f=1-r^4_0/r^4$. The Hawking temperature of the black brane is the temperature
of the field theory.  In the metric \Eq{thermal}, the temperature is given by
\be
T=\frac{r_0}{\pi R^2} \, . 
\ee

\subsection{\label{WADS}Wilson Lines in AdS/CFT}
According to the correspondence, an external heavy probe in the fundamental 
representation of the gauge group (``quark'') is described in the dual 
theory by a classical string
which pends from the boundary of the AdS space \cite{Maldacena:1998im}.
The string is a two dimensional surface parametrized by the internal coordinates
$\tau, \sigma$
\footnote{The string also has coordinates in the 5-dimensional sphere. However in all the applications below these remain constant and, thus, we will suppress
them for simplicity.}
\be
\left(\tau,\sigma \right)\rightarrow X^M=\left(t(\tau,\sigma),x^i(\tau,\sigma),r(\tau,\sigma)\right) \, ,
\ee
where the index $i=1,2,3$.
In analogy with the relativistic point particle, the action of the string is given by its area;
this is known as the 
the Nambu-Goto action 
\be
\label{NG}
S_{NG}=-\frac{1}{2\pi\alpha'}\int d\sigma d\tau \sqrt{\rm{-det}\left\{G_{MN}\del_a X^M \del_b X^N\right\}} \, ,
\ee
where $a,b=\tau,\sigma$ and $G_{MN}$ is the background metric \Eq{vac} (\Eq{thermal} if we are interested in a finite
temperature system)

 As we have obtained in the previous sections, the propagation
of a heavy probe in the field theory  is described in terms of a Wilson line in the 
fundamental representation along the path of the 
quark. This path is also the boundary condition to solve the equations of motion derived from  \Eq{NG}. After the solution of the classical equation of motion 
is found $X^M_{cl}$,
 the expectation value of the corresponding Wilson line is given by 
\be
\left<W\right>\propto e^{-i S_{NG}\left(X^M_{cl}\right)}
\ee

As a simple example, let us consider a single massive quark at rest. Its space 
time trajectory is describe by a line at constant position ${\bf x}_0$. Due to
symmetry, the solution of the Nambu-Goto action with the lowest action is 
the surface
\be
X^M_{cl}=\left(t,x^i_0,r\right)\, .
\ee
According to the correspondence, the expectation value of the Wilson line is
\be
\left< W\right>\propto \exp
                 \left\{{-\frac{i}{2\pi\alpha'}\int^{\mT/2}_{-\mT/2}dt \int^{r_{max}}_{r_o} d r } 
\right\} \,,
\ee
where $r_{max}$ is an upper cut off. 
At zero temperature we can set $r_0=0$. In this case, the divergent prefactor
is naturally interpreted as the mass of the quark
\be
\label{rofmass}
M=\frac{r_{max}}{2\pi\alpha'} \, .
\ee
At finite temperature, the free energy of a single quark is given by
\be
F_1=-\frac{r_0}{2\pi\alpha'}=-\frac{\sqrt{\lambda}T}{2}\,
\ee
It has been also shown in \cite{Herzog:2006gh} that as long as $M>> -F_1$
the thermal mass of the quark is given by
\be
M_{th}=M-\frac{\sqrt{\lambda}T}{2}
\ee

\subsection{Computing the Jet-Quenching parameter}
We now turn our attention to the computation of two parallel Wilson lines
in the fundamental representation \cite{Liu:2006ug,Liu:2006he}. 
\be
\left<\rm{Tr}\left[W_\mC^\dagger\left(x_1=-\frac{y}{2}\right)W_\mC
\left(x=\frac{y}{2}\right)\right]\right> \, ,
\ee 
where $\mC$ is the line of constant rapidity 
\be
t_\mC=t'\cosh \eta \,, \,\,\, x_{3\mC}=-t'\sinh \eta \,.
\ee
Lightlike trajectories are obtained by taking the limit $\eta\rightarrow \infty$
The line cover the interval $t'=\left(-\mT'/2,\mT'/2\right)$ with $\mT'>>y$. In 
this limit, we can assume that the dynamics are independent of $t'$. Due to 
this, it is convenient to introduce the following change of coordinate (boost 
to the rest frame)
\be
dt&=&dt'\cosh \eta -dx'_3 \sinh\eta
\\
dx_3&=&-dt'\sinh\eta+dx'_3\cosh\eta
\ee

To continue further, we need to choose a particular set of coordinates 
$\left(\tau,\sigma\right)$. Since the string solution we are interested on is
independent of $t'$, we choose $\tau=t$; it is also convenient to chose
$\sigma=x_1$. It is also clear that the action is smaller if the 
the string world-sheet does 
not extend in the longitudinal direction $x'_3$. Thus, we need to find 
a solution 
\be
X^M_{cl}=\left(t',x_1,0,0,r(x_1)\right),
\ee
which should be symmetric under the change $x_1\rightarrow-x_1$. Thus,
if the string connects the two endpoints, it should have turning point at 
$x_1=0$, {\it i. e.}
\be
\dot{r}(x_1=0)=0 \, ,
\ee
where $\dot{r}=dr/dx_1$.

The Nambu-Goto action reads,
\be
S=-\frac{1}{2\pi\alpha'}\mT' 2\int^{y/2}_{0} dx_1
\sqrt{\frac{1}{R^2}\left(r^4-r^4_0 \cosh^2\eta \right)
\left(\frac{1}{R^2}+\frac{R^2\dot{r}^2}{r^4-r^4_0}\right)} \,.
\ee

Note that this expression remains real only if $r>r_0\cosh\eta$ ($r>r_0$
since the string is outside the horizon). As we have seen, the maximum
value of $r$, $r_{max}$ is proportional to the mass of the probe
\Eq{rofmass}. Thus, for a given value of the mass, there is a 
critical value of the rapidity 
\be
\label{critg}
\cosh\eta_c =\frac{M^2}{\left(\frac{\sqrt{\lambda}T}{2}\right)^2}\, ,
\ee 
above which, the Wilson loop is exponentially suppressed. It is clear, then, 
that the limits $M\rightarrow\infty$ and $\eta\rightarrow\infty$ do not 
commute. According to \cite{Liu:2006he}, the lightlike Wilson loops are 
obtained by taking the $\eta\rightarrow\infty$ first and then 
$M\rightarrow\infty$. In this limit, the action is simplified to 
\be
\label{Action}
S=-\frac{i}{\pi\alpha'}\mT' \cosh\eta\int^{y/2}_{0} dx_1
\frac{r^2_0}{R}
\sqrt{
\frac{1}{R^2}+\frac{R^2\dot{r}^2}{r^4-r^4_0}} \,.
\ee
Since the action does not depend on $x_1$, we have a conserved Hamiltonian
\be
h=\mL-\dot{r}\frac{\del \mL}{\dot{\del r}}\, ,
\ee
where 
\be
\mL=\sqrt{\frac{1}{R^2}+\frac{R^2\dot{r}^2}{r^4-r^4_0}}\,.
\ee
From this expression, we obtain
\be
\label{dotr}
\dot{r}=\frac{1}{R^3 h}\sqrt{\left(r^4-r^4_0\right)\left(1-R^2 h^2\right))}
\ee
Thus, in order to have real coordinates, $R^2 h^2<<1$. 
Note also that in this approximation, the only point where the $\dot{r}$
vanishes in the horizon $r=r_0$. Thus, the string world-sheet always touches
the horizon.

From \Eq{dotr} we obtain an expression for the position of the string 
$x_1$.
\be
\label{x1sol}
x_1(r)=\frac{R^3 h}{\sqrt{1-R^2 h^2}} \int^{r}_0 dr \frac{1}{\sqrt{r^4-r^4_0}}\, .
\ee
The constant of motion is determined by the condition 
\be
x_1(r\rightarrow\infty)=\frac{y}{2} \, ,
\ee
which leads to
\be
\frac{y}{2}=\frac{R^3 h}{r_0\sqrt{1-R^2 h^2}}
\sqrt{\pi}\frac{\Gamma(\frac{5}{4})}
                                            {\Gamma(\frac{3}{4})}
\ee

Substituting \Eq{x1sol} into the action \Eq{Action} we find 
\be
S=-i\sqrt{\lambda}T\sqrt{\pi}
\frac{\Gamma(\frac{5}{4})}{\Gamma(\frac{3}{4})}L
\sqrt{1+\left(\frac{y T}{2}\sqrt{\pi} \frac{\Gamma(\frac{3}{4})}{\Gamma(\frac{5}{4})}\right)^2}\,
\ee
where $L=\mT'\cosh\eta$ is the length travelled by the parton. 

At small transverse distance we obtain 
\be
\left<\rm{Tr}\left[W_\mC^\dagger\left(x_1=-\frac{y}{2}\right)W_\mC
\left(x=\frac{y}{2}\right)\right]\right>
&\propto& 
\nonumber \\
&& 
\hspace*{-5cm}
\exp\left\{-
\sqrt{\lambda}T\sqrt{\pi}
\frac{\Gamma(\frac{5}{4})}{\Gamma(\frac{3}{4})}L
\left(
1+\frac{y^2T^2}{8}\pi\frac{\Gamma^2(\frac{3}{4})}{\Gamma^2(\frac{5}{4})}
\right) \right\}
 \, .
\ee
The term independent of the transverse position $y$ corresponds to two 
independent quarks. In \cite{Liu:2006he} it is argued that this contribution
should be subtracted. The remaining piece has the form of 
\Eq{eq:multiple} and we can read
\be
\hat{q}_F=\sqrt{\lambda} T^3 \frac{\pi^{3/2}}{2}\frac{\Gamma(\frac{3}{4})}{\Gamma(\frac{5}{4})}\, .
\ee
 The index $F$ indicates that the calculation has been performed with 
Wilson lines in the fundamental representation. This is not quite the object 
that appears in the radiative spectrum, where the Wilson lines were evaluated
in the adjoint representation.  However, these two objects are related in 
$SU(N_c)$ theories \cite{Liu:2006ug}
\be
{\rm Tr} W_A = {\rm Tr} W_F {\rm Tr} W^\dagger_F-1 
\ee
and, thus, in the large $N_c$ limit we obtain,
\be
\label{hqads}
\hat{q}=\sqrt{g^2 N_c} T^3 \pi^{3/2}\frac{\Gamma(\frac{3}{4})}{\Gamma(\frac{5}{4})}\, .
\ee

Despite of the fact that the above computation is performed in the 
infinite coupling, infinite number of colours limit, we would 
like to have a numeric estimate as a comparison to perturbative 
expectations. To do so, we will assume typical values for 
$\alpha_s=0.3$ and $N_c=3$ of RHIC physics. Recalling that 
$\alpha_s=g^2/4\pi$, we obtain
\be
\hat{q}=4.5,\, 10.6,\,20.7\, \rm{GeV^2/fm} \, \, \, for \, \,\,
T=300,\, 400,\, 500\, MeV
\ee 
Remarkably, the values obtained are much larger than the typical scale
of the medium, set by the temperature.  This is also what it has been 
observed in the phenomenological fits presented in Section \ref{sec:application}. What
is more, these numeric values
are comparable to what was obtained in the 
phenomenological fits to RHIC data. Once again, this is a remarkable success,
since the theory in which this parameter has been obtained, $\mN=4$ SYM, is not
QCD. This observation, together with the small values of the viscosity at
RHIC, constitute a compelling indication (but, clearly, not an evidence) 
that the relevant physics at RHIC are those of strong coupling. 

However, we should be careful in comparing this values to RHIC
phenomenology.
As we can see, the jet quenching parameter depends explicitly in the 
number of colors and the coupling. This is different from other 
transport parameters computed in AdS/CFT such as the shear viscosity. 
The explicit dependence on the number of colors hints that, unlike the 
$\eta/s$ ratio, the result \Eq{hqads} is not universal. 
This statement has been corroborated by explicit calculation in other 
theories with gravity duals.
In fact, 
the number of colors is the only measurement of the medium density 
in the $\mN=4$ SYM theory. Based on this and on the explicit computations
in other theories, the authors of \cite{Liu:2006he} have proposed the 
following scaling
\be
\hat{q}_{QCD}=\sqrt{\frac{s_{QCD}}{s_{\mN=4}}} \hat{q}_{\mN=4}\approx 0.6 \,\hat{q}_{\mN=4}
\ee
where $s_{QCD}$ and $s_{\mN=4}$ are the entropy density of QCD and 
$\mN=4$ SYM respectively. Even with this correction, the obtained values are
comparable with those obtained from the phenomenological fits. 

\subsection{Momentum Broadening in Gauge Theories}
In the previous discussion, we have elevated the dipole formula \Eq{eq:multiple} to 
a non perturbative definition of the jet quenching parameter, {\it} the medium
scale responsible for the emission of hard gluons. The values obtained for 
this parameter are in surprisingly good agreement with those obtained from 
experimental data. This is a great success, however the correspondence is a 
powerful theoretical tool that allows to address issues that go 
beyond the estimation of
the value of parameters. 

One of those issues is the relation of $\hat{q}$ with the momentum broadening. 
In subsection \ref{RMB} we have shown that in the dipole approximation $\hat{q}$ 
coincides with the mean momentum transferred squared per unit length. Since
this is a quantity that could potentially be explicitly measured, it 
is interesting to check whether this relation still holds at strong coupling. 

The goal is to compute the momentum broadening of a probe within 
the AdS/CFT correspondence. We  study a very heavy (almost) static
probe. In this case, its motion in the plasma can be described by a
classical equation of motion under a random force
\be
\frac{d p_i}{d t}=\mF_i \,
\ee
where, the force $\mF$ is due to the thermal fluctuations of the (chromo) 
electric fields in the bath \cite{CasalderreySolana:2006rq}
\be
 \label{operator}
 \mathcal{F} \equiv \int d^3 x \,
                        Q^{\dagger}(t,{\bf x})T^aQ(t,{\bf x})\,{ E}_a(t,x) \, ,
\ee
with $Q({\bf x},t)$ the heavy quark field. 

When the quark mass is very heavy, the time required to change its momentum is 
much greater than the typical scale of the medium, $T$.
Thus, the distribution of ``kicks'' can be characterized by its 
second moment,
\be
\langle \mF_i(t) \mF_j(t') \rangle = \kappa \delta_{ij}
	\delta(t-t') \, .
\ee
where the constant $\kappa$ is the mean squared momentum transferred to the probe.
From this, it is clear that
\begin{eqnarray}
\label{kdef}
\kappa &=&\int dt  
 \llangle
\mathcal{F}(t)
\,
\mathcal{F}(0)
\rrangle_{HQ} \, ,
\end{eqnarray}
where the average is taken with the partition function of the gauge theory plus
the heavy quark.

In this expression there is a technical subtlety. The correlator in \Eq{kdef}
is not time ordered and, thus, its extraction from the partition function should 
be treated with care. It is convenient to express $\kappa$ in terms of the 
retarded correlator. By using the Kubo-Martin-Schwinger (KMS) \cite{lebellac} 
relations we can
relate it to the retarded correlator
\be
\label{KGret}
\kappa= 
 \lim_{\omega \rightarrow 0} -\frac{2 T}{ \omega} {\rm Im} G_R(\omega) \, ,
\ee
where,
\be
i G_R(t)=\theta(t) \llangle[\mathcal{F}(t),\mathcal{F}(0)]\rrangle_{HQ} \,. 
\ee
This is a rather technical step but it is  motivated by the fact that
there is a well defined prescription to compute retarded correlators in
the AdS/CFT correspondence. 

The partition function of the bath in the presence of a heavy quark is 
given by \cite{McLerran:1981pb}
\begin{eqnarray}
Z_{HQ}&=&\sum_s \llangle s \left| e^{-\beta H} \right | s \rrangle
\nonumber \\
 &=&
    \int d^3 x   \sum_{s'} \llangle  s' \left | Q(\x, -T) e^{-\beta H}  
                                              Q^{\dagger}(\x,-T) 
                  \right | s' \rrangle \, ,  
\end{eqnarray}
where the states $\left | s' \rrangle$ are states of the gauge theory
without the heavy quark. 

\begin{figure}
\begin{center}
\includegraphics[width=0.6\textwidth]{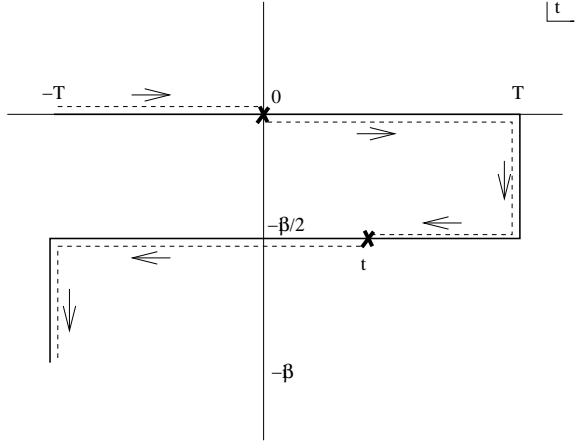}
\end{center}
\caption{
Schwinger-Keldysh contour. The crosses represent the points of insertions of the 
electric field operators while the dashed lines are the paths of the Wilson lines
connecting those insertions. The periodic boudary condition in imaginary time makes
the force-force correlator \Eq{diffWL} gauge invariant.
}
\label{SK}
\end{figure}

As is well know, the Hamiltonian in the partition function can be 
understood as evolution in the imaginary time direction. The 
description of dynamical  processes requires the introduction of 
a time contour that spans the real-time  axis and extends towards
the imaginary direction. This is known as the Schwinger-Keldysh contour 
which is illustrated in Fig. \ref{SK}.
In equilibrium, the partition function is independent of the 
initial time $T$. However, in computing a real time correlator it is necessary 
to specify the time in which the bath was created, since the phase of the
heavy quark depends on it.

 As in the case of the energetic probe, the propagator of the
heavy quarks field leads to a Wilson line. After some simple algebra,
the froce-force correlator can be expressed as
\be
\label{diffWL}
\llangle T_C[\mathcal{F}(t_C)\mathcal{F}(0)]
\rrangle_{HQ}&=&\frac{1}{Z_{HQ} }
\times\nonumber\\
&&
\hspace*{-2cm} 
\llangle {\rm tr}{\left[U(-T -i\beta,t_C)\,  E(t_C) \,
  U(t_C,0)\, E (0)\,  U(0, -T)\right]}\rrangle \; ,
\ee
where $T_C$ denotes order along the time contour. 

The expression above can be obtained from a Wilson line along the time
contour with a deformation in the $\hat{y}$ direction $\delta y$
\be
W_C[\delta y] = T_C \exp{\left\{-i\int_C dt \left(
                               A_0 + \delta \dot{y} A_y
                                    \right)
                                   \right\}} \, ,
\ee
by taking (functional) derivatives on the path.
\be
 \llangle T_C[\mathcal{F}(t_C)\mathcal{F}(0)]
\rrangle_{HQ}= 
  \frac{1}{\llangle W_C[0]\rrangle}
\left.  \llangle \frac{\delta^2 W_C[\delta y]}
{\delta y(t_C)\; \delta y(0)} \rrangle
\right|_{\delta y=0}
\, .
\ee
Thus, the broadening is obtained by studying small fluctuations of the Wilson 
line along the heavy quark path. 

\subsection{\label{StaticB}AdS/CFT Computation of the broadening}
\begin{figure}
\begin{minipage}{0.5\textwidth}
\begin{center}
\includegraphics[width=0.85\textwidth]{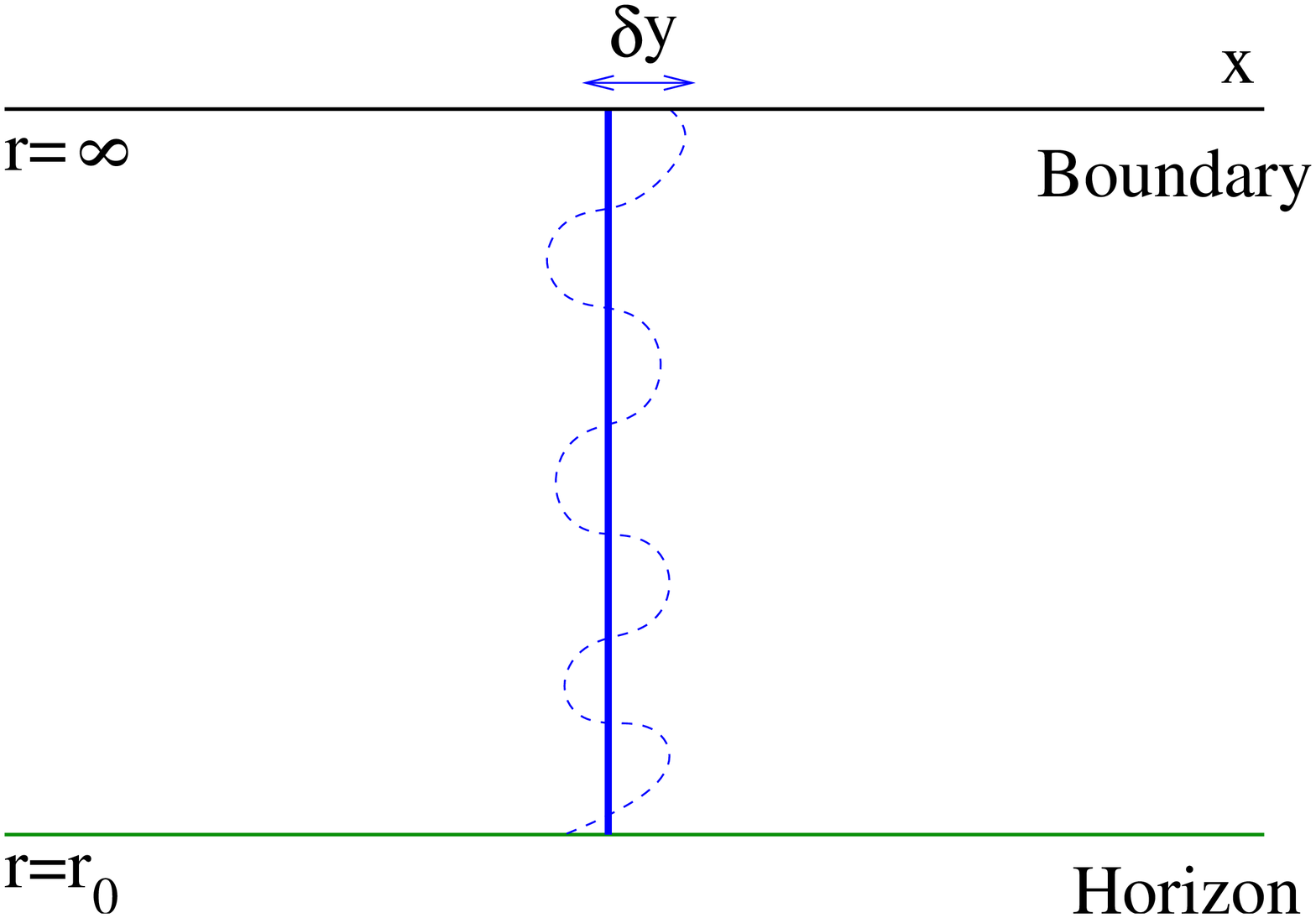}
\end{center}
\end{minipage}
\hfill
\begin{minipage}{0.5\textwidth}
\begin{center}
\includegraphics[width=0.85\textwidth]{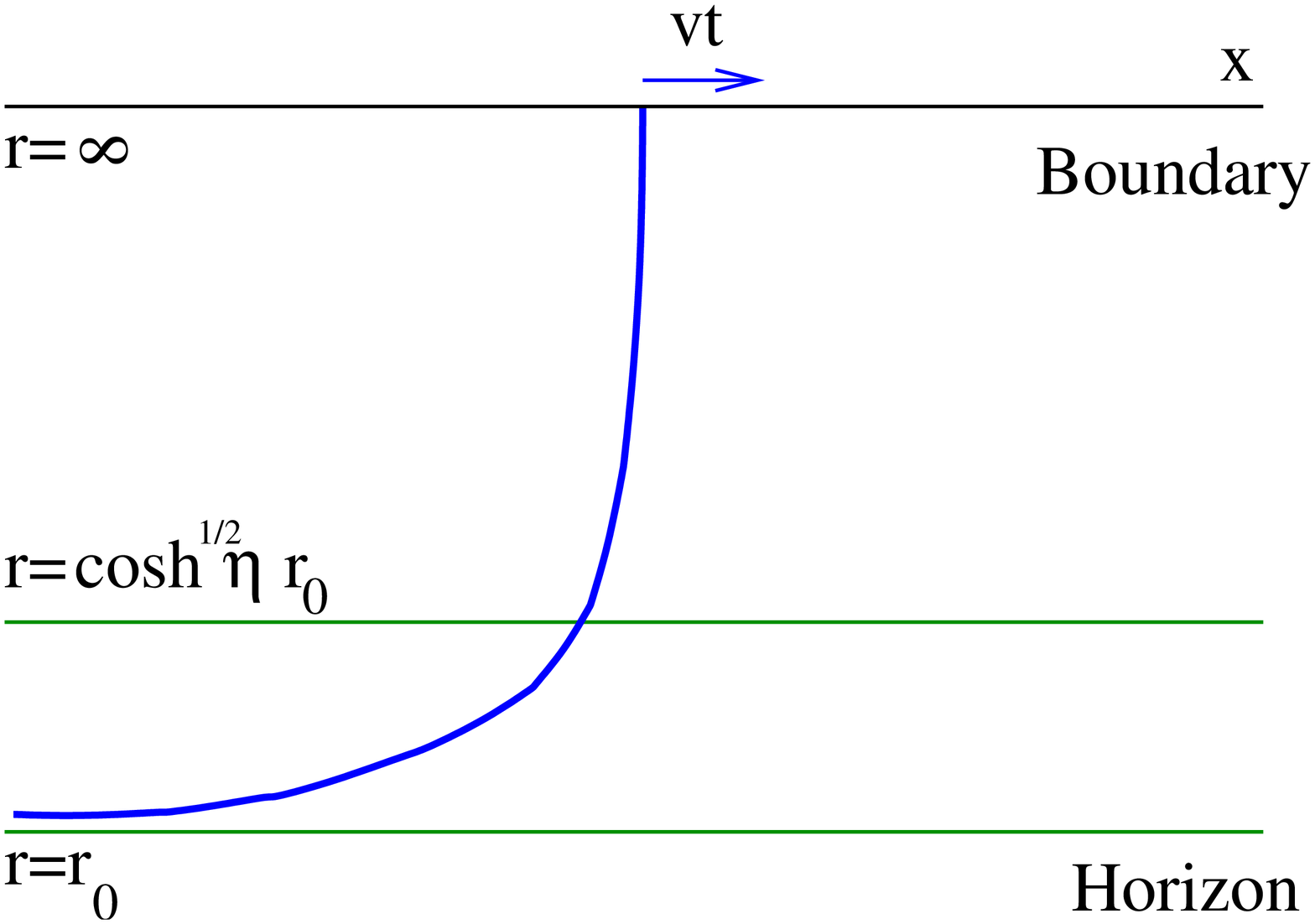}
\end{center}
\end{minipage}
\caption{Left: 
Static string solution. The dashed line represents the fluctuations in the transverse coordinates
 Right: Trailing string correspoding to a probe moving at finite velocity. The string 
approaches logarithmically to the horizon. There is a new scale appearing at $r=\sqrt{\cosh \eta} \, r_0$}
\label{strings}
\end{figure}

As we saw in Subsection \ref{WADS}, a static Wilson line in the gauge theory
is described in the gravity theory by a single string pending down from the 
boundary of AdS. Our task is to study how small fluctuations
in the boundary along a path $\delta y$ propagate into the bulk. This is illustrated
in Fig. \ref{strings}.

The string solution is a surface parametrized by 
\be
X^{M}_{cl}=\left(t,r,0,y(t,r),0 \right) \,,
\ee
which minimizes the Nambu-Gotto action, \Eq{NG}, and with boundary condition
\be
y(t,r\rightarrow\infty)=\delta y(t)\, .
\ee
Since the boundary conditions are set at $r\rightarrow\infty$ it is convenient
to introduce the variable 
\be
u=\frac{r^2_0}{r^2}\, .
\ee
We also introduce the dimensionless variables
\be
\pi T t&=& \bar{t}\\
\pi T y&=& \bar{y}
\ee
Since we are interested in small fluctuations, we expand the action to 
quadratic order and we obtain
\be
S_{NG}=\frac{R^2}{2\pi\alpha'}
\int \frac{d\bar{t}\,d u}{2u^{3/2}}
\left[
1-\frac{1}{2}\left(
             \frac{\dot{\bar{y}}_{\parallel}^2}{f}-
              4 f u \left(\bar{y}'_{\parallel}\right)^2
             \right)
\right]
\, ,
\ee
where $f=1-u^2$. Performing a Fourier transform
\be
\label{ftbc}
  \bar{y}(t,u) = \int e^{-i \w \bar{t}}\, \bar{y}(\w)\, Y_\w(u)  \, \frac{d\w}{2\pi}
\, ,
\ee
where $Y_\w(u=0)=1$ and, thus, $\bar{y}(\w)$ is the Fourier transformed of
the path on the boundary.
The equations of motion are
\be
\label{eomy}
\partial_{u}^2 Y_\w - \frac{(2 + 6 u^2)}{4 u f} \partial_u Y_\w + 
\frac{\w^2}{4 u f^2 } Y_\w = 0
\, .
\ee
This differential equation has a regular singular point at the horizon, $u=1$. Close
to this point its solutions behave as 
\be
\label{inf}
Y_\w=\left(1-u\right)^{\pm i\w/4}F_\pm(u)\, ,
\ee
where the functions $F_\pm(u)$ are regular. These two solutions
correspond to modes that are in-falling (-) and outgoing at the horizon (+).
This can be seen from the fact that, close to the horizon, the lines of constant
phase follow the path $\bar{t}=\pm\ln(1-u) /4$.
 Retarded
correlators are obtained by imposing the physically intuitive condition that all the
modes close to the horizon are in-falling \cite{Policastro:2002se,Herzog:2002pc}. 

The above prescription for the retarded correlators imposes the boundary condition
at the horizon. Our task is to find the regular function $F_-(u)$. Inserting
\Eq{inf} into \Eq{eomy} we find
\be
\partial_{u}^2 F_-(u) + \left(\frac{i\omega}{2 f}-\frac{2+6 u^2}{4 u f}\right)\partial_{u} F_-(u)
+&& 
\nonumber \\
\left(\frac{2u-1-3u^2}{8f^2u}i\w + \frac{4-u}{16f^2u}\w^2\right)F_-(u)&=&0\,.
\ee
Since $F_-(u)$ is regular, we can expand it in a power series in $\w$. Taking into
account the boundary condition \Eq{ftbc} we find 
\be
\label{FSol}
F_-(u)&=&1 +
\nonumber \\
&& 
\frac{i \w}{8} \left\{\pi - 4\tan^{-1}(\sqrt{u}) - 6\log 2 + 4\log(1+\sqrt{u}) + 2\log(1+u) \right\}  
+
\nonumber \\
&& 
 O(\w^2)
\, .
\ee

Finally, the retarded correlator can be computed from the quadratic action. Following
the prescription for Lorentzian correlators \cite{Policastro:2002se} we obtain
\be
\label{Gret}
     G_{R}(\omega) = - A(u) Y_{-\omega}(u) \partial_u Y_\omega(u)|_{u\rightarrow 0} \, ,
  \qquad   A(u) = \frac{R^2\, \left(\pi T \right)^2}{\pi \alpha'}
                  \frac{f}{u^{1/2}}
\, .
\ee
Inserting \Eq{FSol} into \Eq{Gret} and \Eq{KGret} we have
\be
\label{kappa}
  \kappa  &=& \lim_{\omega\rightarrow 0} \,\frac{-2T}{\omega}\, \mbox{Im} G_{R}(\omega)\,, \\
          &=& \sqrt{ \lambda}\, T^3\,\pi  \, ,
\ee
where, as before, $R^2/\alpha'=\sqrt{\lambda}$.

Our computation of the momentum diffusion constant $\kappa$ has the same temperature
and coupling dependence than the jet-quenching parameter. This is not surprising, since
the theory is conformal and the $\sqrt{\lambda}$  dependence is common in the 
observables that involve semiclassical strings in AdS. However, the numeric coefficient
is different. In fact, the relation between $\hat{q}$ and the momentum broadening
derived in \Eq{hatqmb} would imply
\be
\hat{q}_F=2\kappa \, ,
\ee
which does not hold. However, before drawing any conclusion we must point
out that the computation we just performed is in the static limit, while the 
$\hat{q}$ is in the light-like limit. Thus, in the next subsection, we extend the 
above computation to quarks moving with a finite velocity.

\subsection{Drag Force and Broadening at Finite Velocity }
We study now a single heavy quark propagating in the thermal bath 
at a finite rapidity $\eta$ along the $x_1$ direction \cite{Herzog:2006gh,Gubser:2006bz}.
As before, the quark is described in the gravity
theory by a semiclassical string. Assuming that the quark is in a stationary
situation, the string may be parametrized as
\be
X^M=\left(t,vt+\xi(r),0,0,r\right) \,
\ee
with $v=\tanh\eta$ and $\xi(r\rightarrow\infty)=0$. With this parametrization,
the Nambu-Gotto action is
\be
S_{NG}=\frac{1}{2\pi\alpha'}\mathcal{T}\int^\infty_{r_0} dr 
\sqrt{\frac{1/\cosh^2\eta-r_0^4/r^4}{f}+\frac{r^4}{R^4}f\xi^{'2}} \,.
\ee
Since the action only depends on $\xi'$, the solutions to the equations of 
motions compatible with the specified boundary conditions are parametrized
in terms of a constant of motion
\be
\label{Cofmon}
\frac{r^4}{R^4}f\xi'\frac{1}{\sqrt{\frac{1/\cosh^2\eta-r_0^4/r^4}{f}+\frac{r^4}{R^4}f\xi^{'2}}}
=\mathcal{C}
\ee
For these solutions, the classical action is given by
\be
S_{cl}=\frac{1}{2\pi\alpha'}\int^{\infty}_{r_0} dr \frac{f-\mathcal{C}^2 R^4/r^4}{f\left(1/\cosh^2\eta-r_0^4/r^4\right)}\, ,
\ee
Note that the action has an imaginary part unless 
\be
\label{Csol}
\mathcal{C}_s=\pm \frac{r^2_0}{R^2} \sinh\eta \, .
\ee
Thus, for sufficiently long times, all other solutions are (exponentially) suppressed
and the only relevant ones are those fulfilling \Eq{Csol}. Inserting \Eq{Csol} into
\Eq{Cofmon} we have
\be
\xi'=\pm\frac{R^2r^2_0}{r^4 f} v\, ,
\ee
which can be easily solved
\be
\label{xisol}
\xi=\pm\frac{v}{2 \pi T}\left(\arctan\left(\frac{r_0}{r}\right)
                       -{\rm arctanh}\left(\frac{r_0}{r}\right)\right)\,
\ee
From these two solutions, only the (+) one is physical, since the string 
trails back form the quark \footnote{A more rigorous derivation of this fact
can be found in \cite{Herzog:2006gh,Gubser:2006bz}}. This string solution is 
illustrated in Fig. \ref{strings}.  

As opposed to the static case discussed in the previous section, the 
sting bends behind the probe. Since the string has tension,  work must be performed in order to keep the stationary situation. The momentum flux   
from the boundary to the bulk 
along the string is given by \cite{Herzog:2006gh,Gubser:2006bz}
\be
\pi^1_\mu=-\frac{1}{2\pi\alpha'}G_{\mu\nu}
\frac{\left(X'\dot X\right)X^{'\nu}-\left(X'\right)^2\left(\dot X\right)^2}{\sqrt{-h}}
\, ,
\ee
where $X'$, $\dot X$ represent the derivatives with respect to $\sigma$ and $\tau$ 
respectively and 
\be
-h=\left(X'\dot X\right) - \left(X'\right)^2 \left(\dot X\right)^2\,,
\ee
the determinant of the induced metric. Substituting the solution \Eq{xisol} into
this expression we find that there is a non vanishing flux of momentum in the $x$
direction. This means that the probe loses momentum at a rate \cite{Herzog:2006gh,Gubser:2006bz}
\be
\frac{dp}{dt}=\pi^1_x=-\frac{\pi}{2} \sqrt{\lambda} T^2 \frac{v}{\sqrt{1-v^2}}\,.
\ee
The reader may immediately find that the velocity dependence is that of the 
relativistic momentum. Multiplying and dividing by the mass we find 
\be
\label{dragg}
\frac{dp}{dt}=-\eta_D p,  \, \, \, \, \, \, 
 \eta_D=\pi\sqrt{\lambda} T^3 \frac{1}{2 T M} \, .
\ee
At small momentum, the hard probe is expected to follow Langevine dynamics 
\cite{Moore:2004tg} where the average momentum loss has the drag form of \Eq{dragg}.
For slow probes, the fluctuation dissipation theorem relates the value of the 
drag force and the momentum broadening 
\be
\eta_D=\kappa \frac{1}{2 T M} \,.
\ee
From this expression we see that the drag coefficient and the momentum broadening
computed before fulfill this relation. The combination of these two independent 
computations can be taken as an explicit check of the AdS/CFT correspondence. 

Remarkably, \Eq{dragg} is, in principle, valid for all momentum. As in the previous
section, we would like to compute the (transverse)  momentum broadening of the probe
at a finite velocity. As in the previous case, we need to study small transverse 
fluctuation of the string and compute the retarded correlator. The computation 
is greatly simplified if we introduce the following change of 
coordinates \cite{CasalderreySolana:2007qw}:
\be
\hat{t}&=&\frac{1}{\sqrt{\cosh\eta}} 
        \left(t + \frac{1}{2} \arctan\left(\frac{r_0}{r}\right)
                - \frac{1}{2}{\rm arctanh}\left(\frac{r_0}{r}\right)
\right .
\\ \nonumber
&&
        \left.
                - \frac{1}{2}\sqrt{\cosh\eta} \arctan\left(\sqrt{\cosh\eta}\,\frac{r_0}{r}\right)
                + \frac{1}{2}\sqrt{\cosh\eta}\,{\rm arctanh}\left(\sqrt{\cosh\eta}\,\frac{r_0}{r}\right)
        \right) \, ,
\\
\hat{r}&=&\frac{r}{\sqrt{\cosh\eta}} \, .
\ee
In these coordinates, the induced metric in the string $h_{ab}=G_{MN}\del_aX^M\del_bX^N$
is,
\be
h_{\hat{t}\hat{t}}&=&-\frac{\hat{r}^2}{R^2}f(\hat{r}) \, ,\\
h_{\hat{r}\hat{r}}&=&\frac{R^2}{\hat{r}^2}{R^2}\frac{1}{f(\hat{r})} \, ,\\
h_{\hat{t}\hat{r}}&=&0 \, ,
\ee 
where $f(\hat{r})=1-r^4_0/\hat{r}^4$. This is formally the same as the induced metric
in the static string studied in section \ref{StaticB} but replacing the $(t,r)$ 
coordinates with $(\hat{t},\hat{r})$. 

Since the induced metric has the same functional form, the equation of motion for
transverse fluctuations is also the same as in the static case. However, there is 
a subtlety: from the point of view of the moving string the horizon is lifted from
its original location $r_0$ up to a higher scale $r_0 \sqrt{\cosh\eta}$. This is the
scale where the local speed of light in the AdS space coincides with the speed of 
propagation of the probe. Thus, for excitations originated in the boundary, this
scale plays the role of the black hole. 

To obtain the momentum broadening we repeat the same steps as before. The only 
difference is that the in-falling boundary condition should be imposed at the lifted
horizon. 
 Since, as we 
have remarked, the equations in these coordinates are the same, we do not repeat the
computation and write only the final result \cite{CasalderreySolana:2007qw}. 
\be
\kappa_T=\sqrt{\cosh\eta}  \sqrt{\lambda} T^3 \pi \,.
\ee 

The obtained transverse momentum broadening grows with the rapidity of the particle
propagating in the medium. This is an unexpected result, since the jet quenching 
parameter $\hat{q}$ yields an energy independent result. As mentioned, these quantities
coincide in perturbation theory but they do not seem to do at strong coupling. 
However, before drawing this conclusion we should note that the computation 
presented in this section is not valid for arbitrarily large values of the 
rapidity. 

Since the quark loses momentum, the stationary situation is only reached
when there is an external source. This can be thought of as a non vanishing
electric field $\mathcal{E}$ coupled to the quark living on the brane. The magnitude of the 
electric field grows with the velocity of the quark
\be
\mathcal{E}=\frac{\pi}{2} \sqrt{\lambda} T^2 \frac{v}{\sqrt{1-v^2}} \, .
\ee
However, the brane does not support arbitrary values of the electric field, since 
it becomes unstable under pair creation. This occurs at a critical value of the
electric field \cite{CasalderreySolana:2007qw}
\be
\mathcal{E}_c=\frac{M^2}{\sqrt{\lambda}}\, .
\ee
As a result, there is an upper limit on the rapidity of the probe described by the 
solution \Eq{xisol}
\be
\cosh\eta<\left(\frac{M}{\sqrt{\lambda}T}\right)^2 \,.
\ee

The limiting value of the rapidity coincides (parametrically) with the one encountered
in the computation of the jet quenching parameter, \Eq{critg}. Thus, the computations
based on the trailing string cover a different range of rapidities than the computation
of the jet quenching parameter. 

Thus, we cannot yet resolve the issue of whether the broadening and the $\hat{q}$ are
related beyond perturbation theory. However, it is remarkable that we have obtained
an energy dependent result for ``small'' values of the rapidity, while the 
ultra-relativistic limit is rapidity independent, as indicated by $\hat{q}$. Having
a consistent dynamical picture that is valid in both regimes would be desirable and
could lead to new observables able to point out whether the matter created at RHIC is 
indeed strongly coupled. 

The computation of the broadening is also interesting on its own. Slowly moving
quarks follow Langevine dynamics and the heavy quark diffusion constant obtained
from our computations 
\be
D=\frac{2 T^2}{\kappa}\approx\frac{0.9}{2\pi T}\left(\frac{1.5}{\alpha_s N_c}\right)^{1/2}\, ,
\ee  
is comparable with the value obtained from fits to the single electron suppression and 
$v_2$ (originated from charm and bottom quarks) within a Langevine model \cite{Moore:2004tg}. The energy dependence of the 
broadening constitutes a new observable which could, in principle, be measured
by measuring the jet acoplanarity with fully reconstructed jets 
\cite{CasalderreySolana:2007sw}.

As a final remark, we would like to mention that the drag forced obtained for
the heavy quarks is limited to the same constraint in the value of the 
rapidity. At very large rapidities (close to the light-cone) the mass of 
the probe should be irrelevant for its dynamics. Even though an explicit 
calculation has not been performed, it is clear that the mass dependence obtained
in \Eq{dragg} is not valid in the ultra-relativistic limit.

\section{Last comments}

The formalism described in Sections \ref{sec:propagation} to \ref{sec:application} and others \cite{Wang:2001if,Arnold:2002ja}  have been very successful for the description of high-p$_T$ data from RHIC, specially for the inclusive yields. We have already commented on some limitations which would need to be addressed for the applications to the LHC, where the reach in the $p_T$ spectrum will be enhanced by more than one order of magnitude and other observables as jets will be available for the first time in heavy-ion collision experiments.

The main conclusion from the phenomenological studies of jet quenching is the characterization of the produced medium as extremely dense. We have quoted the values from \cite{Eskola:2004cr, Dainese:2004te} of $\hat q\simeq 5...15$ GeV$^2$/fm. A perturbative calculation of the transport coefficient \cite{Baier:2002tc} underestimates this result by a factor of at least 5. This seems to indicate that the typical cross sections in the medium averages described in Section \ref{sec:propagation} are much larger than the perturbative estimates based on a free gas of quarks and gluons. 

Several results from RHIC point, in fact, to this direction, as the small viscosity needed to reproduce the soft bulk data. Most of these phenomenologically accessible quantities are dynamical parameters for which the lattice calculations at finite temperature are not well suited. For these reasons, a flourishing  activity on the relation of these findings with theoretically computable quantities in String Theory by the AdS/CFT correspondence has recently started. Although still highly speculative, this connection is opening new ways of facing the challenges on the study of collective properties at the fundamental level.

\section*{Acknowledgements}
We thank N\'estor Armesto and Xin-Nian Wang for critical reading of the manuscript.
CAS would like to thank the organizers of the XLVII Cracow School of Theoretical Physics in Zakopane, Poland, for the nice atmosphere during the meeting. 
The work of JCS was supported by the Director, 
Office of Science, Office of High Energy and Nuclear Physics, 
Division of Nuclear Physics, and by the Office of Basic Energy
Sciences, Division of Nuclear Sciences, of the U.S. Department of Energy 
under Contract No. DE-AC03-76SF00098.
CAS is supported by the FP6  of the European Community under the contract MEIF-CT-2005-024624.

%
%

\section{Appendix A}
Let $W(x_+,\xperp)$ be the Wilson line at $\xperp$ in the fundamental 
represention. Thus, it verifies

\be
-i\del_+W(x_+,\xperp)=A^a_-(x_+,\xperp)T^a W(x_+,\xperp)
\ee
We now define the object
\be
W^{bc}_A(x_+,\xperp)=\frac{1}{N} {\rm Tr} \left[
                         W^\dagger(x_+,\xperp) T^b W(x_+,\xperp) T^c
                         \right] 
\ee
Taking the derivative with respect to $x_+$ we find
\be
-i\del_+ W^{bc}_A\left(x_+,\xperp\right)=A^a_-\frac{1}{N}
                                      {\rm Tr}
                                      \left[
                                      W^\dagger(x_+,\xperp)\left[T^b,T^a\right]
                                      W(x_+,\xperp) T^c
                                      \right]
\ee
Thus
\be
-i\del_+ W^{bc}_A\left(x_+,\xperp\right)=A^a_- if^{bad} W^{dc}_A\left( x_+,\xperp\right)
\ee
Using that 
\be
\left(T^a_A\right)^{bd}=-if^{abd}
\ee
we find
\be
-i\del_+ W_A\left(x_+,\xperp\right)=A_- T_A  W_A\left(x_+,\xperp\right)
\ee
Thus, it is the Wilson line in the adjoint. The normalization guaranties that W at the origin
is the identity.

%
%

\section{Appendix B}

For consistency with previous approaches, used, in particular in Section \ref{sec:application}, we present  here the analytical results in ordinary variables -- this means making the changes $\omega\simeq k_+/\sqrt{2}$, $x\simeq x_+/\sqrt{2}$ 
\begin{eqnarray}
\omega\frac{dI}{d\omega d^2{\bf k_\perp}}=\frac{\alpha_S C_R}{(2\pi)^2\omega}2{\rm Re}\int_{x_{0}}^{L+x_0} \hspace{-0.35cm} dx\int d^2{\bf x}\ e^{-i{\bf k_\perp\cdot x}} \Bigg[\frac{1}{\omega}\int_{x}^{L+x_0}\hspace{-0.35cm} d\bar x\ e^{-\frac{1}{2}\int_{\bar x}^{L} d\xi n(\xi) \sigma({\bf x})}\times\nonumber\\ 
\times \frac{\partial}{\partial{\bf y}}\cdot\frac{\partial}{\partial{\bf x}}{\cal K}({\bf y}=0,x;{\bf x},\bar x)
-2\frac{\bf k_\perp}{{\bf k}_\perp^2}\cdot \frac{\partial}{\partial {\bf y}}{\cal K}({\bf y}=0,x;{\bf x},{L})\Bigg]+\frac{\alpha_S C_R}{\pi^2}\frac{1}{{\bf k}_\perp^2}\hspace{0.7cm}
\label{eq:MIGRov}
\end{eqnarray}
Notice that the definition of the transport coefficient, $\hat q$, differs also by a factor $\sqrt{2}$ 
\cite{Liu:2006ug}, so the path integral is  \cite{Zakharov:1996fv, Baier:1998yf, Wiedemann:2000tf, Salgado:2003gb}
\begin{equation}
{\cal K}\left({\bf r}(x),x;{\bf r}(\bar x),\bar x|\omega\right)=\int {\cal D}{\bf r}\exp\left[i\frac{\omega}{2} \int_{x}^{\bar x}d\xi\left(\dot{\bf{r}}^2+i\frac{\hat q(\xi)}{2 \omega} {\bf r}^2\right)\right]
\label{eq:kpropap}
\end{equation}
which corresponds to a 2-dimensional harmonic oscillator with time-dependent imaginary frequency
\begin{equation}
\frac{\Omega_\alpha^2(\xi_0)}{\xi^\alpha}=  -i\frac{\hat{q}(\xi)}{\,2\, \omega}
  = -i \frac{\hat{q}_0}{2\, \omega} \left( \frac{\xi_0}{\xi}\right)^\alpha\, 
  \label{c.2}
\end{equation}
and {\it mass} $\omega$. The solution of (\ref{eq:kpropap}) can be written
in the form~\cite{Baier:1998yf, Salgado:2003gb}
\begin{eqnarray}
  {\cal K}({\bf r}_1,x;{\bf r}_2,\bar x|\omega) =
  \frac{\omega}{2\pi\, i\, D(x,\bar x)}\,
  \exp\left[ i S_{\rm cl}({\bf r}_1,x;{\bf r}_2,\bar x) \right]\, .
  \label{c.3}
\end{eqnarray}
Here, the classical action $S_{\rm cl}$
in (\ref{c.3}) takes the form
\begin{eqnarray}
  S_{\rm cl}({\bf r}_1,x;{\bf r}_2,\bar x)
  = \frac{\omega}{2} \left[ {\bf r}_{\rm cl}(\xi)
    \cdot \frac{d}{d\xi} {\bf r}_{\rm cl}(\xi)\right]
  \Bigg\vert^{x}_{\bar x}\, ,
  \label{c.4}
\end{eqnarray}
where the classical path ${\bf r}_{\rm cl}(\xi)$ satisfies
the homogeneous differential equation
\begin{equation}
  \left[ \frac{ d^2}{d\xi^2}
          - \frac{\Omega_\alpha^2(\xi_0)}{\xi^\alpha}\right]\,
        {\bf r}_{\rm cl}(\xi)
        =0,
        \label{c.5}
\end{equation}
with initial conditions
\begin{equation}
  {\bf r}_{\rm cl}(x) = {\bf r}_1
  \hspace{20pt} \hbox{\rm and}\hspace{20 pt}
  {\bf r}_{\rm cl}(\bar x) = {\bf r}_2\, .
\label{c.6}
\end{equation}
The fluctuation determinant $D(\xi,\xi')$ in (\ref{c.3}) satisfies
\begin{equation}
  \left[ \frac{ d^2}{d\xi^2}
          - \frac{\Omega_\alpha^2(\xi_0)}{\xi^\alpha}\right]\,
        D(\xi,\xi')=0,
        \label{c.7}
\end{equation}
with initial conditions
\begin{equation}
  D(\xi, \xi)=0 \hspace{20pt} \hbox{\rm and}\hspace{20pt}
  \frac{ d}{d\xi}   D(\xi,\xi')|_{\xi=\xi'}=1\, .
\label{c.8}
\end{equation}
In practice, $D(\xi,\xi')$ is found by combining the two independent
(scalar) solutions $f_1$, $f_2$ of (\ref{c.5}),
\begin{equation}
  D(\xi,\xi') = {\cal N}\, \left( f_1(\xi)\, f_2(\xi')
              - f_2(\xi)\, f_1(\xi') \right)\, ,
            \label{c.9}
\end{equation}
and fixing the norm ${\cal N}$ by the initial condition (\ref{c.8}).
The solution of (\ref{eq:kpropap}) can be written in terms of $D(\xi,\xi')$
and two $\xi$- and $\xi'$-dependent variables $c_1$, $c_2$,
\begin{eqnarray}
  {\cal K}\left({\bf r}(x),x;{\bf r}(\bar x),\bar x|\omega\right)=
  \frac{i\, \omega}{2\pi D(x,\bar x)}\times
  \nonumber \\
  \times \exp\left[-
    \frac{-i \omega}{2\, D(x,\bar x)}
      \left(c_1 {\bf r}_1^2+ c_2 {\bf r}2^2 -
            2 {\bf r}_1\cdot {\bf r}_2\right)\right]\, .
  \label{c.10}
\end{eqnarray}
The particular form of the coefficient variables $c_1$ and $c_2$ depend on $\alpha$. Most of the phenomenologically relevant values are for $\alpha<2$  for which the two independent solutions of the homgeneous differential equation (\ref{c.5}) are  ~\cite{Baier:1998yf}.
    \begin{eqnarray}
      f_1(\xi) &=& \sqrt{\xi}\,
         I_\nu\left( 2\nu\, \Omega_\alpha(\xi_0)\,
             \xi^{\frac{1}{2\nu}}\right)\, ,
         \label{c.11} \\
      f_2(\xi) &=& \sqrt{\xi}\,
         K_\nu\left( 2\nu\, \Omega_\alpha(\xi_0)\,
                     \xi^{\frac{1}{2\nu}}\right)\, ,
         \label{c.12}
    \end{eqnarray}
    where $I_\nu$ and $K_\nu$ are modified Bessel functions
    with argument
\begin{equation}
  \nu = \frac{1}{2-\alpha} \, .
  \label{c.13}
\end{equation}
In terms of the variable
\begin{eqnarray}
  z(\xi) &=& 2 \nu\Omega_\alpha(\xi_0)\xi^{\frac{1}{2\nu}}\, ,
       \label{c.14}
\end{eqnarray}
the solution (\ref{c.10}) is given by [we use $z \equiv z(\xi)$,
$z' \equiv z(\xi')$]
\begin{eqnarray}
   D(\xi,\xi') &=&
   \frac{2\nu}{(2\nu\Omega_\alpha(\xi_0))^{2\nu}}\, (z z')^\nu
   \left[I_\nu(z)K_\nu(z')-K_\nu(z)I_\nu(z')\right]\, ,
   \label{c.15}\\
  c_1 &=& z\, \left(\frac{z'}{z}\right)^\nu\,
  \left[I_{\nu-1}(z)K_\nu(z')+K_{\nu-1}(z)I_\nu(z')\right]\, ,
  \label{c.16}\\
  c_2 &=& z'\, \left(\frac{z}{z'}\right)^\nu\,
  \left[K_\nu(z)I_{\nu-1}(z') + I_\nu(z)K_{\nu-1}(z')\right]\, .
  \label{c.17}
\end{eqnarray}
We now write the three contributions from (\ref{eq:MIGRov}) as 
\begin{equation}
\omega{dI\over d\omega d^2{\bf k}_\perp}= {\alpha_s\over\pi^2} C_F (I_4+I_5+I_6)
=  \omega{dI^{\rm med}\over d\omega d^2{\bf k}_\perp}+\omega{dI^{\rm vac}\over d\omega d^2{\bf k}_\perp} \, .
\label{c.25}
\end{equation}
with, $I_6=1/k_\perp^2$. $I_4$ and $I_5$ define the medium-induced
part $I^{\rm med}$ which, using  (\ref{c.10}) -- (\ref{c.17}) leads to
\begin{eqnarray}
  I_4 &=& \frac{1}{4\omega^2}\,
  2{\rm Re} \int_{x_0}^{L+x_0} dx \int_{x}^{L+x_0} d\bar{x}
  \left( \frac{-4A^2_4 \bar{D}_4}{(\bar{D}_4-iA_4B_4)^2}
         + \frac{iA^3_4B_4\, {\bf k}_\perp^2}{(\bar{D}_4-iA_4B_4)^3} \right)
       \nonumber \\
       && \times
       \exp\left[{-\frac{{\bf k}_\perp^2}{4\, (\bar{D}_4 -
                  i\, A_4\, B_4)}}\right]\, ,
  \label{c.26}\\
  I_5 &=& \frac{1}{2\omega}\, {\rm 2Re} \int_{x_0}^{L+x_0} dx\,
        \frac{-i}{B_5^2}\,
        \exp\left[{-i\frac{{\bf k}_\perp^2}{4\, A_5\, B_5}}\right]\, ,
          \label{c.27}
\end{eqnarray}
where
\begin{eqnarray}
    A_4&=&{\omega\over 2D(\bar x,x)}\, , \hspace{0.5cm}
    B_4=c_1(\bar x,x)\, , \hspace{0.5cm}
    \bar D_4 = \frac{1}{2}\int_{\bar x}^{L+x_0} d\xi
     n(\xi)\sigma({\bf r})\hspace{0.3cm}
    \label{c.28}\\
    A_5&=&{\omega\over 2D(L+x_0,x)}\, , \hspace{0.5cm}
    B_5=c_1(L+x_0,x)\, .
    \label{c.29}
\end{eqnarray}
In the case of a static medium, $\alpha=0$,  the functions
$I_{\pm 1/2}(z)$ and $K_{\pm 1/2}(z)$ entering (\ref{c.10})
have explicit expressions in terms of exponentials and the results are simply
\begin{eqnarray}
  A_4={\omega\Omega\over 2
      \sin(\Omega(\bar x-x))}\, , \hspace{0.5cm}
  B_4=\cos(\Omega(\bar x-x))\, , \hspace{0.5 cm}
 \bar D_4=\frac{1}{2}n_0C(L-\bar x)\nonumber\, .
\end{eqnarray}



\begin{thebibliography}{99}


\bibitem{Hatsuda:2007rt}
  T.~Hatsuda,
  J.\ Phys.\ G {\bf 34} (2007) S287
  [arXiv:hep-ph/0702293].

\bibitem{Karsch:2001cy}
  F.~Karsch,
  Lect.\ Notes Phys.\  {\bf 583} (2002) 209
  [arXiv:hep-lat/0106019].

\bibitem{Cheng:2007jq}
  M.~Cheng {\it et al.},
  arXiv:0710.0354 [hep-lat].

\bibitem{Mocsy:2004bv}
  A.~Mocsy and P.~Petreczky,
  Eur.\ Phys.\ J.\  C {\bf 43} (2005) 77
  [arXiv:hep-ph/0411262];
  Eur.\ Phys.\ J.\  C {\bf 43} (2005) 77
  [arXiv:hep-ph/0411262].


\bibitem{Landau:1953gs}
  L.~D.~Landau,
  Izv.\ Akad.\ Nauk Ser.\ Fiz.\  {\bf 17} (1953) 51.

\bibitem{Bjorken:1982qr}
  J.~D.~Bjorken,
  Phys.\ Rev.\  D {\bf 27}, 140 (1983).

\bibitem{Rischke:1998fq}
  D.~H.~Rischke,
  arXiv:nucl-th/9809044.

\bibitem{d'Enterria:2006su}
  D.~d'Enterria,
  J.\ Phys.\ G {\bf 34} (2007) S53
  [arXiv:nucl-ex/0611012].

\bibitem{Bjorken:1982tu}
  J.~D.~Bjorken,
Fermilab-Pub-82-059-THY, Batavia (1982); Erratum unpublished.

\bibitem{RHIC}
K.~Adcox {\it et al.}  [PHENIX Collaboration],
Nucl.\ Phys.\ A {\bf 757}, 184 (2005);
B.~B.~Back {\it et al.} [PHOBOS Collaboration],
Nucl.\ Phys.\ A {\bf 757}, 28 (2005);
I.~Arsene {\it et al.}  [BRAHMS Collaboration],
Nucl.\ Phys.\ A {\bf 757}, 1 (2005);
J.~Adams {\it et al.}  [STAR Collaboration],
Nucl.\ Phys.\ A {\bf 757}, 102 (2005).

\bibitem{SPS}
  M.~M.~Aggarwal {\it et al.}  [WA98 Collaboration],
  Eur.\ Phys.\ J.\  C {\bf 23} (2002) 225
  [arXiv:nucl-ex/0108006];
  F.~Antinori {\it et al.}  [NA57 Collaboration],
  Phys.\ Lett.\  B {\bf 623} (2005) 17
  [arXiv:nucl-ex/0507012].


  \bibitem{dglap}
  Y.~L.~Dokshitzer,
  Sov.\ Phys.\ JETP {\bf 46}, 641 (1977)
  [Zh.\ Eksp.\ Teor.\ Fiz.\  {\bf 73}, 1216 (1977)];
  V.~N.~Gribov and L.~N.~Lipatov,
  Sov.\ J.\ Nucl.\ Phys.\  {\bf 15}, 438 (1972)
  [Yad.\ Fiz.\  {\bf 15}, 781 (1972)];
675 [1218];
  G.~Altarelli and G.~Parisi,
  Nucl.\ Phys.\  B {\bf 126}, 298 (1977).

  \bibitem{ellis} R.~K.~Ellis, W.~J.~Stirling and B.~R.~Webber, {\it QCD and
collider physics}, Cambridge University Press, Cambridge 1996.

\bibitem{Salgado:2007rs}
  C.~A.~Salgado,
  arXiv:0706.2264 [hep-ph].


\bibitem{Matsui:1986dk}
  T.~Matsui and H.~Satz,
  Phys.\ Lett.\  B {\bf 178} (1986) 416.

  \bibitem{jpsiSPS}
  C.~Baglin {\it et al.}  [NA38],
  Phys.\ Lett.\  B {\bf 220} (1989) 471;
  Phys.\ Lett.\  B {\bf 255} (1991) 459;
  M.~C.~Abreu {\it et al.}  [NA50],
  Phys.\ Lett.\  B {\bf 410} (1997) 337;
  Phys.\ Lett.\  B {\bf 477} (2000) 28;
  R.~Arnaldi {\it et al.}  [NA60],
  Nucl.\ Phys.\  A {\bf 774} (2006) 711;

    \bibitem{jpsiRHIC}
  S.~S.~Adler {\it et al.}  [PHENIX],
  Phys.\ Rev.\  C {\bf 69} (2004) 014901;
  arXiv:nucl-ex/0611020.


    \bibitem{Gribov:1984tu}
  L.~V.~Gribov, E.~M.~Levin and M.~G.~Ryskin,
  Phys.\ Rept.\  {\bf 100} (1983) 1.

  \bibitem{Mueller:1985wy}
  A.~H.~Mueller and J.~w.~Qiu,
  Nucl.\ Phys.\  B {\bf 268}, 427 (1986).

\bibitem{Kovchegov:1999yj}
  Y.~V.~Kovchegov,
  Phys.\ Rev.\  D {\bf 60} (1999) 034008;
  Phys.\ Rev.\  D {\bf 61} (2000) 074018.

\bibitem{jimwlk}
 Recent reviews: 
  E.~Iancu and R.~Venugopalan,
 hep-ph/0303204;
  D.~N.~Triantafyllopoulos,
  Acta Phys.\ Polon.\ B {\bf 36}, 3593 (2005);
  A.~Kovner,
  Acta Phys.\ Polon.\ B {\bf 36} (2005) 3551.


\bibitem{Eskola:1998iy}
  K.~J.~Eskola, V.~J.~Kolhinen and P.~V.~Ruuskanen,
  Nucl.\ Phys.\ B {\bf 535} (1998) 351.

\bibitem{Eskola:1998df}
  K.~J.~Eskola, V.~J.~Kolhinen and C.~A.~Salgado,
  Eur.\ Phys.\ J.\ C {\bf 9} (1999) 61.

\bibitem{Hirai:2001np}
  M.~Hirai, S.~Kumano and M.~Miyama,
  Phys.\ Rev.\ D {\bf 64} (2001) 034003.

\bibitem{Hirai:2004wq}
  M.~Hirai, S.~Kumano and T.~H.~Nagai,
  Phys.\ Rev.\ C {\bf 70} (2004) 044905.

\bibitem{deFlorian:2003qf}
  D.~de Florian and R.~Sassot,
  Phys.\ Rev.\ D {\bf 69} (2004) 074028.

\bibitem{Eskola:2007my}
  K.~J.~Eskola, V.~J.~Kolhinen, H.~Paukkunen and C.~A.~Salgado,
    JHEP {\bf 0705} (2007) 002
  [arXiv:hep-ph/0703104];
  arXiv:0707.0060 [hep-ph];
  arXiv:0709.4525 [hep-ph].

\bibitem{Hirai:2007sx}
  M.~Hirai, S.~Kumano and T.~H.~Nagai,
  arXiv:0709.3038 [hep-ph].

\bibitem{Vogt:2004hf}
  R.~Vogt,
  Phys.\ Rev.\  C {\bf 70} (2004) 064902;
  Phys.\ Rev.\  C {\bf 71} (2005) 054902.


  \bibitem{satur}
  D.~Kharzeev, E.~Levin and L.~McLerran,
  Phys.\ Lett.\  B {\bf 561} (2003) 93;
  R.~Baier, A.~Kovner and U.~A.~Wiedemann,
  Phys.\ Rev.\ D {\bf 68}, 054009 (2003);
  D.~Kharzeev, Y.~V.~Kovchegov and K.~Tuchin,
  Phys.\ Rev.\ D {\bf 68} (2003) 094013;
  J.~L.~Albacete, N.~Armesto, A.~Kovner, C.~A.~Salgado and U.~A.~Wiedemann,
  Phys.\ Rev.\ Lett.\  {\bf 92}, 082001 (2004).



\bibitem{Hebecker:1999ej}
  A.~Hebecker,
  Phys.\ Rept.\  {\bf 331} (2000) 1
  [arXiv:hep-ph/9905226].

\bibitem{Mueller:2001fv}
  A.~H.~Mueller,
  arXiv:hep-ph/0111244.

\bibitem{feynmhibbs} R.~P.~Feynman and A.~R.~Hibbs, {\it Quantum Mechanics and Path Integrals}, McGraw-Hill (1965).

\bibitem{Nikolaev:1990ja}
  N.~N.~Nikolaev and B.~G.~Zakharov,
  Z.\ Phys.\  C {\bf 49} (1991) 607.

\bibitem{Kovchegov:1998bi}
  Y.~V.~Kovchegov and A.~H.~Mueller,
  Nucl.\ Phys.\  B {\bf 529} (1998) 451
  [arXiv:hep-ph/9802440].
  
\bibitem{Kovchegov:1999kx}
  Y.~V.~Kovchegov and L.~D.~McLerran,
  Phys.\ Rev.\  D {\bf 60} (1999) 054025
  [Erratum-ibid.\  D {\bf 62} (2000) 019901]
  [arXiv:hep-ph/9903246].
  
\bibitem{Baier:1996sk}
  R.~Baier, Y.~L.~Dokshitzer, A.~H.~Mueller, S.~Peigne and D.~Schiff,
  Nucl.\ Phys.\  B {\bf 484}, 265 (1997)
  [arXiv:hep-ph/9608322].

\bibitem{Salgado:2003gb}
  C.~A.~Salgado and U.~A.~Wiedemann,
  Phys.\ Rev.\  D {\bf 68} (2003) 014008
  [arXiv:hep-ph/0302184].

\bibitem{Zakharov:1996fv}
  B.~G.~Zakharov,
  JETP Lett.\  {\bf 63}, 952 (1996)
  [arXiv:hep-ph/9607440].

\bibitem{Baier:1998yf}
  R.~Baier, Y.~L.~Dokshitzer, A.~H.~Mueller and D.~Schiff,
  Phys.\ Rev.\  C {\bf 58} (1998) 1706
  [arXiv:hep-ph/9803473].

\bibitem{Wiedemann:2000tf}
  U.~A.~Wiedemann,
  Nucl.\ Phys.\  A {\bf 690}, 731 (2001)
  [arXiv:hep-ph/0008241].

\bibitem{Zakharov:1998sv}
  B.~G.~Zakharov,
  Phys.\ Atom.\ Nucl.\  {\bf 61} (1998) 838
  [Yad.\ Fiz.\  {\bf 61} (1998) 924]
  [arXiv:hep-ph/9807540].

\bibitem{Wiedemann:1999fq}
  U.~A.~Wiedemann and M.~Gyulassy,
  Nucl.\ Phys.\  B {\bf 560} (1999) 345
  [arXiv:hep-ph/9906257].

\bibitem{MehtarTani:2006xq}
  Y.~Mehtar-Tani,
  Phys.\ Rev.\  C {\bf 75} (2007) 034908
  [arXiv:hep-ph/0606236].

\bibitem{Gyulassy:1993hr}
  M.~Gyulassy and X.~n.~Wang,
  Nucl.\ Phys.\  B {\bf 420} (1994) 583
  [arXiv:nucl-th/9306003].

\bibitem{Armesto:2003jh}
N.~Armesto, C.~A.~Salgado and U.~A.~Wiedemann,
Phys. Rev. D {\bf 69} (2004) 114003;

\bibitem{Dokshitzer:2001zm}
Y.~L.~Dokshitzer and D.~E.~Kharzeev,
Phys.\ Lett.\ B {\bf 519}  (2001) 199;

\bibitem{Zhang:2003wk}
B.~W.~Zhang, E.~Wang and X.~N.~Wang,
Phys.\ Rev.\ Lett.\  {\bf 93} (2004) 072301;

\bibitem{Djordjevic:2003zk}
M.~Djordjevic and M.~Gyulassy,
Nucl.\ Phys.\ A {\bf 733} (2004) 265.

\bibitem{Eskola:2002kv}
  K.~J.~Eskola and H.~Honkanen,
  Nucl.\ Phys.\  A {\bf 713} (2003) 167
  [arXiv:hep-ph/0205048].

\bibitem{Dokshitzer:1978hw}
  Y.~L.~Dokshitzer, D.~Diakonov and S.~I.~Troian,
  Phys.\ Rept.\  {\bf 58}, 269 (1980).
%

\bibitem{Wang:2001if}
X.~N.~Wang and X.~f.~Guo,
Nucl.\ Phys.\ A {\bf 696}, 788 (2001).

\bibitem{Baier:2001yt}
  R.~Baier, Y.~L.~Dokshitzer, A.~H.~Mueller and D.~Schiff,
  JHEP {\bf 0109}, 033 (2001).

\bibitem{Armesto:2007dt}
  N.~Armesto, L.~Cunqueiro, C.~A.~Salgado and W.~C.~Xiang,
  arXiv:0710.3073 [hep-ph].

  \bibitem{Borghini:2005em}
  N.~Borghini and U.~A.~Wiedemann,
  arXiv:hep-ph/0506218;
  S.~Sapeta and U.~A.~Wiedemann,
  arXiv:0707.3494 [hep-ph].

  \bibitem{Kniehl:2000fe}
  B.~A.~Kniehl, G.~Kramer and B.~Potter,
  Nucl.\ Phys.\  B {\bf 582}, 514 (2000).

    \bibitem{Airapetian:2007vu}
  A.~Airapetian {\it et al.}  [HERMES Collaboration],
  Nucl.\ Phys.\  B {\bf 780} (2007) 1.

\bibitem{Eskola:2004cr}
  K.~J.~Eskola, H.~Honkanen, C.~A.~Salgado and U.~A.~Wiedemann,
  Nucl.\ Phys.\ A {\bf 747} (2005) 511.

\bibitem{Dainese:2004te}
  A.~Dainese, C.~Loizides and G.~Paic,
  Eur.\ Phys.\ J.\  C {\bf 38}, 461 (2005)
  [arXiv:hep-ph/0406201].

\bibitem{Adcox:2001jp}
%
S.~S.~Adler {\it et al.}  [PHENIX],
Phys.\ Rev.\ C {\bf 69} (2004) 034910;
%
%
J.~Adams {\it et al.}  [STAR],
Phys.\ Rev.\ Lett.\  {\bf 91} (2003) 172302;
%
B.~B.~Back {\it et al.}  [PHOBOS],
Phys.\ Lett.\ B {\bf 578} (2004) 297;
%
I.~Arsene {\it et al.}  [BRAHMS],
Phys.\ Rev.\ Lett.\  {\bf 91} (2003) 072305.
%
  M.~Shimomura  [PHENIX],
  nucl-ex/0510023.

\bibitem{Armesto:2005mz}
  N. Armesto, M.~Cacciari, A.~Dainese, C.~A.~Salgado and U.~A.~Wiedemann,
  Phys. Lett. B {\bf 637} (2006) 362;
  C.~A.~Salgado
  Nucl .Phys. A{\bf 783} (2007) 225.

\bibitem{Abelev:2006db}
  B.~I.~Abelev {\it et al.}  [STAR],
  Phys.\ Rev.\ Lett.\  {\bf 98} (2007) 192301.

\bibitem{Adare:2006hc}
  A.~Adare {\it et al.}  [PHENIX],
  Phys.\ Rev.\ Lett.\  {\bf 97} (2006) 252002.

\bibitem{Cacciari:2005rk}
  M.~Cacciari, P.~Nason and R.~Vogt,
  Phys.\ Rev.\ Lett.\  {\bf 95} (2005) 122001.

\bibitem{Renk:2006pk}
  T.~Renk and K.~J.~Eskola,
  Phys.\ Rev.\  C {\bf 75} (2007) 054910
  [arXiv:hep-ph/0610059].

\bibitem{Abreu:2007kv}
  S.~Abreu {\it et al.},
  arXiv:0711.0974 [hep-ph].

\bibitem{Salgado:2003rv}
  C.~A.~Salgado and U.~A.~Wiedemann,
  Phys.\ Rev.\ Lett.\  {\bf 93} (2004) 042301
  [arXiv:hep-ph/0310079].

\bibitem{Cacciari:2005hq}
  M.~Cacciari and G.~P.~Salam,
  Phys.\ Lett.\  B {\bf 641} (2006) 57
  [arXiv:hep-ph/0512210].
  
    \bibitem{Carminati:2004fp}
  F.~Carminati {\it et al.}  [ALICE Collaboration],
  J.\ Phys.\ G {\bf 30} (2004) 1517.

\bibitem{Alessandro:2006yt}
  B.~Alessandro {\it et al.}  [ALICE Collaboration],
  J.\ Phys.\ G {\bf 32}  (2006) 1295.

\bibitem{D'Enterria:2007xr}
  D.~d'Enterria {\it et al.} [CMS Collaboration],
  J.\ Phys.\ G {\bf 34} (2007) 2307.

\bibitem{atlas}
  P.~Steinberg  [ATLAS Collaboration],
  J.\ Phys.\ G {\bf 34} (2007) S527.

\bibitem{Adams:2006yt}
  J.~Adams {\it et al.}  [STAR],
  Phys.\ Rev.\ Lett.\  {\bf 97} (2006) 162301.

\bibitem{Adler:2005ee}
  S.~S.~Adler {\it et al.}  [PHENIX],
  Phys.\ Rev.\ Lett.\  {\bf 97} (2006) 052301;
  arXiv:0705.3238 [nucl-ex].

\bibitem{Polosa:2006hb}
  A.~D.~Polosa and C.~A.~Salgado,
  Phys.\ Rev.\  C {\bf 75} (2007) 041901.

\bibitem{conical}
 H.~Stoecker,
  Nucl.\ Phys.\ A {\bf 750}, 121 (2005);
J.~Casalderrey-Solana, E.~V.~Shuryak and D.~Teaney,
  hep-ph/0411315.

\bibitem{cherenkov}
  V.~Koch, A.~Majumder and X.~N.~Wang,
Phys.\ Rev.\ Lett.\  {\bf 96} (2006) 172302;
  I.~m.~Dremin,
  Sov.\ Phys.\ Usp.\  {\bf 25} (1982) 634
  [Usp.\ Fiz.\ Nauk {\bf 137} (1982) 749].

\bibitem{Maldacena:1997re}
  J.~M.~Maldacena,
  Adv.\ Theor.\ Math.\ Phys.\  {\bf 2}, 231 (1998)
  [Int.\ J.\ Theor.\ Phys.\  {\bf 38}, 1113 (1999)]
  [arXiv:hep-th/9711200].

\bibitem{Aharony:1999ti}
  O.~Aharony, S.~S.~Gubser, J.~M.~Maldacena, H.~Ooguri and Y.~Oz,
  Phys.\ Rept.\  {\bf 323}, 183 (2000)
  [arXiv:hep-th/9905111].

\bibitem{D'Hoker:2002aw}
  E.~D'Hoker and D.~Z.~Freedman,
  arXiv:hep-th/0201253.

\bibitem{Maldacena:1998im}
  J.~M.~Maldacena,
  Phys.\ Rev.\ Lett.\  {\bf 80}, 4859 (1998)
  [arXiv:hep-th/9803002].

\bibitem{Herzog:2006gh}
  C.~P.~Herzog, A.~Karch, P.~Kovtun, C.~Kozcaz and L.~G.~Yaffe,
  JHEP {\bf 0607}, 013 (2006)
  [arXiv:hep-th/0605158].

\bibitem{Liu:2006ug}
  H.~Liu, K.~Rajagopal and U.~A.~Wiedemann,
  Phys.\ Rev.\ Lett.\  {\bf 97}, 182301 (2006)
  [arXiv:hep-ph/0605178].

\bibitem{Liu:2006he}
  H.~Liu, K.~Rajagopal and U.~A.~Wiedemann,
  JHEP {\bf 0703}, 066 (2007)
  [arXiv:hep-ph/0612168].

\bibitem{CasalderreySolana:2006rq}
  J.~Casalderrey-Solana and D.~Teaney,
  Phys.\ Rev.\  D {\bf 74}, 085012 (2006)
  [arXiv:hep-ph/0605199].

\bibitem{lebellac}
  M. ~Le ~Bellac,
  ``Thermal Field Theory'', Cambridge University Press (1996)

\bibitem{McLerran:1981pb}
  L.~D.~McLerran and B.~Svetitsky,
  Phys.\ Rev.\  D {\bf 24}, 450 (1981).


\bibitem{Policastro:2002se}
  G.~Policastro, D.~T.~Son and A.~O.~Starinets,
  JHEP {\bf 0209}, 043 (2002)
  [arXiv:hep-th/0205052].

\bibitem{Herzog:2002pc}
  C.~P.~Herzog and D.~T.~Son,
  JHEP {\bf 0303}, 046 (2003)
  [arXiv:hep-th/0212072].

\bibitem{Gubser:2006bz}
  S.~S.~Gubser,
  Phys.\ Rev.\  D {\bf 74}, 126005 (2006)
  [arXiv:hep-th/0605182].

\bibitem{Moore:2004tg}
  G.~D.~Moore and D.~Teaney,
  Phys.\ Rev.\  C {\bf 71}, 064904 (2005)
  [arXiv:hep-ph/0412346].

\bibitem{CasalderreySolana:2007qw}
  J.~Casalderrey-Solana and D.~Teaney,
  JHEP {\bf 0704}, 039 (2007)
  [arXiv:hep-th/0701123].

\bibitem{CasalderreySolana:2007sw}
  J.~Casalderrey-Solana and X.~N.~Wang,
  arXiv:0705.1352 [hep-ph].
  
\bibitem{Baier:2002tc}
  R.~Baier,
  Nucl.\ Phys.\  A {\bf 715} (2003) 209
  [arXiv:hep-ph/0209038].
  
  \bibitem{Arnold:2002ja}
  P.~Arnold, G.~D.~Moore and L.~G.~Yaffe,
  JHEP {\bf 0206} (2002) 030;
S.~Jeon and G.~D.~Moore,
Phys.\ Rev.\ C {\bf 71} (2005) 034901.


\end{thebibliography}
\end{document}